\documentclass[final,3p,times,authoryear]{elsarticle}
\usepackage{amssymb}
\usepackage{amsmath}
\usepackage{array}
\usepackage{subcaption} 
\usepackage{float}
\usepackage{placeins}
\usepackage{xcolor}
\definecolor{mycolor}{RGB}{80,131,175}
\usepackage{subcaption}
\usepackage[
  colorlinks=true,   
  linkcolor=myblue,  
  urlcolor=myblue,  
  citecolor=myblue,  
  filecolor=myblue,  
  menucolor=myblue,  
  runcolor=myblue,   
  anchorcolor=myblue  
]{hyperref}
\usepackage{comment} 
\usepackage{enumitem} 
\usepackage{pifont} 
\usepackage{tabularx}
\usepackage{longtable}
\newcolumntype{P}[1]{>{\centering\arraybackslash}p{#1}}
\begin{document}
\begin{frontmatter}
\title{Multi-state Modeling of Delay Evolution in Suburban Rail Transports\tnoteref{nota}}
\author[label2]{Stefania Colombo}
\author[label2]{Alfredo Gimenez Zapiola}
\author[label2,label3]{Francesca Ieva}
\author[label2]{Simone Vantini}
\affiliation[label2]{organization={MOX, Department of Mathematics, Politecnico di Milano},
            addressline={Piazza Leonardo da Vinci 32}, 
            city={Milan},
            postcode={20133},
            country={Italy}}

\affiliation[label3]{organization={HDS, Health Data Science Center, Human Technopole},
            addressline={Viale Rita Levi-Montalcini 1}, 
            city={Milan},
            postcode={20157}, 
            country={Italy}}
\tnotetext[nota]{Corresponding author at: MOX, Department of Mathematics, Politecnico di Milano, Piazza Leonardo da Vinci 32, Milan, 20133, Italy.\\
\textit{E-mail address}: simone.vantini@polimi.it (S. Vantini)}

\begin{abstract}
Train delays are a persistent issue in railway systems, particularly in suburban networks where operational complexity is heightened by frequent services and high passenger volumes. Traditional delay models often overlook the temporal and structural dynamics of real delay propagation.

This work applies continuous-time multi-state models to analyze the temporal evolution of delay on the S5 suburban line in Lombardy, Italy. Using detailed operational, meteorological, and contextual data, the study models delay transitions while accounting for observable heterogeneity.

The findings reveal how delay dynamics vary by travel direction, time slot, and route segment. Covariates such as station saturation and passenger load are shown to significantly affect the risk of delay escalation or recovery. The study offers both methodological advancements and practical results for improving the reliability of rail services.
\end{abstract}

\end{frontmatter}

\section{Introduction}
Railway transport represent a fundamental component of urban and suburban mobility, especially in densely populated areas where it supports thousands of daily commuters. Among the most critical issues and persistent challenges that railway systems around the world continue to face is the occurrence of delays, which undermine service reliability, hinder passenger satisfaction, and disrupt wider economic and logistical systems. Delays are not isolated anomalies, but rather complex phenomena influenced by interdependent operational, infrastructural, environmental, and human factors (\cite{delayspec}). In recent years, the increasing demand for sustainable mobility has made the minimization and  management of delays a key strategic priority for public transport systems, particularly in large metropolitan regions. 

In the Italian context, the Lombardy region presents a particularly compelling case study. With over 10 million residents and a dense concentration of economic and commuting activity, it is home to the most extensive regional rail network in Italy. Trenord, the main regional operator, manages over 2,000 daily train services across suburban and regional lines, serving more than 460 stations along a 2,000 Km network and transporting approximately 200 million passengers per year (\cite{trenord}). Despite this strategic role, punctuality remains a persistent concern. Although some progress has been observed in recent years — with trains arriving within five minutes of the scheduled time, improved from 78\% in 2018 to 83\% in 2022 — Lombardy's suburban and regional services continue to fall short of Italy's national average of approximately 95\% (\cite{trenordreport}). More granular data from 2022 show that 99.79\% of trains were delayed by less than 60 minutes (same as in 2021), 0.19\% between 60 and 119 minutes, and 0.02\% experienced extreme delays beyond 120 minutes. These scenarios suggest that while extreme delays are rare, moderate delays and cancellations continue to occur at non-negligible rates, especially during peak hours and on congested segments. The persistence of such disruptions confirms the need for granular, context-sensitive analytical tools to model how delays evolve, escalate, or recover across time and space — motivating the work done in this paper.

Existing literature has addressed the problem of train delays using a wide variety of methodological approaches (\cite{categories, review_treni}). Traditional studies have focused on descriptive statistics or regression-based models to analyze historical delay patterns and identify correlations between delay events and potential influencing factors (\cite{descrittivo}). Over time, the scope of delay research has evolved encompassing more advanced techniques capable of capturing the non-linear and time-varying aspects of railway disruptions. For instance, data-driven approaches — particularly those based on supervised machine learning — have been widely applied for classification and prediction tasks (\cite{trees}). Meanwhile, event-driven models, such as those based on Markov chains (\cite{mc}), Bayesian networks (\cite{bayes}), or delay propagation graphs (\cite{graph}), have aimed to simulate the chain-reaction effects that one delay may have on subsequent services. Additionally, simulation methods have been used to replicate realistic delay scenarios in controlled environments, allowing analysts to test scheduling and re-routing interventions (\cite{optmicro}). Optimization-based frameworks, particularly in operational research, have been employed to improve the rescheduling of trains, minimize total system delay, and allocate infrastructure resources more efficiently (\cite{opt1}).

Building on these modeling frameworks, this work contributes to the study of train delays by applying a continuous-time multi-state modeling approach that has so far received limited attention in the railway domain. Several aspects characterize the contribution of this work. First, the proposed methodology considers a semi-Markovian formulation for arrival delays, explicitly modeling and predicting their progression across a finite set of discrete states. Two alternative state-space formulations are considered, involving three and four delay categories, respectively. Second, all analyses are conducted separately for each direction of travel, applying a further segmentation based on temporal and spatial characteristics of the line, to account for structural heterogeneity. Third, the study explores two levels of model complexity: nonparametric and semi-parametric. In the former case, models are estimated separately for each time interval and each route segment, without including covariates. In the latter, two semi-parametric models are developed: one incorporating time slots covariates, and the other incorporating route segments, along with additional observed variables such as passenger load, weather conditions, and station saturation. These models allow a more detailed understanding of delay dynamics across different temporal and spatial contexts, assuming the same baseline hazards. To the best of our knowledge, this is among the first studies to apply multi-state models in this specific setting, particularly in the context of suburban rail services in Italy. The proposed method offers both methodological value and practical insights into the temporal dynamics of delays, providing a flexible and interpretable tool for further research in railway performance modeling and for supporting data-informed improvements in service reliability.

The rest of the paper is organized as follows: Section~\ref{Literature review} provides a critical overview of existing modeling approaches for railway delays, with emphasis on stochastic frameworks; Section~\ref{Data} details the dataset and the segmentation strategies adopted in this study; Section~\ref{Methods} introduces the theoretical basis for multi-state models, including the mathematical formalism, assumptions, and estimation procedures for the nonparametric and the Cox-based variants; Section~\ref{Results} presents both empirical findings, organized by model type and segmentation, and discusses practical applications of the models including potential use in real-time planning and decision making; Section~\ref{Conclusion} summarizes key findings,  highlighting their strengths and limitations and suggesting directions for future work.

\section{Literature Review}
\label{Literature review}
This section provides a critical overview of existing modeling frameworks for train delays, highlighting their strengths and limitations, and justifying the introduction of the approach presented in this work. A structured examination of the theoretical background on this topic can be undertaken by considering three main dimensions: the level of analysis adopted, the main objectives pursued, and the methodological approaches used.

The first feature covers the extent between microscopic and macroscopic levels. While the former focuses  on individual components — such as specific trains, trips, or stations — to study punctuality and delay propagation at the local level (\cite{micro1, micro2, micro3, micro4}), the latter considers aggregate performance indicators across networks, time-frames, or service types to provide valuable insights into structural inefficiencies and system-wide trends (\cite{macro1, macro2}).

The purpose guiding each analysis contributes to further differentiating the literature. A first group of studies is primarily descriptive, examining the distribution of delays across time, space and train typologies, often with the aim of monitoring performance and informing long-term planning (\cite{descrittivo}). A second stand emphasizes predictive modeling, leveraging historical data to anticipate future delays and improve real-time decision-making (\cite{predizione}). Closely related are investigations into causal mechanisms — both endogenous (e.g., congestion, rolling stock availability) and exogenous (e.g., weather events or incidents) — which provides the basis for targeted interventions (\cite{cause}). An additional line of inquiry explores the propagation of delays through interconnected networks, capturing their temporal and spatial dynamics (\cite{macro2}). Finally, a more applied stream of research focuses on optimization, developing models to improve timetabling and resource allocation in order to strengthen system resilience and mitigate the consequences of delays (\cite{micro1}).

From a methodological standpoint, a broad range of tools has been employed, mirroring the multifaceted nature of railway operations (\cite{review_treni, categories}). 
Data-driven approaches, and in particular supervised machine learning techniques, have become predominant in predictive applications, thanks to the availability of extensive operational datasets. Regression models (\cite{micro4}), decision trees (\cite{trees}), and support vector machines (\cite{svm}) stand out for their predictive accuracy, even though this often comes at the expense of interpretability.
\\At the same time, event-driven models focus on examine causal dependencies and the mechanisms of delay propagation. Techniques such as time-event graphs (\cite{graph}), Markov chains (\cite{mc}), and Bayesian networks (\cite{bayes}) are adopted to disruptions unfold through the system by capturing interactions among train arrivals, departures, and incidents. While these methods tend to be more interpretable than data-driven ones, they may face difficulties in handling nonlinear or high-dimensional dynamics.
\\Simulation tools — including OpenTrack, RailSys, and PROTON — are also widely used to reproduce the operational behavior of railway systems under a variety of scenarios (\cite{optmicro, optmacro}). They allow analysts to explore issues such as network congestion, train scheduling, and risk assessment, offering a realistic representation of system dynamics. Nonetheless, their use is often constrained by heavy computational demands and the simplifications required to keep models tractable.
\\Finally, optimization-based methods, such as mixed-integer programming and max-plus algebra, are commonly applied to enhance timetabling, resource allocation, and overall system robustness (\cite{opt1, opt2, opt3}). These models are designed to minimize delays and improve efficiency, particularly during planning phases. However, their applicability is frequently limited by computational complexity and by assumptions that may reduce their scope.

Building on this overview, the discussion now focuses on methodological frameworks that share a similar foundation to the approach adopted in this work, with particular emphasis on Markov chains and survival models.
\\In train delay prediction, Markov models are widely used to describe the stochastic progression of delays across a train's journey. Early applications employed stationary Markov chains to estimate steady-state delay probabilities and improve timetable robustness by defining states based on arrival and departure times (\cite{assessing}). However, the assumption of stationarity limits applicability in dynamic and complex railway systems. Non-stationary Markov chains have therefore been introduced to account for time-varying transition probabilities, showing improved predictive accuracy and reflecting real-world patterns of delay prediction, including deterioration, recovery, and state maintenance (\cite{delay, exploring}).
\\Further refinements focus on enhancing model granularity. Some approaches model deviations in running and dwelling times rather than absolute delays, allowing more accurate predictions in scenarios with cumulative delays, albeit with reduced interpretability (\cite{process}). The definition of the state space is also critical: flexible or "elastic" state spaces enable models to capture more detailed information, significantly improving predictive performance (\cite{elastic}).
\\Besides improving delay predictions, there is growing interest in understanding the causes of delays, especially those driven by external factors such as weather conditions. In (\cite{top1}), a Cox model is employed to analyze primary delays while a Markov chain model describes arrival delays, allowing the assessment of transition intensities between delay and non-delay states. Environmental variables such as temperature and humidity are found to significantly influence both the occurrence of primary delays and the dynamics of delay propagation, highlighting the value of including weather in predictive models. Building on this approach, (\cite{top2}) extends the analysis to investigate high-speed passenger trains in northern Sweden, using both a stratified Cox model and a heterogeneous Markov chain model to capture time-varying risks. The study confirms the substantial impact of weather variables — temperature, humidity, snow depth and ice/snow precipitation — on train performance and demonstrates how these combined models more effectively capture variability in delays.

Overall, the literature demonstrates that all the presented methods offer powerful tools for analyzing and predicting train delays, particularly when incorporating external factors and dynamic system characteristics. Nevertheless, several methodological and practical challenges remain, motivating the methodological positioning and contributions presented in this work.

\section{Data}
\label{Data}
This section reports the data employed in this study, providing a detailed account of their origins, structural characteristics, and the rationale behind their selection.

\subsection{Data Overview and Provenance}
\label{Data Overview and Provenance}
\begin{figure}[H]
    \centering
    \includegraphics[width=1\textwidth]{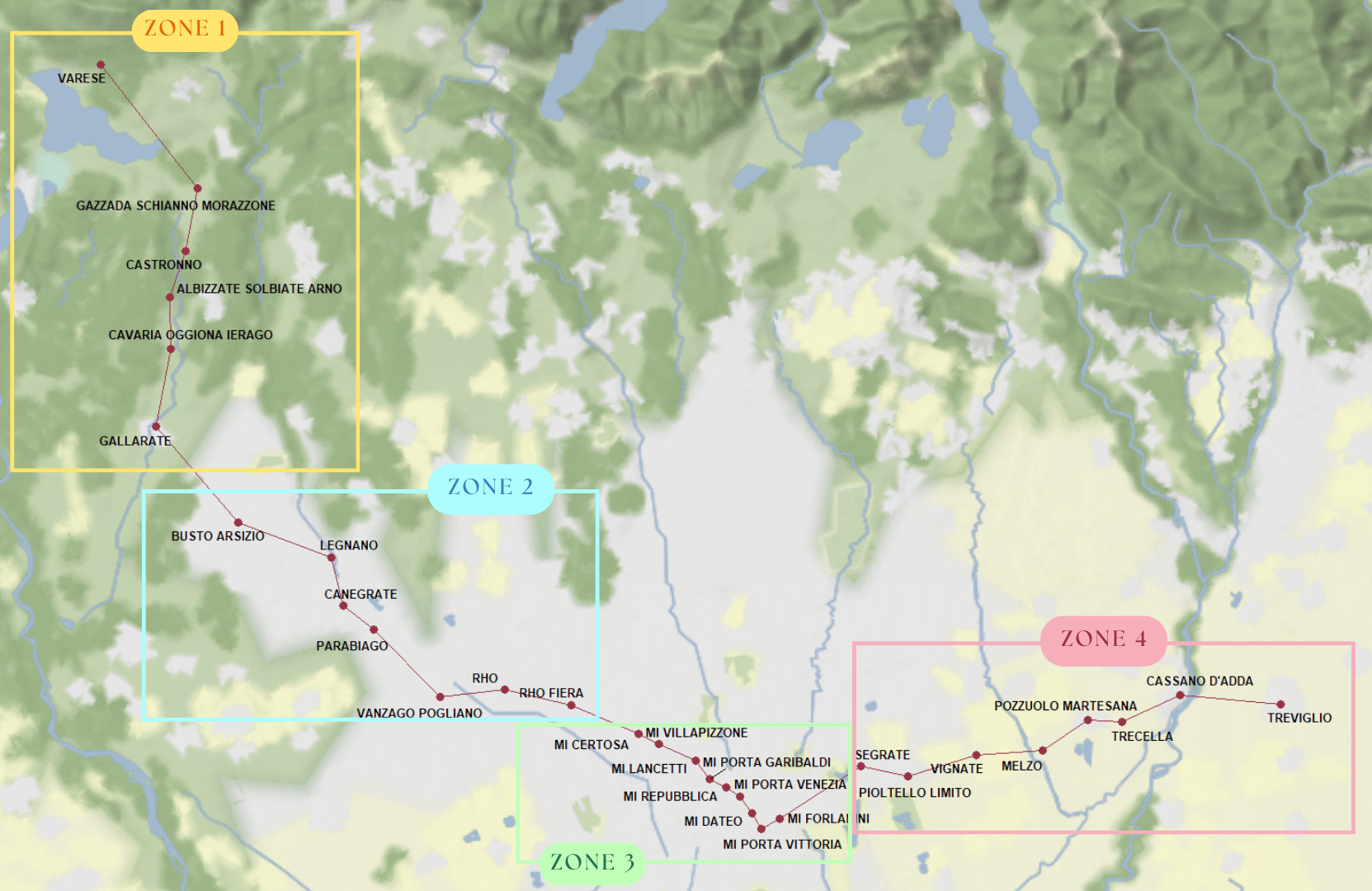}
    \caption{Overview of the S5 line. In Zone 3, the stations whose names start with "MI" (e.g., MI Certosa) are located within Milano and represent the internal stations of the Milan core. For more details about the zone classification see Section~\ref{Data Segmentation Strategy}.}
    \label{fig1}
\end{figure}

\paragraph{Trenord Dataset}
Based in Lombardy, Northern Italy, Trenord is the primary operator of suburban and regional railway services in the region. Beyond its core activities within Lombardy, the company also manages selected lines that extend into adjacent regions. The dataset used in this study is obtained directly from Trenord and contains detailed information comprising scheduled train departure and arrival times, delay records, and passenger flow data gathered from the Automatic Passenger Counting (APC) system, which monitors the number of passengers boarding and alighting at each station of the route.
\\The scope of the analysis is restricted to the S5 subruban railway line, the longest within the Lombardy suburban network, covering approximately 97 kilometers from Varese to Treviglio. Starting in Varese — an Italian city of about 79,000 inhabitants, located 55 km northwest of Milano near the Swiss border and known for its lakes, hills, and strong industrial tradition (\cite{varese1, varese2}) — the line traverse eastward across the region. It passes through central Milano, the main urban and economic core in Lombardy, where it serves nine stations, before reaching Treviglio, a town of around 31,000 residents in the province of Bergamo. Treviglio occupies a strategic position at the intersection of major road and railway routes connecting it to other key Lombard cities (\cite{treviglio1}). Hence, serving as a crucial component of the regional transport system, the S5 provides a valuable setting for investigating passenger flows and service dynamics. With a total of thirty stations distributed along a diverse commuter corridor, and operating every 30 minutes in both directions, it offers a convenient and reliable connection across one of the most densely populated and economically dynamic areas of Northern Italy (\cite{s5trenord}).
A geographical layout of the route is reported in Figure~\ref{fig1}.

The empirical focus is placed on data collected during September, October, and November 2023, considering only arrival delays at each station, as they encapsulate the effects of earlier disruptions and correspond to the metrics most relevant for both passengers and operators. To reflect typical commuting behavior, the study considers only weekdays, explicitly excluding weekends and Italian public holidays. Furthermore, the analysis is confined to the interval between 6:00 a.m. and 8:00 p.m., corresponding to the period of highest passenger demand and operational relevance. Moreover, only regular, scheduled train stops are included in the study, deliberately omitting temporary or special services in order to minimize variability and ensure a consistent foundation for the analysis.

\paragraph{Meteorological Data}
Meteorological data are retrieved from \textit{IlMeteo.it}, a widely used Italian weather forecasting service. Covering the period from September to November 2023, the dataset encompasses daily values of temperature, wind speed, and visibility, as well as the presence of disruptive weather events such as rain, fog, or storms. \\However, weather information does not refer to individual stations, but is instead linked to broader segments of the S5 line. Specifically, conditions are assigned to the line segments according to three main airports situated alond the S5 corridor: Malpensa (Varese-Legnano), Linate (Canegrate-Pioltello Limito), and Orio al Serio (Vignate-Treviglio). Each stop is therefore assigned to the airport closest in distance. Although this approach reduces spatial precision, the fact that many stations are located relatively close to each other makes it reasonable to assume that airport-level data provide a reliable proxy for the actual weather conditions experienced along the line.

\paragraph{Train Frequency Records}
A third, custom-built dataset is developed to capture train frequency at each stop. For every station along the S5 line, the number of trains passing within a one-hour window is recorded. This measure acts as a proxy for service intensity, complementing Trenord's records with additional contextual information on operational pressure and its potential effects in delays and passenger flows.
\\A deeper description of all the variables involved in this study is available in the~\ref{app1}.

\subsection{Data Segmentation Strategy}
\label{Data Segmentation Strategy}
To improve both the interpretability and flexibility of the subsequent multi-state modeling, the dataset is partitioned along multiple dimensions. This approach, also adopted in the literature (\cite{exploring}), enables separate analyses across distinct operational contexts, allowing heterogeneous delay dynamics to emerge and localized behaviors to be detected that might otherwise remain hidden in an aggregated setting.

The primary and most fundamental categorization divides train runs based on their travel direction:
\begin{itemize}
     \item Direction 0: from Varese to Treviglio;
    \item Direction 1: from Treviglio to Varese.
\end{itemize}

All subsequent partitioning are applied independently within each directional subset.
To account for temporal variations in train performance related to traffic volume and passenger load, each train run is classified according to its departure time from the origin station into one of the following time bands:
\begin{enumerate}
    \item [$-$]  Morning-Peak Time: from 6:00 a.m. to 10:00 a.m.;
    \item [$-$] Off-Peak Time: from 10:00 a.m. to 4:00 p.m.;
    \item [$-$] Evening-Peak Time: from 4:00 p.m. to 8:00 p.m..
\end{enumerate}

\newpage
In addition, since the S5 line traverses a geographically and operationally diverse territory, the route is partitioned into four macro-areas (see Figure~\ref{fig1}):
\begin{enumerate}
    \item [$-$] Zone 1: from Varese to Gallarate;
    \item [$-$] Zone 2: from Busto Arsizio to Rho Fiera;
    \item [$-$] Zone 3: stations within the municipality of Milano, from Milano Certosa to Milano Forlanini;
    \item [$-$]
    Zone 4: from Segrate to Treviglio. 
\end{enumerate}
To be more precise, the line was initially divided into three macro-areas — Pre-Milano, Milano-core, and Post-Milano — based on a high-level geographical intuition. However, preliminary analysis revealed distinct delay patterns within the segments between Varese and Gallarate, and between Busto Arsizio and Rho Fiera, indicating sufficiently different behaviors to justify a further refinement. Consequently, the original partition was adjusted to better capture this local heterogeneity. 

\begin{figure}[H]
    \centering
    \includegraphics[width=0.9\textwidth]{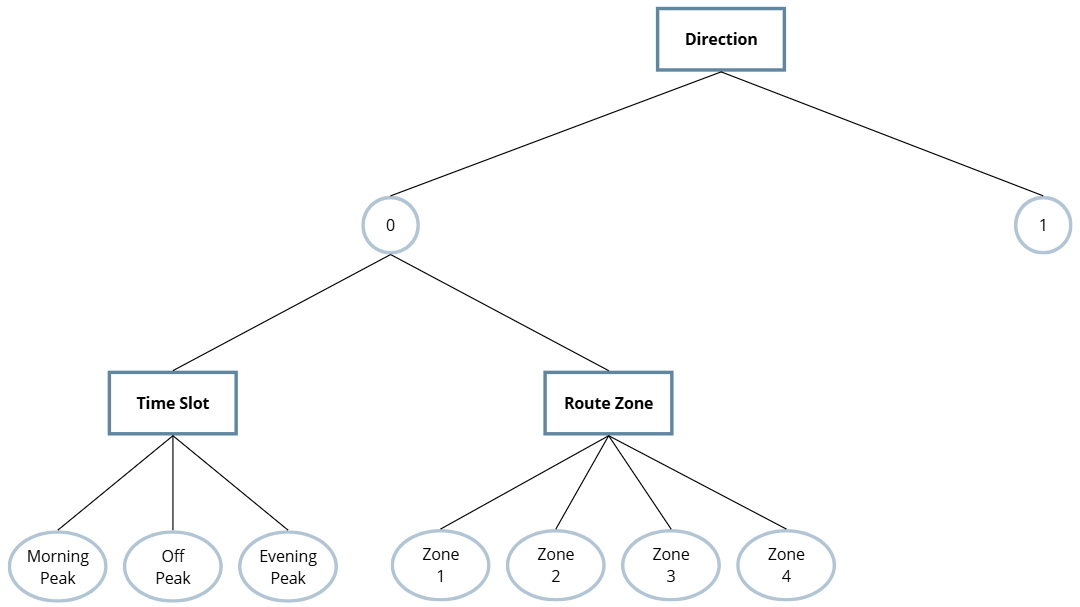}
    \caption{Diagram of the segmentations adopted in this study.}
    \label{fig2}
\end{figure}

In order to provide a comprehensive overview of the adopted segmentation strategy, Figure~\ref{fig2} offers a visual summary of the categories introduced above.

\section{Methods}
\label{Methods}
This section introduces the modeling framework adopted for the analysis of train delays as a multi-state process. It provides both the conceptual setup — through the definition of states and the observational unit — and the mathematical formulation of the model in its nonparametric and semi-parametric specifications.

\subsection{State Space Definition}
\label{State Space Definition}
To operationalize the multi-state framework in the context of delay classification, it is necessary to define the discrete states that characterize the underlying process. In this analysis, two alternative state spaces are considered, each reflecting a distinct categorization scheme for delay severity. The aim is to examine the implications of adopting a coarser versus a more granular discretization when modeling the temporal evolution of delay patterns. 
\newpage
Specifically, we define:
\begin{itemize}
     \item [\ding{111}] The first specification as a three-state model:
    \[\mathcal{S}^{(1)} = \{\text{On} \hspace{0.10cm} \text{Time}, \hspace{0.10cm} \text{Mild} \hspace{0.10cm} \text{Delay}, \hspace{0.10cm} \text{Severe} \hspace{0.10cm} \text{Delay}\}\]
    where states are defined according to a delay threshold parameter \(\tau\):
    \begin{itemize}
        \item \(\text{On} \hspace{0.10cm} \text{Time}\): delay \(\leq \tau_1\);
        \item \(\text{Mild} \hspace{0.10cm} \text{Delay}\): \(\tau_1 <\) delay \(\leq \tau_2\);
        \item \(\text{Severe} \hspace{0.10cm} \text{Delay}\): delay \(> \tau_2\).
    \end{itemize}

    \item [\ding{111}] The second specification introduces an additional intermediate category, yielding a four-state structure:
    \[\mathcal{S}^{(2)} = \{\text{On} \hspace{0.10cm} \text{Time}, \hspace{0.10cm} \text{Mild} \hspace{0.10cm} \text{Delay}, \hspace{0.10cm} \text{Medium} \hspace{0.10cm} \text{Delay}, \hspace{0.10cm} \text{Severe} \hspace{0.10cm} \text{Delay}\}\]
    with the following classification:
    \begin{itemize}
        \item \(\text{On} \hspace{0.10cm} \text{Time}\): delay \(\leq \tau_1\);
        \item \(\text{Mild} \hspace{0.10cm} \text{Delay}\): \(\tau_1 <\) delay \(\leq \tau_2\);
        \item \(\text{Medium} \hspace{0.10cm} \text{Delay}\): \(\tau_2 <\) delay \(\leq \tau_3\);
        \item \(\text{Severe} \hspace{0.10cm} \text{Delay}\): delay \(> \tau_3\).
    \end{itemize}
\end{itemize}
In this application, threshold are fixed at \(\tau_1 = 5\), \(\tau_2 = 10\), and \(\tau_3 = 15\) minutes. These values are consistent with operational benchmarks commonly adopted in railway punctuality assessment.

\begin{figure}[H]
    \centering    \includegraphics[width=0.6\textwidth]{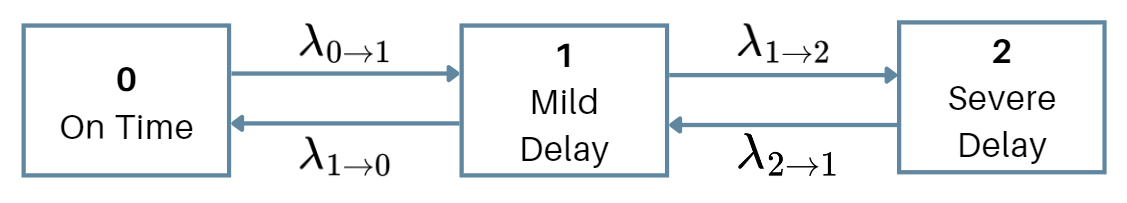}
    \caption{Graphical representation of the three-state space model.}
    \label{fig3}
\end{figure}

\begin{figure}[H]
    \centering    \includegraphics[width=0.85\textwidth]{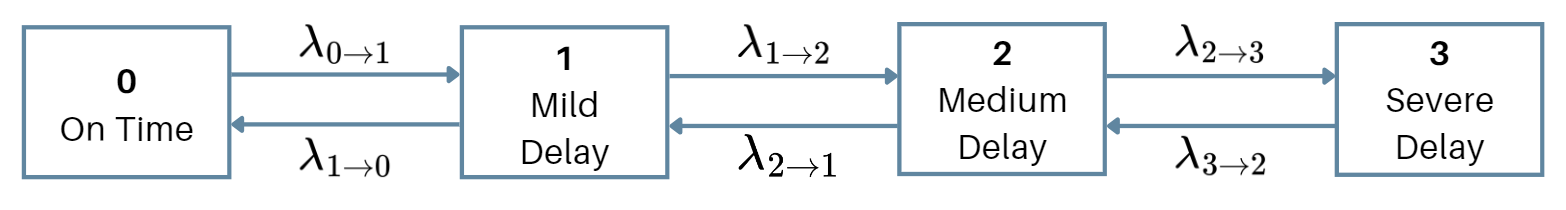}
    \caption{Graphical representation of the four-state space model.}
    \label{fig4}
\end{figure}

In both cases, the states are totally ordered, representing increasing levels of disruption, and the corresponding models allow for both forward and backward transitions, with no absorbing state. Figures~\ref{fig3} and~\ref{fig4} provide a schematic illustration of the two specifications.

These two formulations reflect complementary modeling perspectives: the three-state model prioritizes parsimony and interpretability, whereas the four-state model affords a more detailed representation of delay dynamics. Each model is estimated separately to assess the sensitivity of results to the state space definition. Notably, while the three-state formulation is applied across all models considered in this work, the four-state version is used exclusively within the nonparametric setting, where greater model flexibility accommodates the increased number of transitions resulting from the finer state resolution.

\subsection{Model Setup}
\label{Model Setup}
\begin{figure}[H]
    \centering    \includegraphics[width=0.8\textwidth]{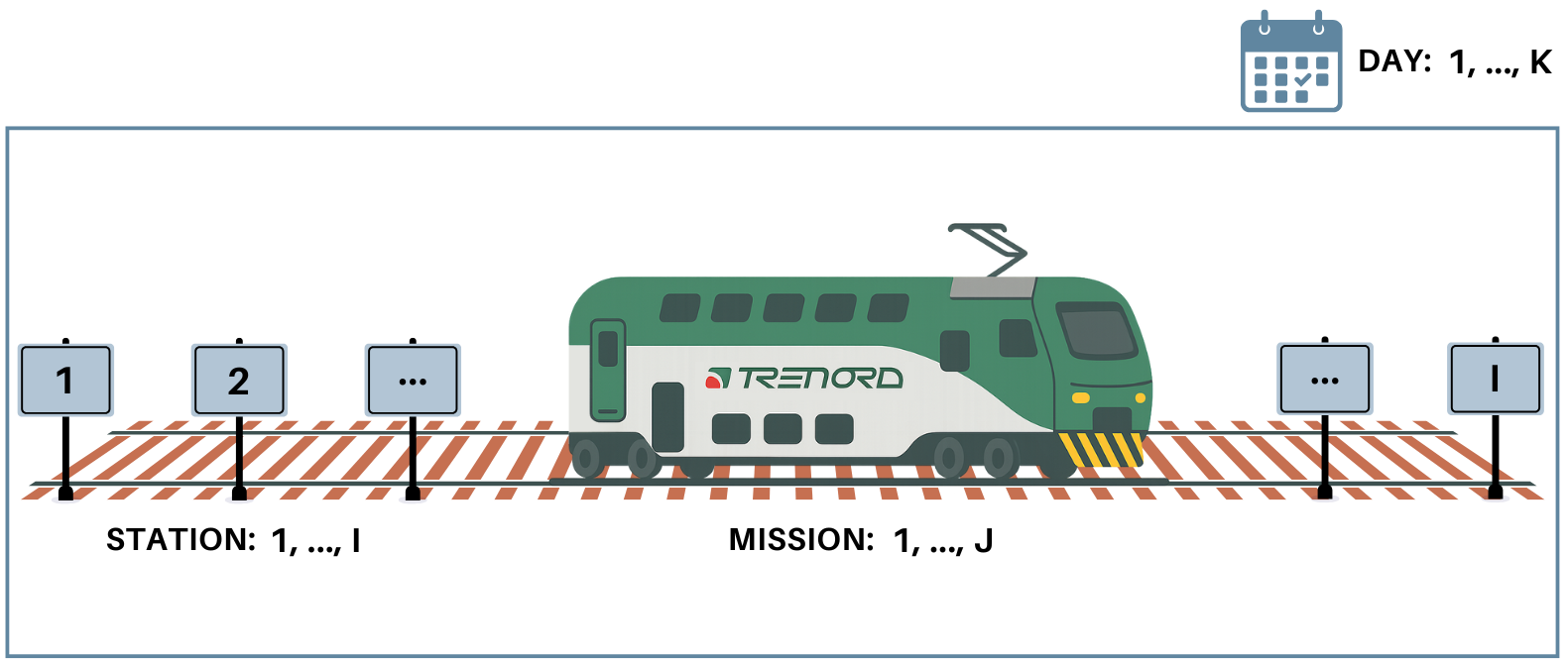}
    \caption{Schematic representation of the observational unit.}
    \label{fig5}
\end{figure}

Rather than considering a train as the basic observational entity, the analysis adopts a stop-oriented granularity. Observations are defined by the triplet \((i, j, k)\), where:
\begin{itemize}
    \item \(i\) denotes the station (or stop) at which the observation is recorded;
    \item \(j\) identifies the mission, i.e., a specific train journey;
    \item \(k\) refers to the calendar day of the observation.
\end{itemize}
Accordingly, each unit \((i,j,k)\) represents a unique instance of a train arriving at a given station on a specific day, as shown in Figure~\ref{fig5}. Such a formulation enables the model to account simultaneously for spatial and temporal heterogeneity in delay dynamics. This unit definition is carried forward consistently across all modeling strategies.

\subsection{Mathematical Formulation}
\label{Mathematical Formulation}
Multi-state models extend classical survival analysis to processes in which individuals or units transition among a finite set of discrete states over time. Each state typically represents a conceptually or clinically distinct condition, while transitions correspond to events of interest. These models are particularly useful when multiple event types can occur or when intermediate states are relevant before reaching a terminal or absorbing state. Foundational treatments and theoretical details are available in (\cite{bibbia, eventhistory, review, tutorialbio, mstate}).

\begin{figure}[H]
    \centering    \includegraphics[width=1\textwidth]{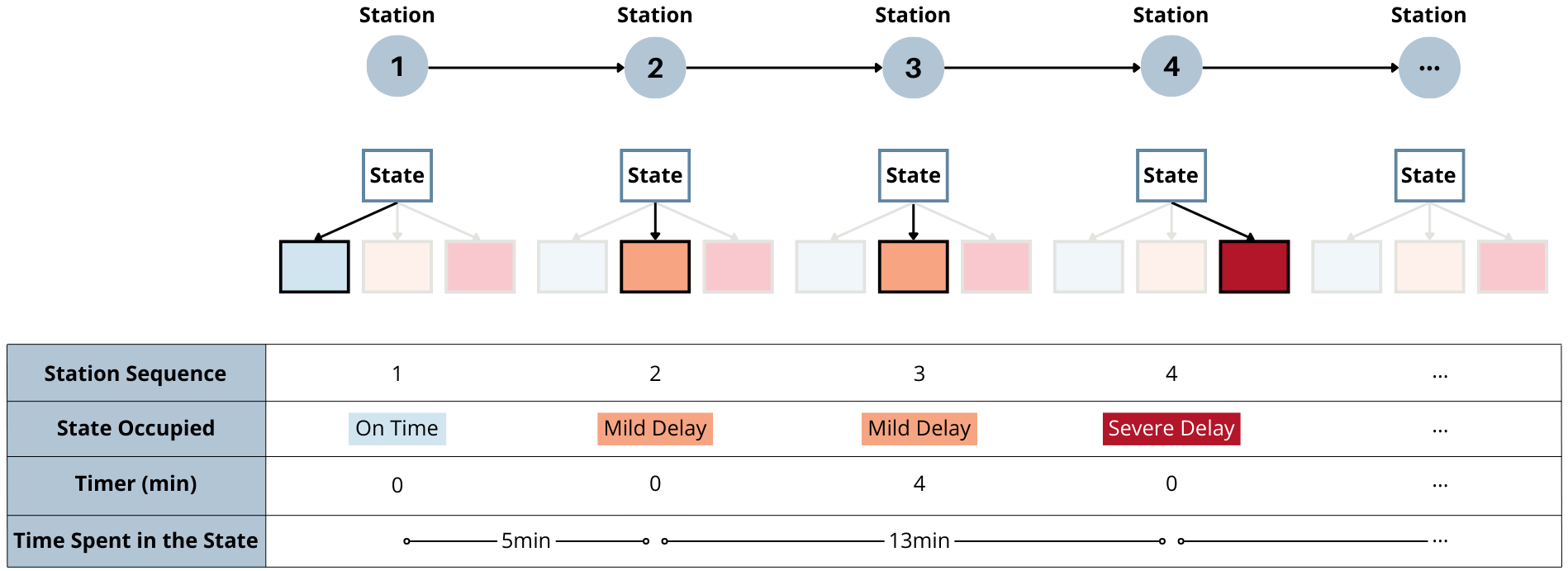}
    \caption{Schematic example of the station and state sequences for a specific train path on a specific day, highlighting the clock-reset time scale. The time counter resets to zero upon entering new state, and the duration spent in each state is measured relative to the corresponding entry time.}
    \label{fig6}
\end{figure}

Formally, let \(X(t)\) denote the state occupied by a unit at time \(t\). A multi-state model is defined as a stochastic process \((X(t), t\in \mathcal{T})\) characterized by:
\begin{itemize}
    \item A finite state space \(\mathcal{S}=\{1, \dots, S\}\), representing all possible states of the system;
    \item Right-continuous sample paths, \(X(t^+)=X(t)\);
    \item A time domain \(\mathcal{T}=[0, \tau]\) or \([0, \tau)\), where \(\tau \leq \infty\) denotes the end of follow-up;
    \item An initial distribution \(\pi_r(0) = \mathbb{P}(X(0) = r)\), specifying the probability of starting in each state;
    \item A history process \(\mathcal{X}_t\), representing the accumulated information about the path up to time \(t\).
\end{itemize}
Having specified the structure of the multi-state process, an important modeling choice concerns the time scale used to describe state transitions. As different scales lead to different interpretations of transition intensities, it is necessary to clarify how time is measured in our framework. Following the classical distinction in (\cite{tutorialbio}), two conventions are commonly adopted: the clock-forward and the clock-reset approaches. In the clock-forward setting, time \(t\) represents the elapsed time since the beginning of follow-up, or train run in this specific scenario; the clock increases monotonically regardless of state transitions. Conversely, in the clock-reset framework — which is the one adopted in this work — time is measured relative to the moment of entry into the current state. Thus, every transition triggers a reset of the temporal counter, and the duration accumulated in each state corresponds to the backward recurrence time. Figure~\ref{fig6} provides a schematic example: the upper row shows the sequence of visited states (and the corresponding stations), while the lower panel highlights how the clock resets to zero at each new state entry and records only the duration time of each state.

This general framework provides the basis for the subsequent specification of nonparametric and semi-parametric multi-state models for the S5 line, where transition probabilities and intensities are formally defined and estimated.

\subsubsection{Nonparametric Models}
\label{Nonparametric Models}
Let a transition from state \(r\) to state \(s\) be denoted by \(r \to s\), and let \(u\) represent the time spent in the current state \(r\). The transition intensity corresponding to a specific station \(i\), mission \(j\), and calendar day \(k\) is denoted by \( \lambda_{r \to s}^{(i,j,k)}(u) \). We assume homogeneity across statistical units, meaning that:
\begin{equation}
    \lambda_{r \to s}^{(i,j,k)}(u) = \lambda_{r \to s}(u), \qquad \forall\,i,j,k.
\end{equation}
In the semi-Markov framework adopted here, also referred to as clock-reset model (see Section~\ref{Mathematical Formulation}), the transition intensity depends on the current state \(r\) and the time \(u\) already spent there, and is defined as:
\begin{equation}
    \lambda_{r \to s}(u) := \lim_{\Delta u \to 0} \frac{\mathbb{P}(X(T + u + \Delta u) = s\,\big|\, X(T) = r,\, X(v) = r, \forall v \in [T, T + u])}{\Delta u},
\end{equation}
where \(T\) is the entry time into state \(r\). The cumulative hazard associated with this transition is computed as the integral of the transition intensity over time spent in state \(r\):
\begin{equation}
    \Lambda_{rs}(u) = \int_0^u \lambda_{r \to s}(v)\, dv.
\end{equation}
Considering \( \boldsymbol{\Lambda}(u) \), the \( S \times S \) matrix, where \( S \) is the number of possible states, the off-diagonal elements are given by \(\Lambda_{rs}(u) \) for \( r \neq s \), while the diagonal elements satisfy the mass conservation:
\begin{equation}
    \Lambda_{rr}(u) = - \sum_{s \neq r} \Lambda_{rs}(u).
\end{equation}

As discussed in Section~\ref{Data Segmentation Strategy}, directional heterogeneity is addressed by estimating separate cumulative hazards for each travel direction. Then, within each direction, a further segmentation is introduced along two dimensions:
\begin{itemize}
    \item Time Slot. Direction- and time-specific cumulative hazards are defined as:
    \begin{equation}
    \Lambda_{rs}^{(d, t)}(u) = \int_0^u \lambda_{r \to s}^{(d, t)}(v) \, dv, \qquad d\in\{0,1\}, \qquad t\in\{\text{Morning-Peak}, \text{Off-Peak}, \text{Evening-Peak}\}.
    \end{equation}
    \item Route Section. Direction- and zone-specific cumulative hazards are defined as:
    \begin{equation}
    \Lambda_{rs}^{(d, z)}(u) = \int_0^u \lambda_{r \to s}^{(d, z)}(v) \, dv, \qquad d\in\{0,1\}, \qquad z\in\{\text{Zone 1}, \text{Zone 2}, \text{Zone 3}, \text{Zone 4}\},
    \end{equation}
\end{itemize}
In both specifications, estimation is carried out independently for each direction, under the assumption of homogeneity within time slots and route sections, respectively.

\paragraph{Estimation}
For ease of notation, the model-indices are omitted in the following expressions whenever no ambiguity arises. Assuming independent trajectories subject to right censoring with a non-informative scheme, the Nelson-Aalen estimator for \(\Lambda_{rs}(u) \) is:
\begin{equation}
    \widehat{\Lambda}_{rs}(u) = \sum_{u_k \leq u} \frac{dN_{rs}(u_k)}{Y_r(u_k)},
    \label{cumhaz}
\end{equation} 
where: \( u_k \) denotes the observed event times; \( dN_{rs}(u_k) \) the number of observed \( r \to s\) transitions at time \( u_k \); \( Y_r(u_k) \) the number of individuals still in state \( r \) at time \( u_k \) and, hence, at risk of transitioning. The estimators for the diagonal elements of \( \Lambda(u) \) are given by:
\begin{equation}
    \widehat{\Lambda}_{rr}(u) = - \sum_{s \neq r} \widehat{\Lambda}_{rs}(u).
\end{equation}
Finally, the cumulative hazard estimates obtained for each transition are used to compute the full transition probability matrix \( \mathbf{P}(v, t) \), which describes the probability of moving between states over time. This is done using the Aalen–Johansen estimator, a matrix-valued extension of the Kaplan–Meier estimator, defined as:
\begin{equation}
    \widehat{\mathbf{P}}(v, t) = \prod_{(v, t]} \left( \mathbf{I} + d\widehat{\boldsymbol{\Lambda}}(u) \right),
    \label{prob}
\end{equation}
where \( \widehat{\boldsymbol{\Lambda}}(u) \) is the estimated cumulative hazard matrix. Off-diagonal elements correspond to estimated transition hazards \( \widehat{\Lambda}_{rs}(u) \), while diagonal entries are defined to ensure that each row of the matrix sums to zero, preserving the stochastic structure of the process.

\subsubsection{Semi-parametric Models}
\label{Semi-parametric Models}
The previous specification does not account for potentially relevant factors that may influence transition intensities. To address this limitation, we introduce covariate effects by adopting a semi-parametric specification based on the Cox proportional hazards framework. For each transition \(r \to s\) and for each travel direction \(d\), the hazard function is defined as:
\begin{equation}
    \lambda_{r \rightarrow s}^{(i,j,k)(d)}(u) =\lambda_{r \to s}^{(d)} (u \mid \mathbf{Z}_{i,j,k}) = \lambda_{r \to s}^{0 (d)} (u) \exp\big( \boldsymbol{\beta}_{r \to s}^{(d) \top} \mathbf{Z}_{i,j,k} \big), \qquad d\in\{0,1\},
\end{equation}
where:
\begin{itemize}
    \item \( \lambda_{r \to s}^{0  (d)} (u) \) is the unspecified baseline hazard function specific to each transition and direction;
    \item \( \mathbf{Z}_{i,j,k} \) is the vector of time-fixed covariates associated with the statistical unit;
    \item \( \boldsymbol{\beta}_{r \to s}^{(d)} \) is the transition- and direction-specific vector of regression coefficients.
\end{itemize}
To capture potential asymmetries in delay dynamics across directions, separate hazard models are estimated for the two directions of travel:
\begin{equation}
    \lambda_{r \rightarrow s}^{(i,j,k)(d)}(u) =\lambda_{r \to s}^{(d)} (u \mid \mathbf{Z}_{i,j,k}) = \lambda_{r \to s}^{0 (d)} (u) \exp\big( \boldsymbol{\beta}_{r \to s}^{(d) \top} \mathbf{Z}_{i,j,k} \big), \qquad d\in\{0,1\},
\end{equation}

\paragraph{Estimation} 
As previously done, unless otherwise specified, the model-indices are omitted. The regression coefficients \(\beta_{r \to s}\) are estimated by partial likelihood maximization. For notational convenience, we collapse the station, mission, and day indices \((i,j,k)\) into a single index \(p\), so that each unit is uniquely identified by \(p\). Accordingly, covariates, risk indicators, and counting processes are denoted by \(p\) as \(\mathbf{Z}_p\), \(Y_{rp}(t)\), and \(N_{rp}(t)\), respectively.
In the counting process formulation (\cite{bibbia}), the partial likelihood combining all transitions and units is:
\begin{equation}
    \mathcal{L}(\boldsymbol{\beta}) \propto \prod_{t \in \mathcal{T}} \left[ \prod_{r, p} \left( d\Lambda_{r}^0(t) \exp\left( \boldsymbol{\beta}^\top \mathbf{Z}_{rp}(t) \right) \right)^{\Delta N_{rp}(t)} \cdot \left(1 - \sum_{r=1}^{S} d\Lambda_r^0(t) \cdot S_r^{(0)}(\boldsymbol{\beta}, t) \right)^{1 - \Delta N_{\cdot \cdot}(t)} \right],
\end{equation}
where:
\begin{itemize}
    \item \( \Delta N_{rp}(t) \) is the number of transitions from state \(r\) for unit \(p\) at time \(t\);
    \item \( S_r^{(0)}(\boldsymbol{\beta}, t) \) is the total risk score for state \(r\);
    \item \( \Delta N_{\cdot \cdot}(t) \) represents the total number of events observed at time \(t\).
\end{itemize}
Maximization of the log-partial likelihood yields to the estimator \( \hat{\boldsymbol{\beta}} \). Given \( \hat{\boldsymbol{\beta}} \), the cumulative baseline hazard is estimated using the Breslow estimator:
\begin{equation}
    \widehat{\Lambda}_r^0(t, \hat{ \boldsymbol{\beta}}) = \int_0^t \frac{J_r(u)}{S_r^{(0)}(\hat{\boldsymbol{\beta}}, u)} \, dN_r(u),
\end{equation}
where \( N_r(u) = \sum_{p=1}^{P} N_{rp}(u) \) is the total number of transitions from state \(r\) up to time \(u\), and \( J_r(u) \) is an indicator equal to 1 if at least one unit is at risk of leaving state \(r\) at time \(u\).
Finally, the cumulative hazard estimates feed into the Aalen-Johansen estimator of the transition probability matrix:
\begin{equation}
    \widehat{\mathbf{P}}(v, t) = \prod_{u \in(v, t]} \left( \mathbf{I} + d\widehat{\boldsymbol{\Lambda}}(u) \right),
    \label{probsemi}
\end{equation}
where \( \widehat{\boldsymbol{\Lambda}}(u) \) is the estimated cumulative hazard matrix.

\subsection{Results Synthesis and Interpretation Strategy} 
The synthesis of results will be organized consistently across model specifications. For each multi-state model, the main outputs — transition probabilities and expected length of stay (ELOS) within each state — will be reported and visualized to provide an intuitive understanding of delay mitigation and propagation dynamics along the S5 line. Graphical representations will highlight how transition patterns evolve across time and service segments.
\\For the semi-parametric specification, particular attention will be given to how covariates influence transition intensities, while detailed hazard ratio estimates will be provided in the~\ref{app2} for completeness.
\\The overall synthesis aims to distill complex model outputs into interpretable evidence on delay dynamics, serving as the analytical foundation for the subsequent discussion section,

\section{Results}
\label{Results}
In this section we present the results obtained from the analysis of train delay dynamics along the selected railway route, namely the S5 suburban line connecting Varese to Treviglio, in the three-month period of September-December 2023. 
\\All analyses are conducted using the R software environment (\cite{R}), employing the \texttt{survival} and \texttt{mstate} (\cite{mstate, msdata}) packages to prepare and analyze data in a multi-state framework.

\subsection{Exploratory Analysis}
\label{Exploratory Analysis}
\begin{figure}[H]
    \centering    \includegraphics[width=0.85\textwidth]{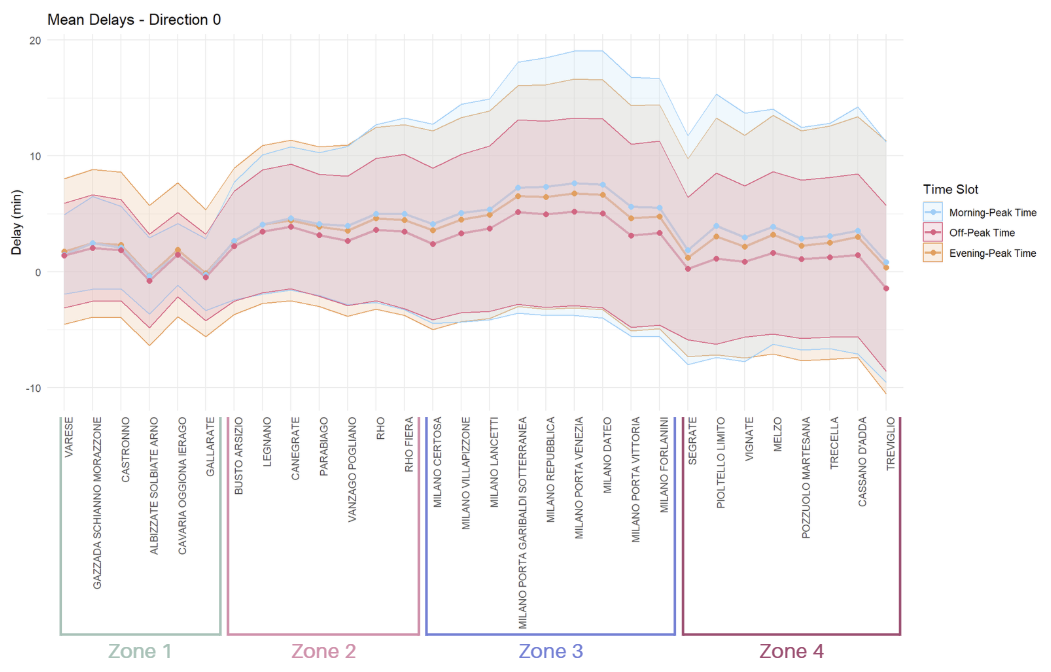}
    \caption{Average arrival delays and variability (expressed as \(\pm2\) standard deviations) are shown for each station along the route in the Varese \(\to\) Treviglio direction. The colors indicate different time periods: light blue represents Morning-Peak hours, pink indicates Off-Peak hours, and orange corresponds to Evening-Peak hours.}
    \label{fig7}
\end{figure}

\begin{figure}[H]
    \centering    \includegraphics[width=0.85\textwidth]{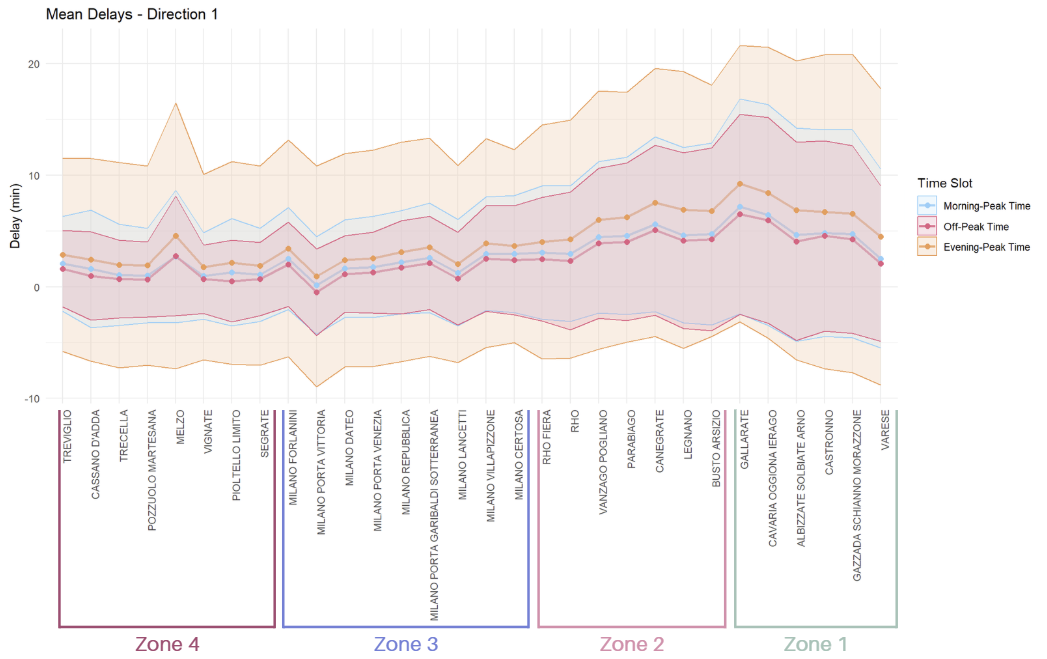}
    \caption{Average arrival delays and variability (expressed as \(\pm2\) standard deviations) are shown for each station along the route in the Treviglio \(\to\) Varese direction. The colors indicate different time periods: light blue represents Morning-Peak hours, pink indicates Off-Peak hours, and orange corresponds to Evening-Peak hours.}
    \label{fig8}
\end{figure}

We begin by presenting an exploratory analysis based on the dataset obtained from the three main sources described in Section~\ref{Data Overview and Provenance}.
Figure~\ref{fig7} and Figure~\ref{fig8} show average arrival delays and variability (expressed as \(\pm 2\) standard deviation bands) for each station along the route, respectively for the Varese \(\to\) Treviglio and the Treviglio \(\to\) Varese directions. In the former case, delays remain moderate in the outer section, but rise sharply approaching Milano — especially during morning peak — reflecting network congestion and commuter demand. The urban core exhibits the highest variability, confirming Milano's role as a systemic bottleneck, while punctuality progressively recovers beyond the city during off-peak hours. Conversely, in the opposite direction, delays accumulate gradually along the route: operations are stable in the eastern section and within central Milano, but become increasingly unstable toward the northwest, particularly during the evening peak. Although the line is not perfectly symmetric, both directions display broadly consistent delay dynamics, with only minor differences. Therefore, the subsequent modeling sections focus on the Varese-Treviglio direction, while results for the reverse direction are omitted for brevity.

Turning to the characterization of the key covariates involved in the semi-parametric procedure outlined in Section~\ref{Semi-parametric Models}, the average number of boarding passengers is around 28 (SD $\approx$ 39), with a right-skewed distribution driven by high-volume urban hubs; the alighting distribution follows a similar pattern with a slightly higher median. Train frequency ranges between 4 and 46 trains per hour, with the first and third quartiles at 7 and 16, respectively. The upper bound corresponds to the high-capacity section of Milano Porta Garibaldi, a major junction that accommodates both regional trains on the upper level and suburban lines on the lower level. A binary covariate identifies adverse meteorological conditions - including heavy precipitation, storms, or fog - present in roughly 52\% of the observations.

Altogether, this exploratory evidence highlights the heterogeneity of both delay patterns and covariates across temporal and spatial dimensions, justifying their inclusion in the transition-specific hazard models developed in the following sections.

\subsection{Spatio-temporal Delay Patterns}
\label{Spatio-temporal Delay Patterns}
Building on the formulation reported in Section~\ref{Nonparametric Models}, we analyze the delay dynamics through a nonparametric method under the three-state and the four-state models, grouped by travel direction (Varese \(\to\) Treviglio and Treviglio \(\to\) Varese), time slot (Morning Peak, Off-Peak, and Evening Peak), and route segment (Zone 1-4). This approach enables a fully flexible estimation of transition intensities within each subgroup, without imposing parametric constraints or structural assumptions across models. 

\subsubsection{Three-state Model}
\begin{figure}[H]
    \centering    \includegraphics[width=0.9\textwidth]{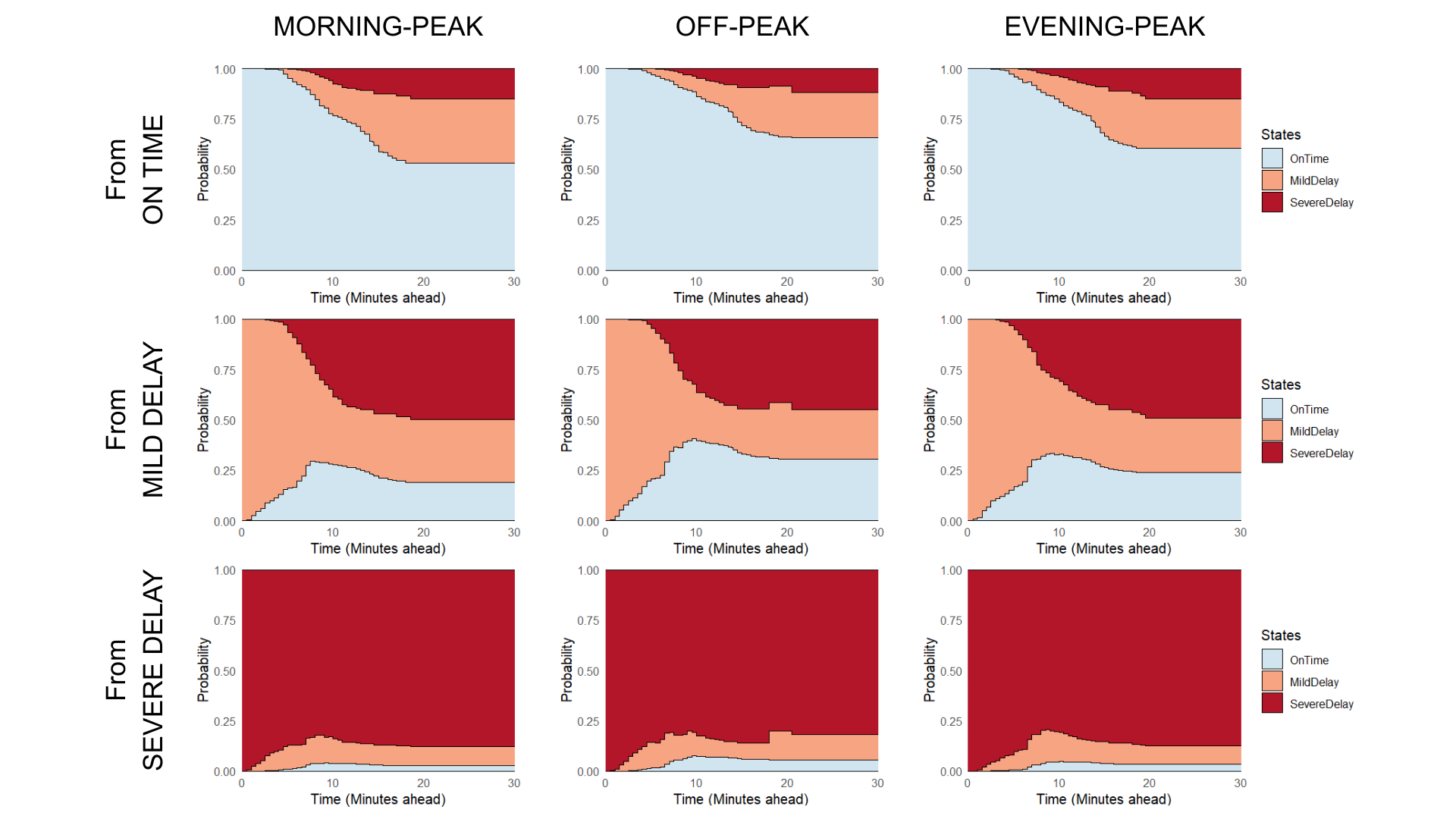}
    \caption{Nonparametric estimates of stacked transition probabilities by time since entry into the current state, for direction Varese \(\to\) Treviglio. Time slots are shown vertically, starting states horizontally. Color legend: light blue = On Time, orange = Mild Delay, red = Severe Delay.}
    \label{fig9}
\end{figure}

To investigate how delay evolve over time, we first consider the temporal resolution adopted for the estimation of transition dynamics. As described in Eq.~\ref{prob}, transition probabilities are computed separately for each time slot: Morning-Peak Time, Off-Peak Time, and Evening-Peak Time.
\\Figure~\ref{fig9} reports the evolution of state probabilities over a 30-minute horizon, conditional on the initial delay state and disaggregated by time slot. As an illustrative example, consider a specific train \(i\) observed at station \(j\) on day \(k\) entering the On Time state during the Morning-Peak (top-left panel in the figure): after 30 minutes, slightly more than 50\% of such instances remain on time, around 30\% transition to a Mild Delay, and nearly 15\% move to a Severe Delay. Overall, the transition patterns across all panels reveal clear temporal regularities. Delay deterioration is most pronounced during the Morning-Peak, when network congestion limits recovery and amplifies propagation effects. The Off-Peak period exhibits the greatest stability, with higher probabilities of remaining or return to the On Time state. The Evening-Peak shows intermediate behavior, marked by moderate persistence of delays but partial recovery potential. Across all time slots, train starting in a Severe Delay state display strong state persistence, confirming the low likelihood of spontaneous recovery once significant delays occur.

Beyond transition probabilities, we also examine the expected duration of stay (ELOS) in each state. In addition, to assess the uncertainty associated with these estimates, we implement a nonparametric bootstrap with 1,000 replications, resampling at the mission level to preserve the integrity of the multi-state structure. Results are displayed in Table~\ref{tab1}. During the Morning-Peak, ELOS ranges from 74 minutes in the On Time state to 115 minutes under the Severe Delay, with lower values Off-Peak and similar patterns in the Evening.
Narrow confidence intervals for On Time and wider ones for Mild Delay reflect stable versus variable estimates.
\\In practical terms, these results indicate that once a train experiences a severe delay, it tends to remain in that condition for an extended period, suggesting limited recovery capacity within the same mission. This persistence is consistent with congested operating conditions, especially during peak hours, where limited buffer time and network interactions reduce the likelihood of regaining punctuality. Conversely, the relatively short sojourns (around 40 minutes) observed in the Mild Delay state highlight its transitory nature, typically evolving either toward recovery to the On Time state or deterioration into Severe Delay. Taken together, these findings provide a valuable input for operational planning and timetable adjustment aimed at enhancing delay recovery. 

\begin{table}[H]
\centering
\begin{tabular}{|P{8em}|c|c|}
\hline
Time Slot & State & Expected Length of Stay (min) \\
\hline \hline
Morning-Peak & On Time     & 74.06 (69.07-78.88)  \\
                  & Mild Delay  & 42.73 (33.64-51.09) \\
                  & Severe Delay& 114.77 (108.81-119.51)\\
\hline
Off-Peak     & On Time     & 89.58 (85.51-93.53)  \\
                  & Mild Delay  & 36.30 (24.62-47.59)  \\
                  & Severe Delay& 107.27 (91.59-117.57) \\
\hline
Evening-Peak     & On Time     & 83.02 (77.89-88.22)  \\
                  & Mild Delay  & 40.17 (30.83-49.80)  \\
                  & Severe Delay& 113.77 (107.24-118.63) \\
\hline
\end{tabular}
\caption{Expected length of stay within each state by time slot and delay state. The total travel time indicated by Trenord is around 2 hours and 10 minutes.}
\label{tab1}
\end{table}

\begin{figure}[H]
    \centering    \includegraphics[width=1\textwidth]{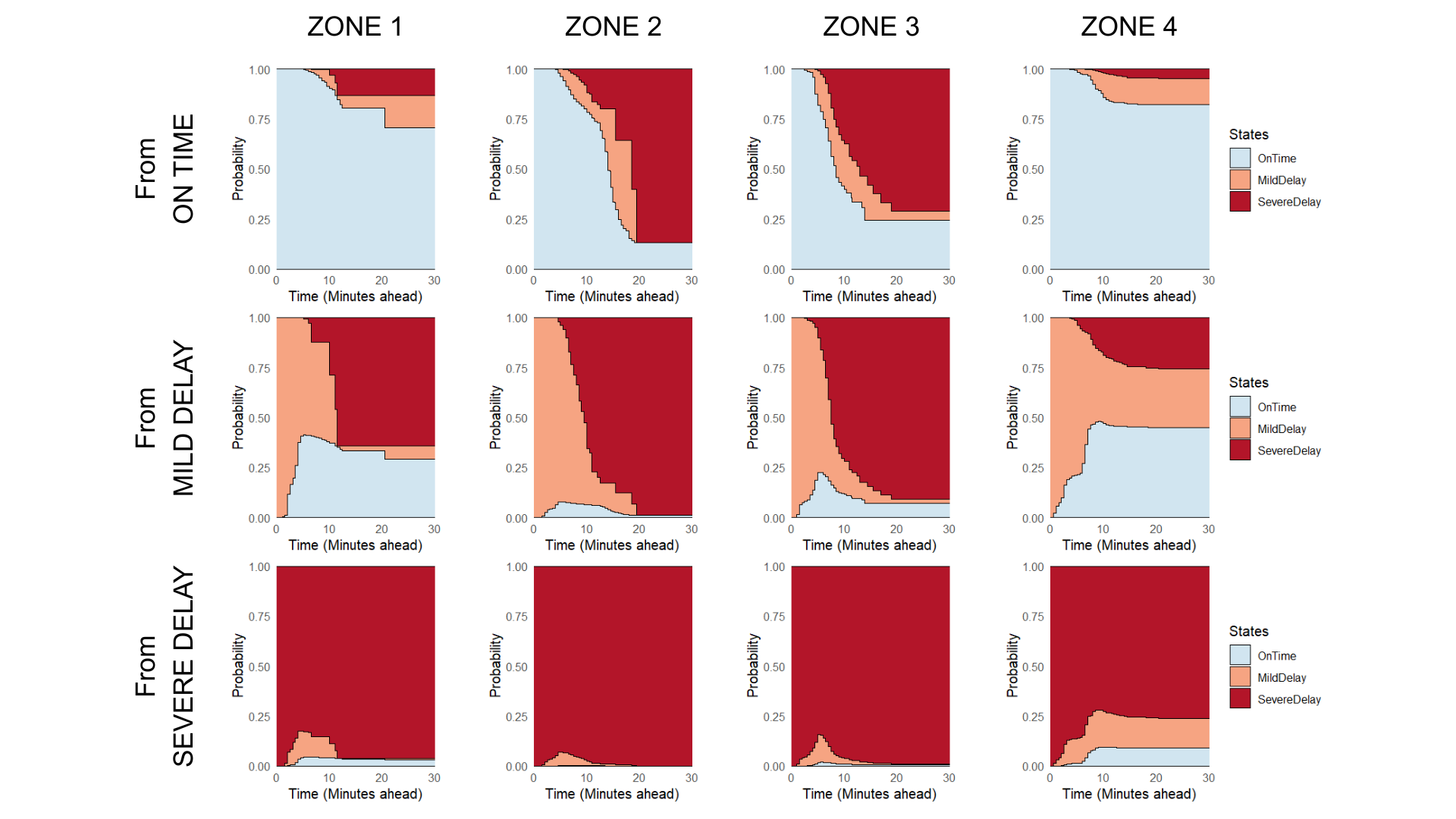}
    \caption{Nonparametric estimates of stacked transition probabilities by time since entry into the current state, for direction Varese \(\to\) Treviglio. Route Sections are shown vertically, starting states horizontally. Color legend: light blue = On Time, orange = Mild Delay, red = Severe Delay.
    Segmentation legend: Zone 1 = Varese-Gallarate, Zone 2 = Busto Arsizio-Rho Fiera, Zone 3 = Milano Certosa-Milano Forlanini, Zone 4 = Segrate-Treviglio.}
    \label{fig10}
\end{figure}

Moving now to the route-section segmentation, Figure~\ref{fig10} presents the estimated transition probabilities along the four sections from Varese to Treviglio: Zone 1 (Varese-Gallarate), Zone 2 (Busto Arsizio-Rho Fiera), Zone 3 (Milano Certosa-Milano Forlanini), and Zone 4 (Segrate-Treviglio). Delay accumulation is most pronounced in the central segments (Zones 1 and 2), where trains entering On Time quickly lose punctuality and transitions from Mild to Severe Delay are frequent. In contrast, the outer sections (Zones 1 and 4) exhibit grater stability, with higher probabilities of remaining On Time and more frequent recovery from Mild Delay. Across all segments, once a train enters a Severe Delay state, recovery is rare, particularly in the Milano core. Expected length of stay, reported in Table~\ref{tab2}, generally aligns closely with planned travel times in the On Time state, confirming the model's consistency with scheduled operations. These results underscore the central urban section as the critical stretch for delay propagation, likely due to higher traffic density and operational complexity.

\begin{table}[H]
\centering
\begin{tabular}{|P{8em}|c|c|c|}
\hline
Route Section & State & ELOS (min) & Traveling Time (min) \\
\hline \hline
Zone 1 & On Time     & 25.14 (22.55-27.00) & 25  \\
                  & Mild Delay  & 8.08 (7.17-15.64) & \\
                  & Severe Delay& 27.96 (26.45-28.51) &  \\
\hline
Zone 2     & On Time     & 15.00 (14.42-15.66) & 28   \\
                  & Mild Delay  & 9.43 (8.82-10.44) &  \\
                  & Severe Delay& 29.51 (29.32-29.65) & \\
\hline
Zone 3     & On Time     & 13.25 (11.05-15.37) & 26  \\
                  & Mild Delay  & 7.57 (6.83-8.43) &  \\
                  & Severe Delay& 29.00 (28.75-29.2) & \\
                  \hline
Zone 4     & On Time     & 26.20 (25.77-26.64) & 32  \\
                  & Mild Delay  & 12.93 (11.89-14.03) &  \\
                  & Severe Delay& 23.59 (22.4-24.73) & \\
\hline
\end{tabular}
\caption{Expected length of stay within each state by route section and delay state. Traveling Time refers to the planned traveling time indicated by Trenord. Segmentation legend: Zone 1 = Varese-Gallarate, Zone 2 = Busto Arsizio-Rho Fiera, Zone 3 = Milano Certosa-Milano Forlanini, Zone 4 = Segrate-Treviglio.}
\label{tab2}
\end{table}

We also explore the full combination of direction, time slot, and route section to assess potential interactions. Overall, the transition probability patterns remain broadly consistent with those observed for individual factors. For this reason, results are reported in~\ref{app2: ProbPlots}.

That said, a few limitations qualify these findings. Cross-model comparisons should be made cautiously, as each framework reflects different aspects of the delay process and assumptions. Estimated transition probabilities may also suffer from data sparsity in certain time slots or route segments, particularly for rare events. Furthermore, while nonparametric models flexibility capture interesting patterns, they do not permit testing covariate effects. For this reason, Section~\ref{When Delay Meets Context: Covariate Effects} introduces a semi-parametric models that combines flexibility with explanatory power for a more structured analysis of the factors influencing delay trajectories.

\subsubsection{Four-state Model}
As introduced in Section~\ref{State Space Definition}, the four-state model adds an intermediate delay state, providing a finer representation of delay dynamics. For the sake of brevity, detailed results are not reported here. Nonetheless, it is worth nothing that the four-state specification largely confirms the patterns observed in the three-state setting, while the introduction of the intermediate Medium Delay state offers a finer description of both deterioration and partial recovery trajectories. This additional granularity reveals gradual transitions that were previously masked in the coarser framework, particularly in peak periods and congestion-prone sections, where delays tend to accumulate progressively rather than abruptly. The model also offers improved interpretability of transient advancements, making it easier to identify phases where operational interventions could be most effective. 

Despite these advantages, also the four-state approach has some limitations. The additional state increases model complexity and reduces sample sizes per transition, particularly in less frequent states, resulting in higher variability and wider confidence intervals for expected durations.  Furthermore, while the four-state models enhances granularity and detail, it does not fundamentally change the overall insights and macro-patterns on system dynamics. For these reasons, only the three-state models is retained for semi-parametric analyses.

\subsection{Impact of Covariates on Delay Transitions}
\label{When Delay Meets Context: Covariate Effects}
To go beyond, we extend the three-state nonparametric model within a semi-parametric Cox proportional hazards framework, as introduced in Section~\ref{Semi-parametric Models}. Estimation is performed separately by travel direction and covariate group to account for directional asymmetries and to isolate the role of different sources of heterogeneity.

The estimated hazard ratios (HRs) with their 95\% confidence intervals, summarizing the covariates effects on the transition intensities between delay states, are reported in~\ref{app2}, together with the complete set of model estimates — including coefficients, confidence intervals, and significance levels.

\subsubsection{Effect of Time Slot Covariates}
\label{TimeSlotCov}
We first examine the effects of time slot covariates on delay transitions, which include passenger volumes (boarding and alighting), train frequency, meteorological conditions, and time-of-day indicators.

To further explore the influence of temporal factors on delay dynamics, we compute future state transition probabilities over a 30-minute horizon and summarize the range of daily-risk dynamics through best-case and worst-case scenarios, defined by contrasting combinations of features. In the best-case passenger volumes correspond to the 15th percentile of the boarded and alighted distributions, train frequency is set to 4 trains per hour — typical of peripheral stations — and no adverse weather is present. Conversely, the worst case reflects high-demand and congested conditions, with passenger loads at the 85th percentile, train frequency of 24 trains/h (as in the Milano urban area), and the occurrence of adverse weather (e.g., rain, fog,, and/or storms). Finally, the analysis explicitly considers three daily time slots (Morning-, Off-, and Evening-Peak) to assess how time of day modulates these effects. Additional results of a systematic one-at-a-time variation of each covariate across its observed range, while holding all others constant at their mean (for numerical factors) or reference category (for categorical ones), are reported in~\ref{app2: Timeslot}.
\\As visible in Figure~\ref{fig11}, performance strongly depends on operating conditions. In the best case (blue boxes), trains largely remain On Time or experience only Mild Delays, even during peak hours. Under adverse conditions (red boxes), deterioration is rapid and escalates within 10-15 minutes almost uniformly across all time-slots. When starting from a Mild Delay, recovery is feasible mainly under favorable settings, while in congested and adverse contexts, escalation becomes dominant. Once in a Severe Delay state, the process is nearly absorbing with improvements that remain rare even in the best case. Table~\ref{tab5} reports the ELOS in each state, under the best- and worst-case scenarios and time slots, as a complementary summary to the probability profiles.

\begin{figure}[H]
    \centering
     \makebox[\textwidth][c]{%
        \includegraphics[width=1\textwidth]{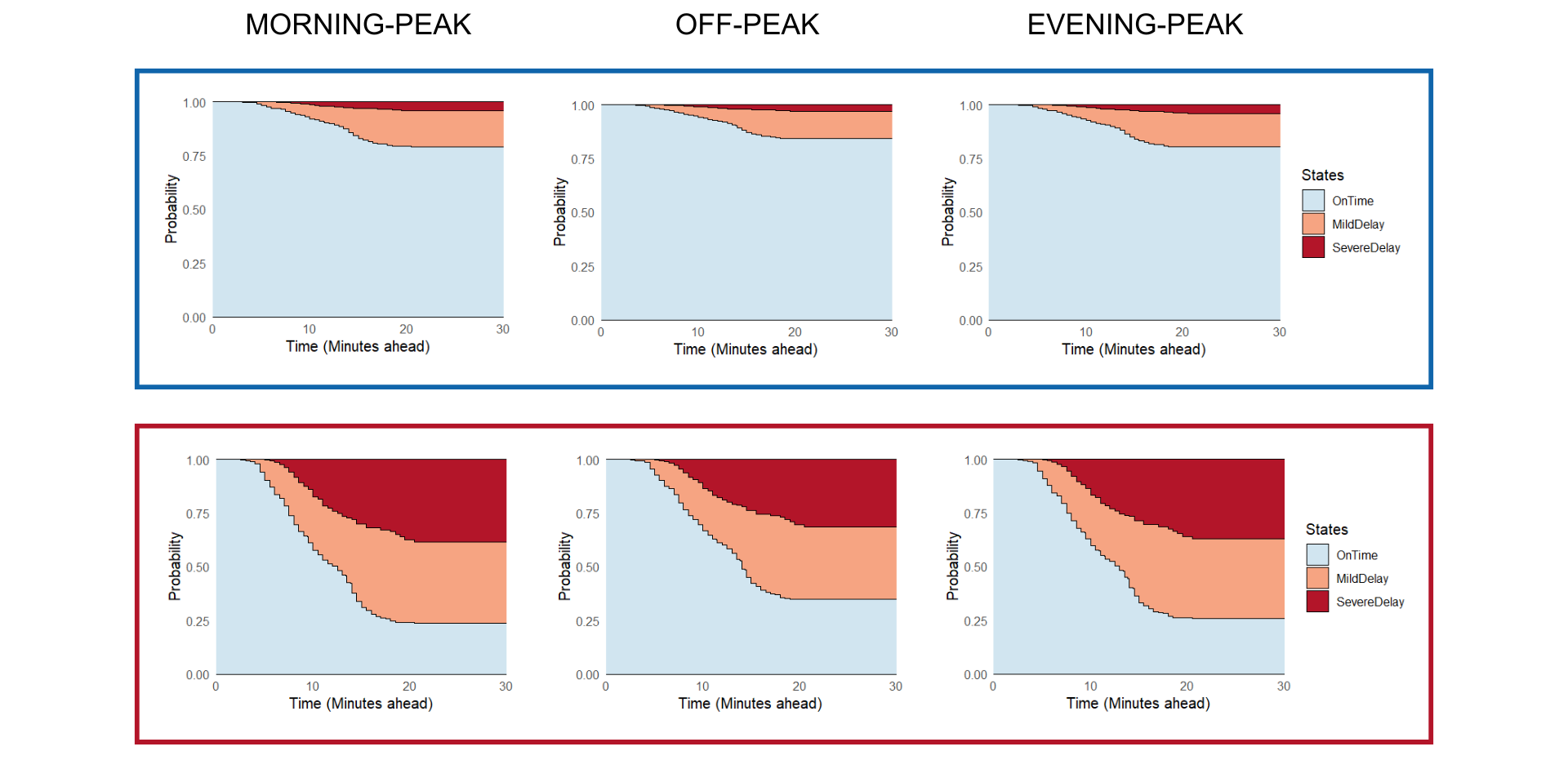}%
    }
    \caption*{(a) From On Time.}
\end{figure}

\begin{figure}[H]
    \centering
     \makebox[\textwidth][c]{%
        \includegraphics[width=1\textwidth]{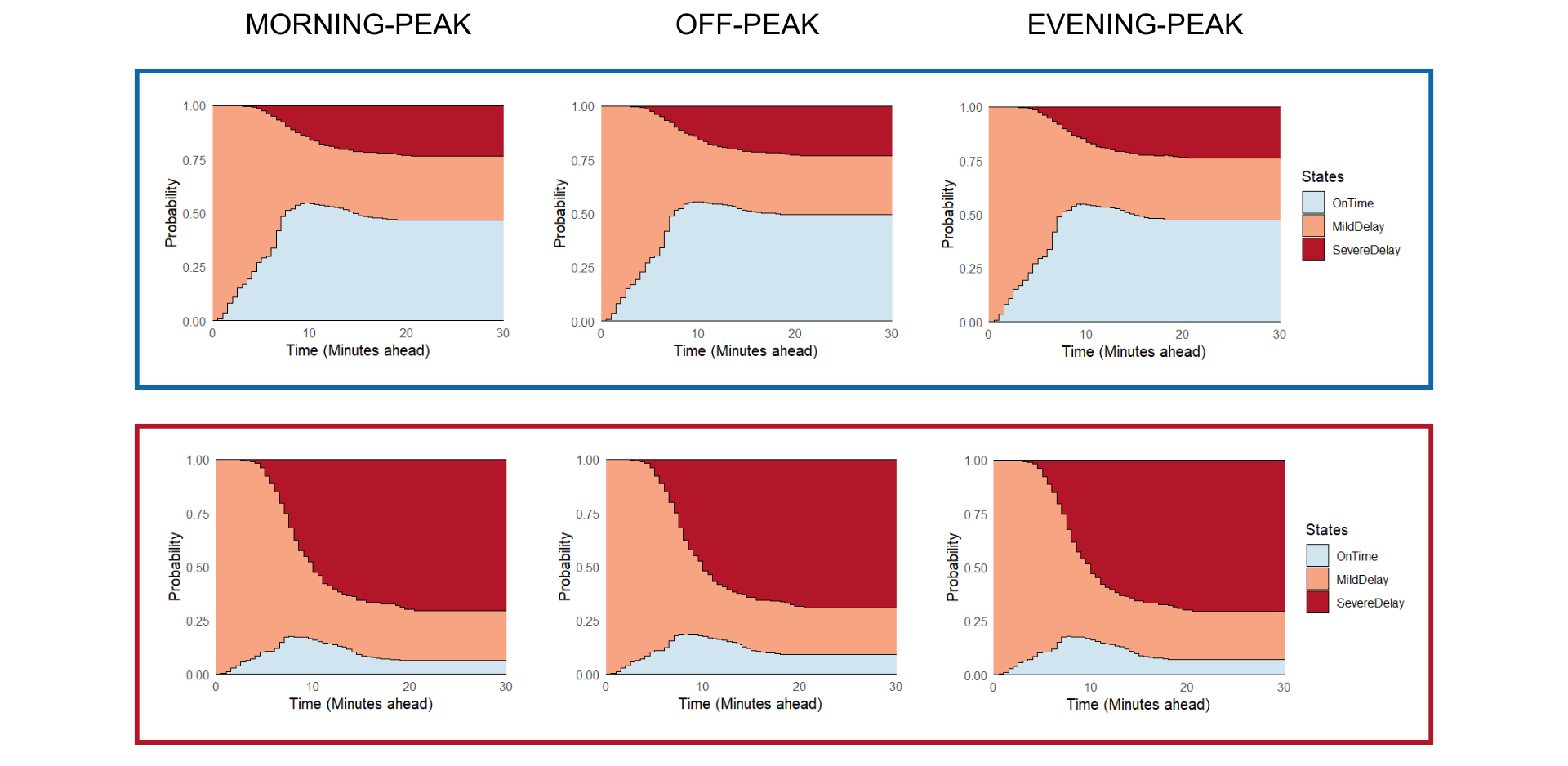}%
    }
    \caption*{(a) From Mild Delay.}
\end{figure}

\begin{figure}[H]
    \centering
    \makebox[\textwidth][c]{%
        \includegraphics[width=1\textwidth]{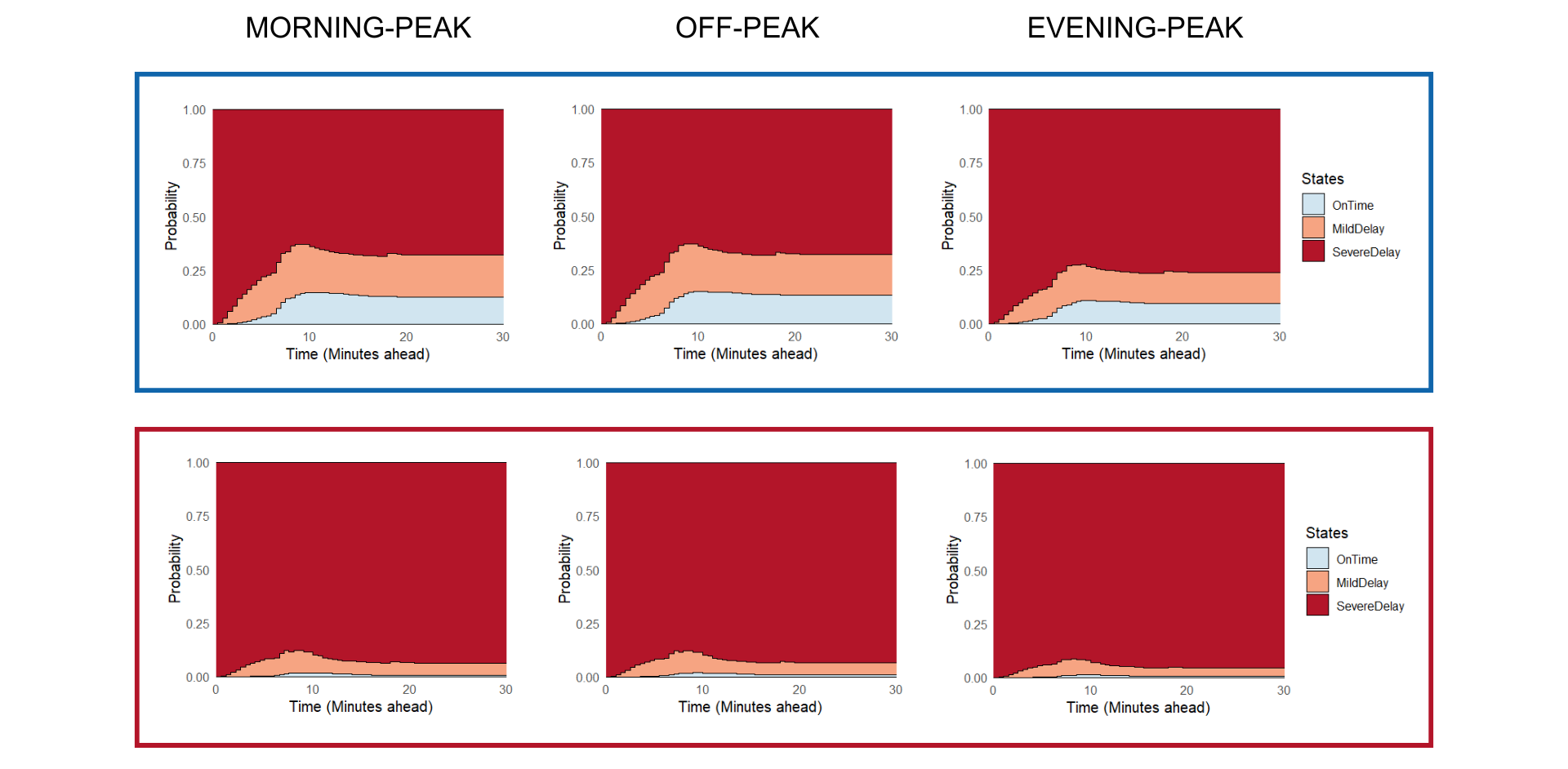}%
    }
    \caption*{(b) From Severe Delay.}
    \caption{Predicted future state probabilities displaying best and worst-case scenarios across time slots. Blue boxes indicate states resulting from favorable conditions — low passenger volumes (15th percentile), low train frequency (e.g., 4 trains/h), and good weather — while red boxes reflect states under unfavorable conditions such as high passenger volumes (85th percentile), high frequency (e.g., 24 trains/h), and adverse weather (e.g., rain, fog, or storms).}
    \label{fig11}
\end{figure}

\begin{table}[H]
\centering
\begin{tabular}{|P{8em}|c|P{8em}|P{8em}|}
\hline
Time Slot & State & ELOS (min) Best & ELOS (min) Worst \\
\hline \hline
Morning-Peak & On Time     & 105 & 39 \\
                  & Mild Delay  & 42 & 35 \\
                  & Severe Delay& 90 & 122 \\
\hline
Off-Peak     & On Time     & 111 & 52\\
                  & Mild Delay  & 39 & 33  \\
                  & Severe Delay & 90 & 122 \\
\hline
Evening-Peak     & On Time     & 107 & 41  \\
                  & Mild Delay  & 41 & 34  \\
                  & Severe Delay & 100 & 124 \\
\hline
\end{tabular}
\caption{Expected Length of Stay in each delay state across time slots, under best-case (favorable combination of covariates) and worst-case (adverse combination of covariates) scenarios. The scheduled travel time on the Varese–Treviglio route is approximately 130 minutes.}
\label{tab5}
\end{table}

\subsubsection{Effect of Route Section Covariates}
\label{Effect of Route Section Covariates on Delay Transitions}
Now the aim is to investigate the influence of factors such as passenger volumes, train frequency, meteorological conditions, and the route zone in the intensities of transitions between delay states.

To examine spatial heterogeneity in delay dynamics, state transition probabilities are computed over a 30-minute horizon (Figure~\ref{fig14}), comparing the best-case and worst case scenarios, defined as in Section~\ref{TimeSlotCov}, while additional results from the one-at-a-time variation of each covariate are reported in~\ref{app2: Routesection}. In this specific context, results are compared across the four route segments (Zone 1-4). Transitions involving Severe Delay indicate limited recovery capacity even under favorable conditions (blue box), especially in the central urban areas (Zone 2 and 3), probably due to structural congestion and minimal operational flexibility. From the On Time state, structural, environmental and operational factors exert only minor effects on the delay dynamics of Zones 1 and 4, with a more pronounced degradation in correspondence of Zones 2 and 3. In addition, transitions from the Mild Delay state reveal strong congestion effects: under the worst-case scenario (red box), the likelihood of escalation rises markedly across all route segments, exhibiting near-absorbing behavior, particularly in peripheral areas (Zones 1 and 4), where recovery probabilities decline from about 60\% to near absorbing levels.
Table~\ref{tab6} reports the expected lengths of stay (ELOS) in each state, under best- and worst-case scenarios and route sections, as a complementary summery to the probability profiles.

\begin{figure}[H]
    \centering
     \makebox[\textwidth][c]{%
        \includegraphics[width=0.95\textwidth]{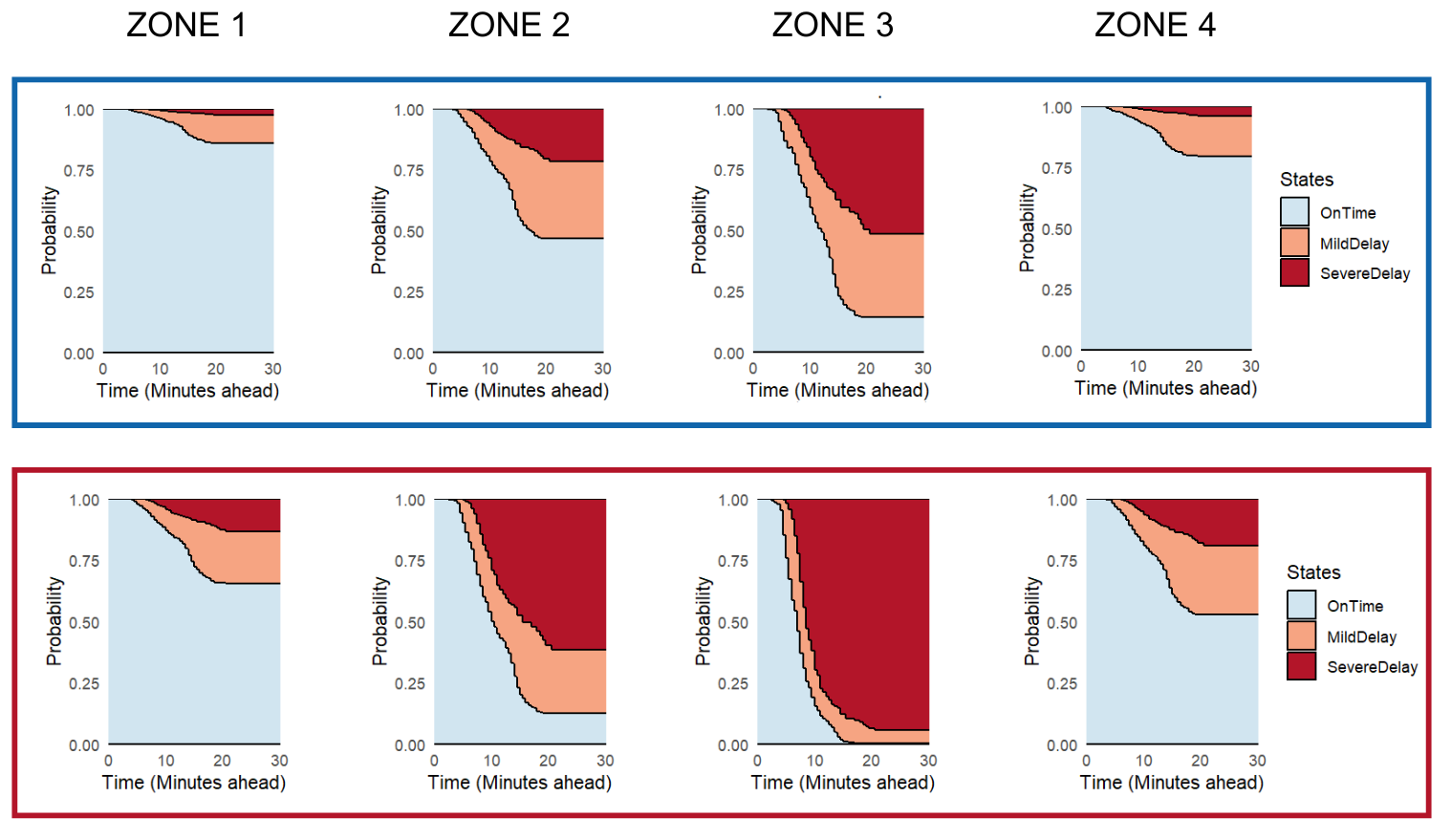}%
    }
    \caption*{(a) From On Time.}
\end{figure}

\begin{figure}[H]
    \centering
     \makebox[\textwidth][c]{%
        \includegraphics[width=0.95\textwidth]{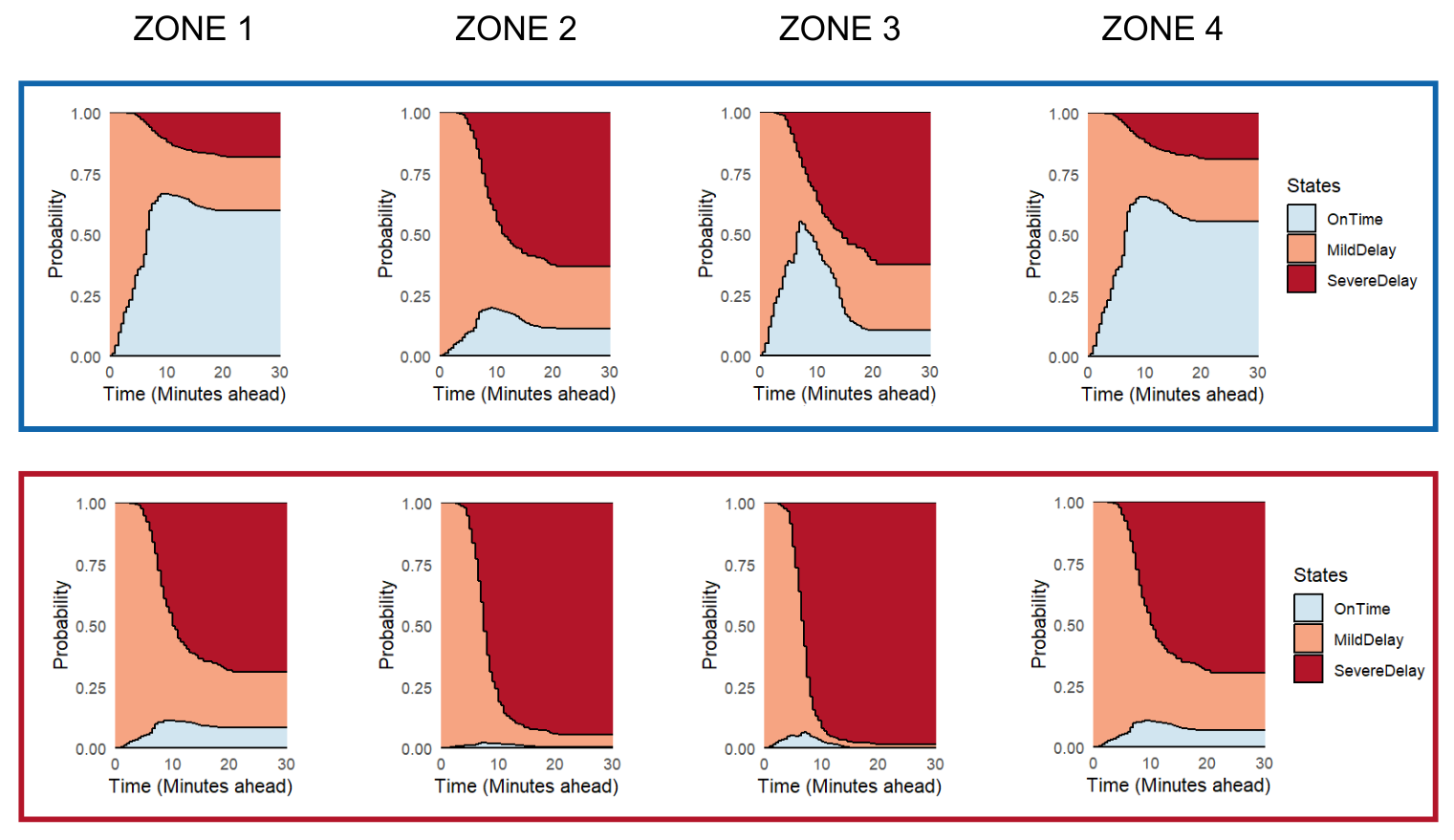}%
    }
    \caption*{(b) From Mild Delay.}
\end{figure}

\begin{figure}[H]
    \centering
    \makebox[\textwidth][c]{%
        \includegraphics[width=0.95\textwidth]{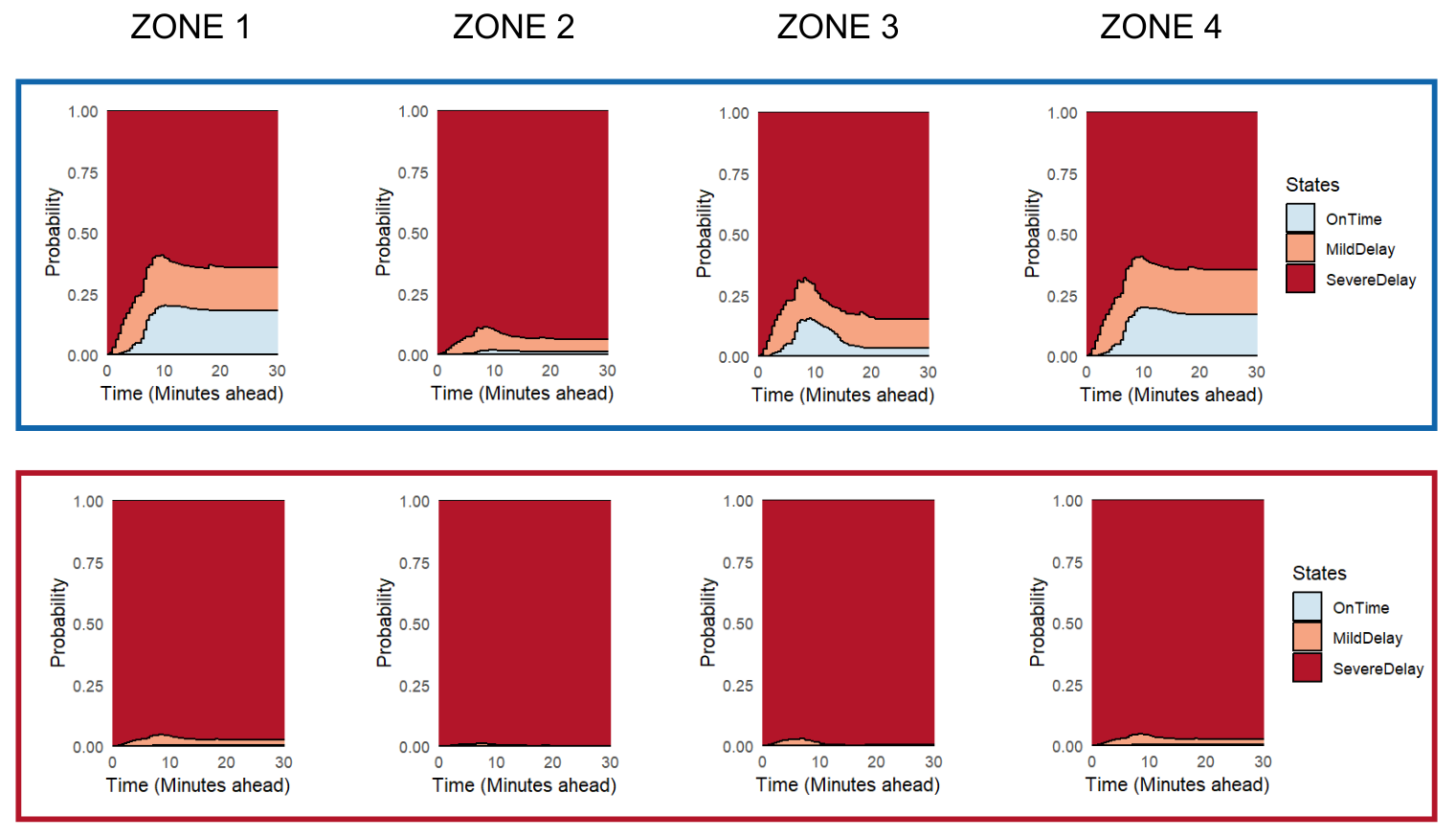}%
    }
    \caption*{(c) From Severe Delay.}
    \caption{Predicted future state probabilities displaying best and worst-case scenarios across route sections. Blue boxes indicate states resulting from favorable conditions — low passenger volumes (15th percentile), low train frequency (e.g., 4 trains/h), and good weather — while red boxes reflect states under unfavorable conditions such as high passenger volumes (85th percentile), high frequency (e.g., 24 trains/h), and adverse weather (e.g., rain, fog, or storms). Segmentation legend: Zone 1 = Varese-Gallarate, Zone 2 = Busto Arsizio-Rho Fiera, Zone 3 = Milano Certosa-Milano Forlanini, Zone 4 = Segrate-Treviglio.}
    \label{fig14}
\end{figure}

\begin{table}[H]
\centering
\begin{tabular}{|P{7em}|c|P{7em}|P{7em}|c|}
\hline
Time Slot & State & ELOS (min) Best & ELOS (min) Worst & Traveling Time (min)\\
\hline \hline
Zone 1 & On Time     & 87 & 24 & 25\\
                  & Mild Delay  & 11 & 13 & \\
                  & Severe Delay& 21 & 29 & \\
\hline
Zone 2     & On Time     & 20 & 17 & 12 \\
                  & Mild Delay  & 13 & 9 & \\
                  & Severe Delay & 28 & 30 & \\
\hline
Zone 3     & On Time     & 13 & 8 & 26 \\
                  & Mild Delay  & 13 & 7 & \\
                  & Severe Delay & 28 & 30 & \\
                  
\hline
Zone 4     & On Time     & 26 & 21 & 32 \\
                  & Mild Delay  & 13 & 15 & \\
                  & Severe Delay & 21 & 29 & \\
\hline
\end{tabular}
\caption{Expected Length of Stay in each delay state across route sections, under under best-case (favorable combination of covariates) and worst-case
(adverse combination of covariates) scenarios. Traveling Time refers to the planned traveling time indicated by Trenord. Segmentation legend: Zone 1 = Varese-Gallarate, Zone 2 = Busto Arsizio-Rho Fiera, Zone 3 = Milano Certosa-Milano Forlanini, Zone 4 = Segrate-Treviglio.}
\label{tab6}
\end{table}

\newpage
However, despite the valuable insights and the strong interpretative foundation derived from this semi-parametric procedure, the results obtained from the analysis of the Schoenfeld residuals suggest that the assumption of proportional hazards is not fully satisfied. 
This limitation points to the need for more flexible and data-adaptive modeling approaches in future research.

\subsection{From Analytical Findings to Operational Integration}
Beyond their analytical interpretation, the estimated transition probabilities visualized in the previous figures can be translated into operationally meaningful transition probability matrices. These matrices summarize the likelihood of moving between delay states and are computed over two distinct time horizons:
\begin{itemize}
    \item 10-minute matrices, reflecting short-term dynamics relevant for short-distance commuters who are particularly sensitive to minor disruptions;
    \item 30-minute matrices, capturing long-term effects that are important for passengers traveling over extended portions of the route.
\end{itemize}
Conditioned on the current delay state, the matrices indicate the probabilities of remaining in the same state, recovering to a lower delay, or deteriorating to a higher delay. Derived from the Aleen-Johansen estimates introduced in Eqs.~\ref{prob} and~\ref{probsemi}, they provide a compact and interpretable representation of delay dynamics, useful for both real-time operational decisions and tactical planning.

\subsubsection{Application of Nonparametric Models}
For nonparametric models, these matrices hold particular value for both passenger information systems and service management. In the first case, they enable the dissemination of probabilistic delay forecasts in real time (e.g., "there is a 65\% chance the train will remain mildly delayed over the next 10 minutes"), enhancing transparency and supporting informed passenger decisions. In the second, they assist operational planning by identifying high-risk periods or route segments where targeted interventions — such as schedule adjustment, insertion of buffer times, or deployment of additional rolling stock — may mitigate the impact of delays.
The visual plots presented in Sections~\ref{Spatio-temporal Delay Patterns} and ~\ref{When Delay Meets Context: Covariate Effects} can be translated into these matrices, providing a numerical complement that enhances transparency, reproducibility, and practical actionable insights. Table~\ref{tab7} summarize the time-based stratification, while Table~\ref{tab8} report the route-segment one.

\begin{table}[H]
\renewcommand{\arraystretch}{1.1}
\centering
\hspace{-1.1cm}
\makebox[\textwidth]{%
\begin{subtable}{0.44\textwidth}
  \centering
  \begin{tabular}{|c|c|c|c|}
    \hline
    \multicolumn{4}{|c|}{Morning-Peak Time} \\ \hline \hline
     & On Time & Mild Delay & Severe Delay \\ \hline
    On Time & 80\% & 15\% & 5\%\\
    Mild Delay & 30\% & 35\% & 35\%\\
    Severe Delay & 5\% & 15\% & 80\%\\
    \hline \hline
    \multicolumn{4}{|c|}{Off-Peak Time} \\ \hline \hline
     & On Time & Mild Delay & Severe Delay \\ \hline
    On Time & 90\% & 5\% & 5\%\\
    Mild Delay & 45\% & 15\% & 40\%\\
    Severe Delay & 10\% & 10\% & 80\%\\
    \hline \hline
    \multicolumn{4}{|c|}{Evening-Peak Time} \\ \hline \hline
     & On Time & Mild Delay & Severe Delay \\ \hline
    On Time & 85\% & 10\% & 5\%\\
    Mild Delay & 35\% & 35\% & 30\%\\
    Severe Delay & 5\% & 15\% & 80\%\\
    \hline
  \end{tabular}
  \caption{After 10 minutes}
  \label{tab:delta10}
\end{subtable}
\hspace{0.06\textwidth}
\begin{subtable}{0.44\textwidth}
  \centering
  \begin{tabular}{|c|c|c|c|}
    \hline
    \multicolumn{4}{|c|}{Morning-Peak Time} \\ \hline \hline
     & On Time & Mild Delay & Severe Delay \\ \hline
    On Time & 55\% & 30\% & 15\%\\
    Mild Delay & 20\% & 30\% & 50\%\\
    Severe Delay & 0\% & 10\% & 90\%\\
    \hline \hline
    \multicolumn{4}{|c|}{Off-Peak Time} \\ \hline \hline
     & On Time & Mild Delay & Severe Delay \\ \hline
    On Time & 70\% & 20\% & 10\%\\
    Mild Delay & 35\% & 25\% & 40\%\\
    Severe Delay & 5\% & 15\% & 80\%\\
    \hline \hline
    \multicolumn{4}{|c|}{Evening-Peak Time} \\ \hline \hline
     & On Time & Mild Delay & Severe Delay \\ \hline
    On Time & 60\% & 30\% & 10\%\\
    Mild Delay & 25\% & 30\% & 45\%\\
    Severe Delay & 0\% & 10\% & 90\%\\
    \hline
  \end{tabular}
  \caption{After 30 minutes}
  \label{tab:delta30}
\end{subtable}
}
\caption{Conditional probability matrices by time slot.}
\label{tab7}
\end{table}

\begin{table}[H]
\renewcommand{\arraystretch}{1.1}
\centering
\hspace{-1.1cm}
\makebox[\textwidth]{%
\begin{subtable}{0.44\textwidth}
  \centering
  \begin{tabular}{|c|c|c|c|}
    \hline
    \multicolumn{4}{|c|}{Zone 1} \\ \hline \hline
     & On Time & Mild Delay & Severe Delay \\ \hline
    On Time & 90\% & 5\% & 5\%\\
    Mild Delay & 40\% & 45\% & 15\%\\
    Severe Delay & 5\% & 10\% & 85\%\\
    \hline \hline
    \multicolumn{4}{|c|}{Zone 2} \\ \hline \hline
     & On Time & Mild Delay & Severe Delay \\ \hline
    On Time & 85\% & 10\% & 5\%\\
    Mild Delay & 10\% & 25\% & 65\%\\
    Severe Delay & 0\% & 5\% & 95\%\\
    \hline \hline
    \multicolumn{4}{|c|}{Zone 3} \\ \hline \hline
     & On Time & Mild Delay & Severe Delay \\ \hline
    On Time & 40\% & 25\% & 35\%\\
    Mild Delay & 15\% & 20\% & 65\%\\
    Severe Delay & 0\% & 5\% & 95\%\\
    \hline \hline
    \multicolumn{4}{|c|}{Zone 4} \\ \hline \hline
     & On Time & Mild Delay & Severe Delay \\ \hline
    On Time & 85\% & 15\% & 0\%\\
    Mild Delay & 50\% & 35\% & 15\%\\
    Severe Delay & 10\% & 25\% & 65\%\\
    \hline
  \end{tabular}
  \caption{After 10 minutes}
  \label{tab:delta10}
\end{subtable}
\hspace{0.06\textwidth}
\begin{subtable}{0.44\textwidth}
  \centering
  \begin{tabular}{|c|c|c|c|}
    \hline
    \multicolumn{4}{|c|}{Zone 1} \\ \hline \hline
     & On Time & Mild Delay & Severe Delay \\ \hline
    On Time & 75\% & 15\% & 10\%\\
    Mild Delay & 35\% & 5\% & 60\%\\
    Severe Delay & 5\% & 0\% & 95\%\\
    \hline \hline
    \multicolumn{4}{|c|}{Zone 2} \\ \hline \hline
     & On Time & Mild Delay & Severe Delay \\ \hline
    On Time & 15\% & 0\% & 85\%\\
    Mild Delay & 0\% & 0\% & 100\%\\
    Severe Delay & 0\% & 0\% & 100\%\\
    \hline \hline
    \multicolumn{4}{|c|}{Zone 3} \\ \hline \hline
     & On Time & Mild Delay & Severe Delay \\ \hline
    On Time & 30\% & 5\% & 65\%\\
    Mild Delay & 10\% & 0\% & 90\%\\
    Severe Delay & 0\% & 0\% & 100\%\\
    \hline \hline
    \multicolumn{4}{|c|}{Zone 4} \\ \hline \hline
     & On Time & Mild Delay & Severe Delay \\ \hline
    On Time & 85\% & 10\% & 5\%\\
    Mild Delay & 50\% & 35\% & 15\%\\
    Severe Delay & 10\% & 20\% & 70\%\\
    \hline
  \end{tabular}
  \caption{After 30 minutes}
  \label{tab:delta30}
\end{subtable}
} 
\caption{Conditional probability matrices by route section. Segmentation legend: Zone 1 = Varese-Gallarate, Zone 2 = Busto Arsizio-Rho Fiera, Zone 3 = Milano Certosa-Milano Forlanini, Zone 4 = Segrate-Treviglio.}
\label{tab8}
\end{table}

\subsubsection{Application of Semi-parametric Models}
Transition probabilities estimated through the semi-parametric Cox framework are computed over fixed horizons of 10 and 30 minutes, considering varying operational and environmental conditions. To make these results actionable, the analysis focuses on four key contrasts — passenger volume, train frequency, weather conditions, and a combined best- vs worst-case scenario — each designed to isolate the marginal contribution of these factors of delay evolution. For each scenario pair, delta matrices are computed as the difference between adverse and favorable conditions. For instance, an increase of +30 percentage points in the Mild Delay \(\to\) Severe Delay transition under high passenger volume indicates a substantial worsening effect associated with crowding. This comparative framework allows for a concise assessment of risk amplification, highlighting transitions and contexts most sensitive to operational pressure. Their results provide practical guidance for prioritizing interventions — such as rescheduling, capacity adjustment, or improved passenger flow management — where their impact is likely to be the greatest. In line with the analytical structure adopted in Sections~\ref{When Delay Meets Context: Covariate Effects}, and for the sake of clarity and conciseness, only the results for the Best and Worst Case scenarios are reported here in Table~\ref{tab9}, focusing on the temporal model. Results for the other covariate contrasts and for the spatial model are omitted for brevity but follow analogous patterns. 

\begin{table}[H]
\renewcommand{\arraystretch}{1.1}
\centering
\hspace{-1.1cm}
\makebox[\textwidth]{%
\begin{subtable}{0.44\textwidth}
  \centering
  \begin{tabular}{|c|c|c|c|}
    \hline
    \multicolumn{4}{|c|}{Morning-Peak Time} \\ \hline \hline
     & On Time & Mild Delay & Severe Delay \\ \hline
    On Time & -35\% & +25\% & +10\%\\
    Mild Delay & -40\% & +5\% & -35\%\\
    Severe Delay & -15\% & +15\% & +30\%\\
    \hline \hline
    \multicolumn{4}{|c|}{Off-Peak Time} \\ \hline \hline
     & On Time & Mild Delay & Severe Delay \\ \hline
    On Time & -25\% & +15\% & +10\%\\
    Mild Delay & -40\% & +5\% & +35\%\\
    Severe Delay & -15\% & +15\% & +30\%\\
    \hline \hline
    \multicolumn{4}{|c|}{Evening-Peak Time} \\ \hline \hline
     & On Time & Mild Delay & Severe Delay \\ \hline
    On Time & -25\% & +15\% & +10\%\\
    Mild Delay & -40\% & +10\% & +30\%\\
    Severe Delay & -10\% & -10\% & +20\%\\
    \hline
  \end{tabular}
  \caption{After 10 minutes}
  \label{tab:delta10}
\end{subtable}
\hspace{0.06\textwidth}
\begin{subtable}{0.44\textwidth}
  \centering
  \begin{tabular}{|c|c|c|c|}
    \hline
    \multicolumn{4}{|c|}{Morning-Peak Time} \\ \hline \hline
     & On Time & Mild Delay & Severe Delay \\ \hline
    On Time & -50\% & +20\% & +30\%\\
    Mild Delay & -40\% & -5\% & +45\%\\
    Severe Delay & -15\% & -15\% & +30\%\\
    \hline \hline
    \multicolumn{4}{|c|}{Off-Peak Time} \\ \hline \hline
     & On Time & Mild Delay & Severe Delay \\ \hline
    On Time & -50\% & +20\% & +30\%\\
    Mild Delay & -45\% & 0\% & +45\%\\
    Severe Delay & -15\% & -15\% & +30\%\\
    \hline \hline
    \multicolumn{4}{|c|}{Evening-Peak Time} \\ \hline \hline
     & On Time & Mild Delay & Severe Delay \\ \hline
    On Time & -50\% & +20\% & +30\%\\
    Mild Delay & -40\% & -5\% & +45\%\\
    Severe Delay & -10\% & -10\% & +20\%\\
    \hline
  \end{tabular}
  \caption{After 30 minutes}
  \label{tab:delta30}
\end{subtable}
} 
\caption{Delta matrices for Best vs Worst Case Scenario by time slot.}
\label{tab9}
\end{table}

\section{Conclusion}
\label{Conclusion}
The analysis of train delays within suburban rail systems poses a multifaceted challenge, requiring models that can account for the dynamic, heterogeneous, and structurally interconnected nature of such phenomena. This study proposes a continuous-time multi-state modeling framework, applied to the S5 suburban rail line operated by Trenord in Italy, with the aim of capturing how delays evolve, escalate, or recover over time and space.

Nonparametric models, estimated independently for combinations of direction, time slot, and route segment, allow for a flexible and interpretable description of delay dynamics. This disaggregated approach captures local specificities: for example, significant differences emerge between urban and peripheral zones. While these models are less suitable for generalization, their value lies in offering an empirical foundation from which derive targeted hypothesis and operational interventions. The subsequent introduction of semi-parametric Cox models enable the inclusion of relevant covariates, offering a more complete view of the mechanisms that influence delay transitions. Key findings indicate that passenger load — especially boarding volumes — emerges as a critical bottleneck, likely due to longer dwell times at stations. In addition, train frequency consistently correlates with increased delay risk, particularly in high-density segments. Finally, both the time of day and the route segment play a critical role, especially in the case of central segments of the line showing a clear delay-prone behavior. 

The methodological framework developed in this paper is not only of theoretical interest but also offers tangible applications for the improvement of suburban railway operations. Several concrete use cases emerge from the estimated models. First of all, transition intensities from both frameworks considered can be translated into probabilistic risk scores for each train journey. These scores can be integrated into control room dashboards to flag trains with a high likelihood of delay escalation, enabling proactive mitigation or prioritization. Insights on covariate effects — such as station saturation and passenger load — can guide dynamic allocation of personnel and rolling stock. For example, additional staff could be deployed at highly saturated stations during critical time slots, where the model indicates an elevated hazard of delay progression. Moreover, by identifying route sections with consistently higher transition rates toward severe delay states, infrastructure managers and planners can prioritize local interventions — such as timetable adjustments, reallocation of buffer times, or micro-scheduling tweaks — to reduce vulnerability.  At the same time, the transition probability estimates can serve as input for simulation environments or decision-support tools. These tools can be used to test the effect of different operational scenarios (e.g., adverse weather, increased demand) and assess the potential impact on delay propagation under realistic assumptions. Finally, the matrices produced through this analysis reveal to be particularly suitable for passenger information systems, offering both short- and long-term delay probabilities enriched though the regulation of factors based on user needs. These probabilities can be easily communicated (e.g., "There is a 65\% change your train will remain mildly delayed in the next 10 minutes"), enhancing transparency and helping passengers to make more informed decisions in real time (e.g., choosing alternate connections).

As acknowledged along the paper, despite the valuable insights and concrete operational applications, the proposed approach suffer from several limitations that should be considered to properly contextualize its findings and guide future developments. A primary constraint lies in the scarcity of observed transitions, which persists even after extending the observation window to three months. This limited number of state changes reduces the statistical efficiency of the models particularly when estimating effects for less frequent transitions. This issue is especially pronounced in the case of more complex modeling structures — most notably the frailty-based models. While the inclusion of unobserved heterogeneity is theoretical appealing, in practice, the estimation process proved unstable and sensitive to data sparsity. The resulting parameter estimates lacked consistency and interpretability, leading to the exclusion of these models from the main results section. A second key limitation concerns the absence of detailed operational conflict data, such as information about other trains occupying the same infrastructure. Since shared track usage is a common cause of delay propagation, the inability to explicitly model such interactions restricts the realism of the proposed framework. Integrating these dependencies would be crucial for any application aimed at real-time delay forecasting or disruption management. From a methodological standpoint, the multi-state modeling framework itself introduces structural assumptions that may not fully reflect the underlying complexity of railway operations. In particular, the choice of a semi-Markov process — where the transition intensities depend only on the current state and the time spent in it — implicitly assumes that the hazard is memoryless beyond the sojourn time. This assumption may oversimplify delay processes that exhibit cumulative effects or path-dependent dynamics, such as those caused by earlier disruptions or prior routing decisions. Furthermore, the model presumes independent evolution of individual train runs, neglecting potential cross-dependencies, for instance, when one delayed train causes cascading effects on others. Finally, the quality and the granularity of contextual factors, especially weather data, pose a further limitation. In this study, meteorological variables are assigned based on proximity to three major airports, which — while reasonable — does not capture localized conditions along the route. Moreover, the limited variety and resolution of weather data restricts the ability to detect finer-grained effects such as microclimate variations, extreme conditions, or interactions with infrastructure features. Future analysis would benefit from more informative and reliable external data sources, including high-resolution weather forecast, real-time occupancy rates, or incident reports. 

To overcome these limitations, future work should aim to build upon the current framework by collecting larger and more diverse datasets, potentially spanning multiple lines and extended time periods. The integration of infrastructure-level data, such as track occupancy or real-time traffic conditions, would enable a more realistic representation of operational constraints. Finally, the exploration of hierarchical models could improve the treatment of nested heterogenenity and local variability, offering a richer understanding of delay propagation mechanisms.

\section*{CRediT authorship contribution statement}
\textbf{Stefania Colombo}: Writing - review \& editing, Writing - original draft, Software, Methodology, Investigation, Formal analysis, Data curation.
\textbf{Alfredo Gimenez Zapiola}: Supervision, Methodology, Investigation, Formal analysis, Conceptualization.
\textbf{Francesca Ieva}: Writing - review \& editing, Supervision, Resources, Project administration, Methodology, Investigation, Formal analysis, Conceptualization. 
\textbf{Simone Vantini}: Writing - review \& editing, Supervision, Project administration, Methodology, Investigation, Formal analysis, Conceptualization.

\section*{Declaration of competing interest}
The authors declare that they have no known competing financial interests or personal relationships that could have appeared to influence the work reported in this paper.

\section*{Declaration of Generative AI and AI-assisted technologies in the writing process}
During the preparation of this work, the authors used ChatGPT in order to edit the written article and rewrite some pieces aiming for fluency and clarity. After using this tool/service, the authors reviewed and edited the content as needed and take full responsibility for the content in the publication.

\section*{Data availability}
Code to replicate all results in this paper can be accessed at \url{https://github.com/StefaniaColombo/Train-Delays.git}, together with a synthetic version of the dataset.

\section*{Acknowledgments}
S.C., A.G.Z., F.I. and S.V. thank Trenord for the collaboration and for sharing the data used in this work. In particular we thank Dr. Marta Galvani and Dr. Giovanni Chiodi for their support and valuable insights. We also thank Prof. Piercesare Secchi for the insightful suggestions provided. The authors acknowledge the support by MUR, Italy, grant Dipartimento di Eccellenza 2023-2027. This study was founded by the European Union- NextGenerationEU, in the framework of the GRINS- Growing Resilient, INclusive and Sustainable project (GRINS PE00000018– CUP D43C22003110001). The views and options expressed are solely those for the authors and do not necessarly reflect those of the European Union, nor can the European Union be held responsible for them.

\appendix
\section{Dataset Details}
\label{app1}
This appendix provides basic background information and supporting details relevant to the analysis.
\subsection{Description of the Trenord Dataset}
Table~\ref{tabA1} reports a description of all the variables considered for the analysis.
\begin{table}[H]
\centering 
    \begin{tabularx}{\textwidth}{|>{\raggedright\arraybackslash}p{5cm}|>{\raggedright\arraybackslash}X|}
    \hline
    \textbf{Variable} & \textbf{Description}  \\
    \hline \hline
     \textit{Current station} & Representative code for the current stop (e.g., "S00248") \\ \hline
     \textit{Date} & Scheduled departure date, format "YYYY-MM-DD" (e.g., "2023-09-01")\\ \hline
     \textit{Day of the week} & \begin{itemize}
         \item [$-$] 0: Monday;
         \item [$-$] 1: Tuesday;
         \item [$-$] 2: Wednesday;
         \item [$-$] 3: Thursday;
         \item [$-$] 4: Friday;
         \item [$-$] 5: Saturday;
         \item [$-$] 6: Sunday and holidays;
         \item [$-$] -1: race with invalid data.
     \end{itemize} \\ \hline
     \textit{Mission code} & Code of the specific mission (e.g., "24697") \\ \hline     
     \textit{Main route code} & Code of the main route (e.g., "D001") \\ \hline
    \end{tabularx}
\end{table}

\begin{table}[H]
\centering 
    \begin{tabularx}{\textwidth}{|>{\raggedright\arraybackslash}p{5cm}|>{\raggedright\arraybackslash}X|}
     \hline
     \textit{Line code} & Code of the specific line (e.g., "S6") \\ \hline
     \textit{Departure station} & Representative code for the departure station \\ \hline
     \textit{Arrival station} & Representative code for the arrival station \\ \hline
     \textit{Scheduled departure} & Date and time of departure, format "YYYY-MM-DD hh:mm:ss" (e.g., "2023-09-01 06:10:00") \\ \hline
    \textit{Scheduled arrival} & Date and time of arrival\\ \hline
     \textit{Progressive index of the stop} & Index describing the progressive order of stops in the line \\ \hline
     \textit{Station scheduled entry} & Date and time of entry in the current station \\ \hline
     \textit{Station entry delay} & Delay between scheduled and actual entry times in the current station, expressed in minutes \\ \hline
     \textit{Station scheduled exit} & Date and time of exit from the current station \\ \hline
     \textit{Station exit delay} & Delay between scheduled and actual exit times from the current station \\ \hline
     \textit{Boarded} & Number of the passengers boarded at the current stop \\ \hline
     \textit{Alighted} & Number of the passengers alighted at the current stop \\ \hline
     \textit{Cancelled Stop} & \begin{itemize}
         \item [$-$] True: cancelled stop;
         \item [$-$]False: otherwise.
     \end{itemize} \\ \hline
    \end{tabularx}
    \\[10pt]
    \caption{Description of the variables used in the study.}
    \label{tabA1}
\end{table}

\subsection{Description of the Meteorological Dataset}
Table~\ref{tabA2} reports a description of the variables considered for the analysis.
\begin{table}[H]
\centering 
    \begin{tabularx}{\textwidth}{|>{\raggedright\arraybackslash}p{5cm}|>{\raggedright\arraybackslash}X|}
    \hline
    \textbf{Variable} & \textbf{Description} \\
    \hline \hline
     \textit{Location} & Name of the city where the train station is located (used for weather mapping) \\ \hline
     \textit{Date} & Date of observation, format "YYYY-MM-DD" (e.g., "2023-09-01") \\ \hline
     \textit{Mean temperature} & Temperature averaged over the day (measured in Celsius Degrees) \\ \hline
     \textit{Visibility} & Daily visibility level (measured in Km) \\ \hline     
     \textit{Mean wind velocity} & Average wind speed (measured in Km/h) \\ \hline
     \textit{Weather event type} & Kind of adverse weather, if present - rain, fog, storm, none\\
    \hline
    \end{tabularx}
    \\[10pt]
    \caption{Description of the meteorological variables used in the study.}
    \label{tabA2}
\end{table}

\section{Additional Results}
\label{app2}
\subsection{Supplementary Transition Probability Plots}
\label{app2: ProbPlots}
Figure~\ref{figB1} includes the detailed transition probability plots segmented by direction, time slot, and route section. While most patterns remain consistent, some variability is observed in Zone 1 (Varese-Gallarate), likely due to sample size limitations.
\begin{figure}[H]
    \centering
     \makebox[\textwidth][c]{%
        \includegraphics[width=0.95\textwidth]{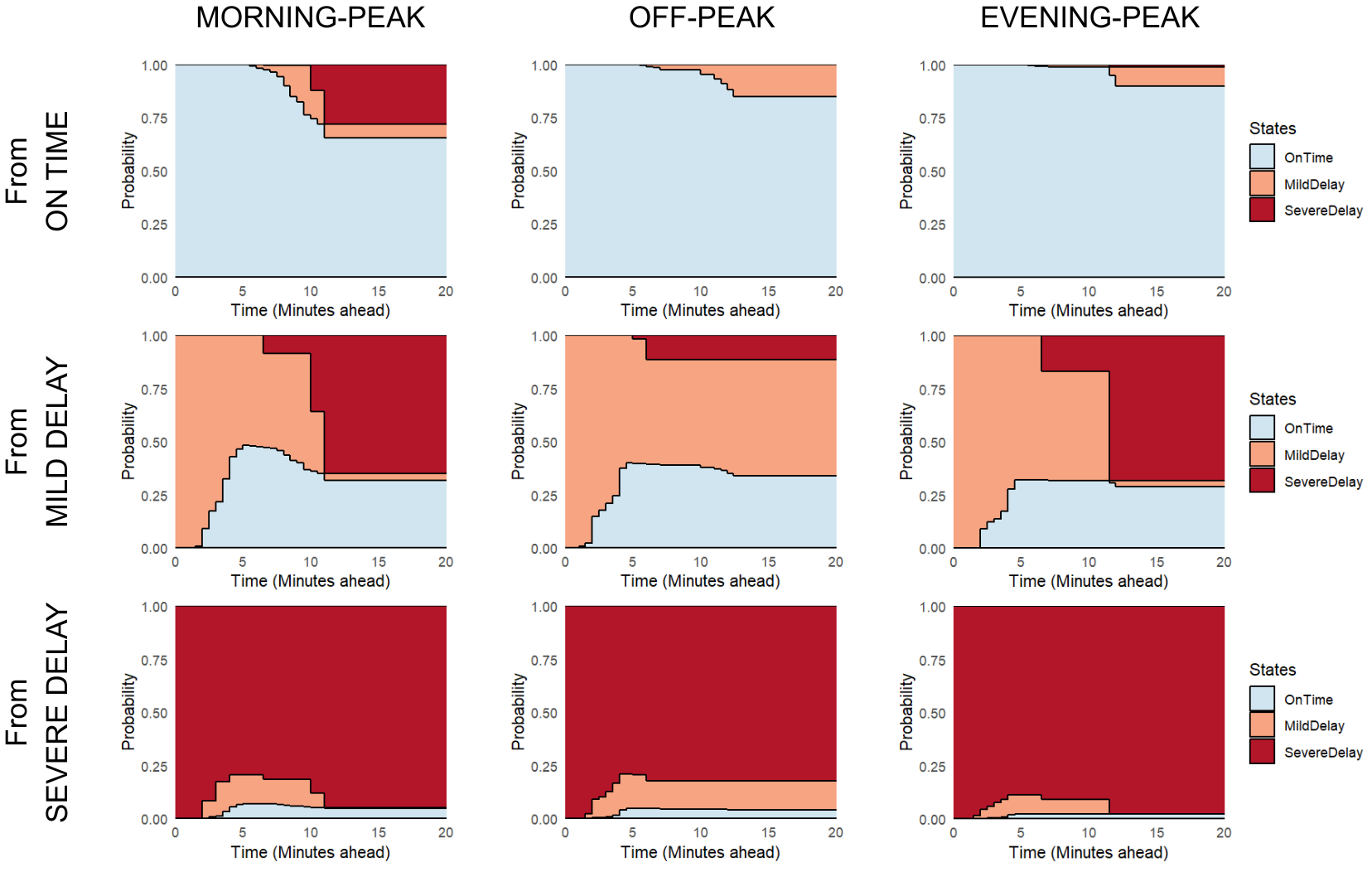}%
    }
    \caption*{(a) Zone 1 (Varese-Gallarate).}
\end{figure}

\begin{figure}[H]
    \centering
     \makebox[\textwidth][c]{%
        \includegraphics[width=0.95\textwidth]{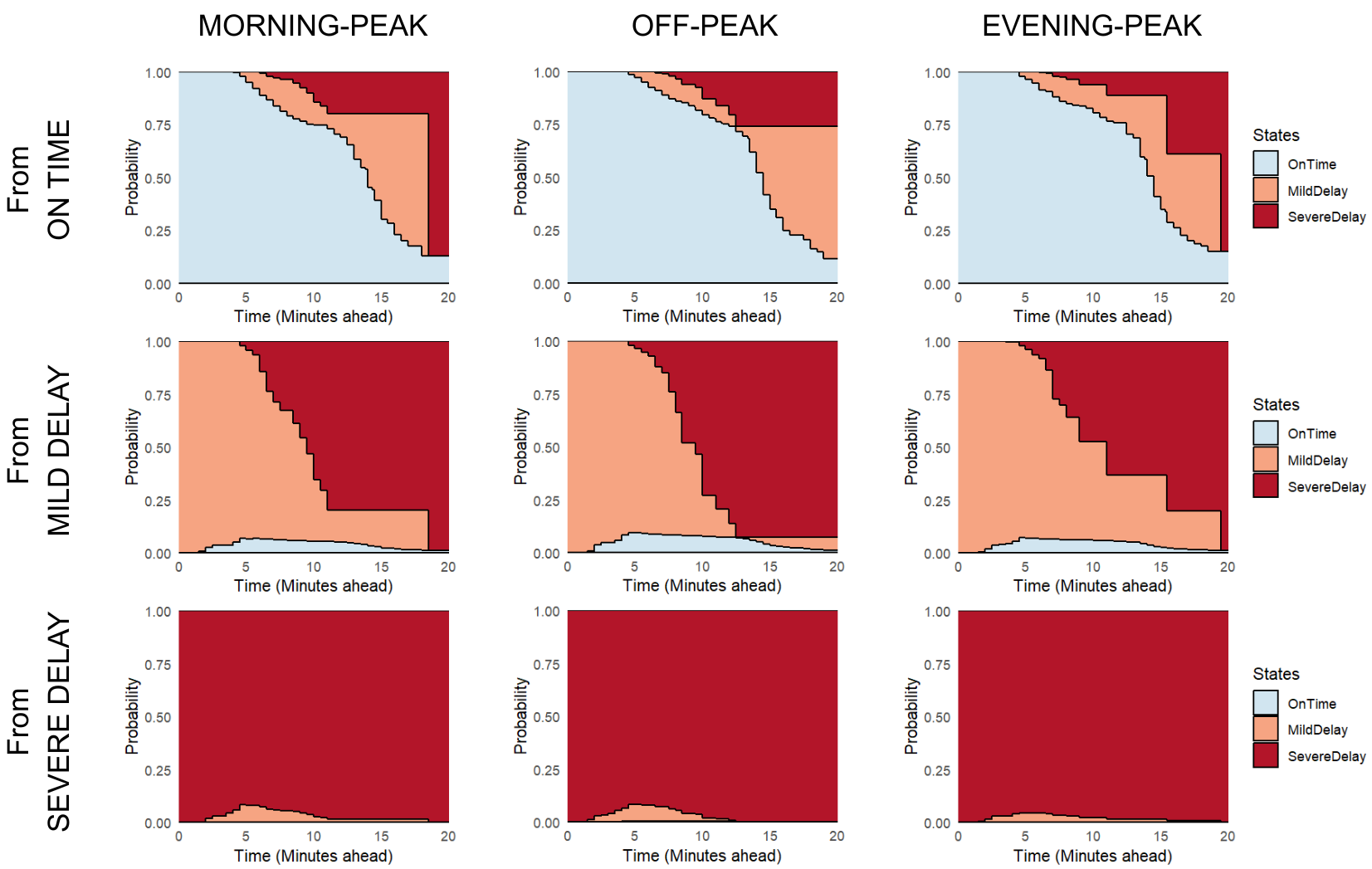}%
    }
    \caption*{(b) Zone 2 (Busto Arsizio-Rho Fiera).}
\end{figure}

\begin{figure}[H]
    \centering
     \makebox[\textwidth][c]{%
        \includegraphics[width=0.95\textwidth]{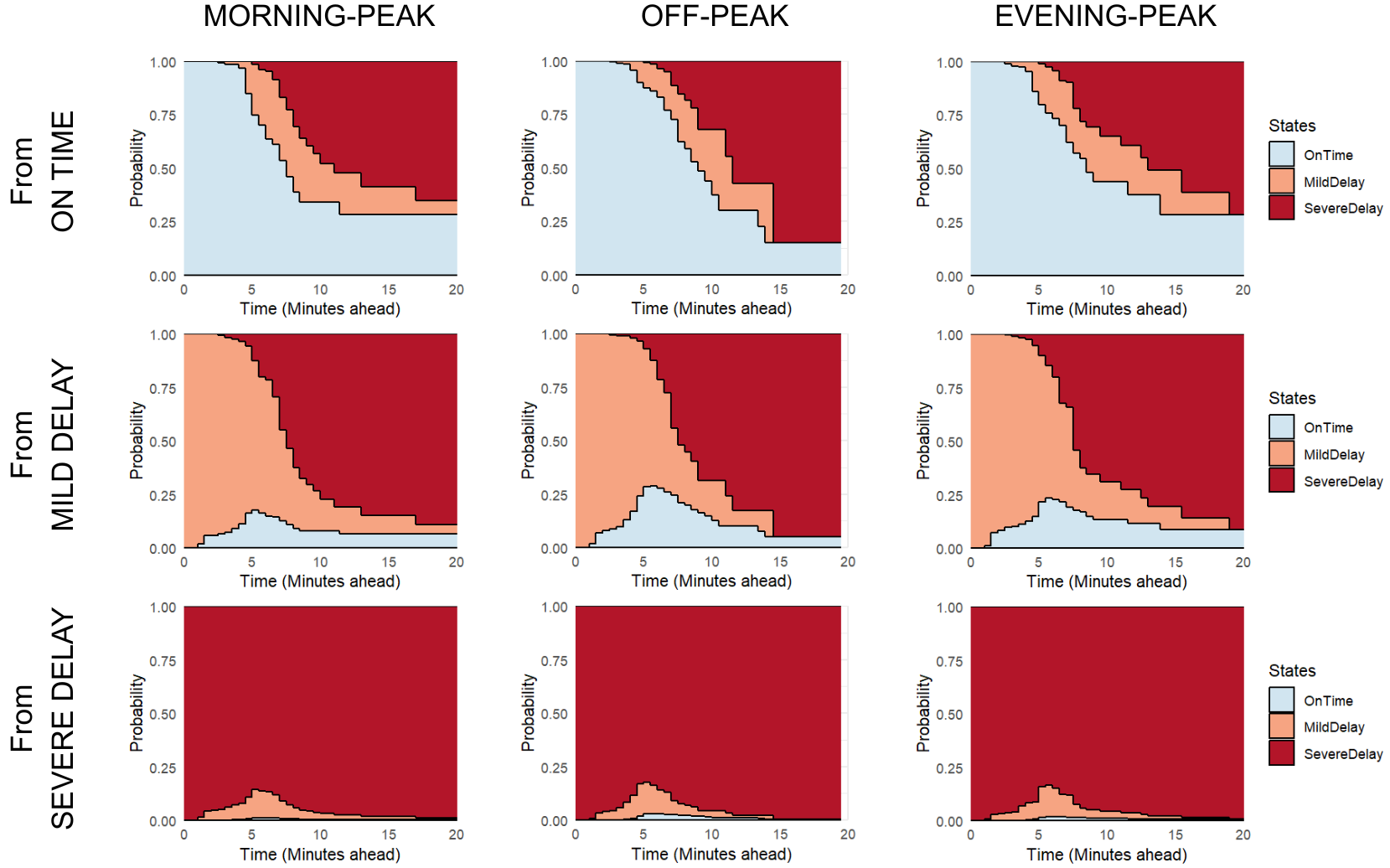}%
    }
    \caption*{(c) Zone 3 (Milano Certosa-Milano Forlanini).}
\end{figure}

\begin{figure}[H]
    \centering
    \makebox[\textwidth][c]{%
        \includegraphics[width=0.95\textwidth]{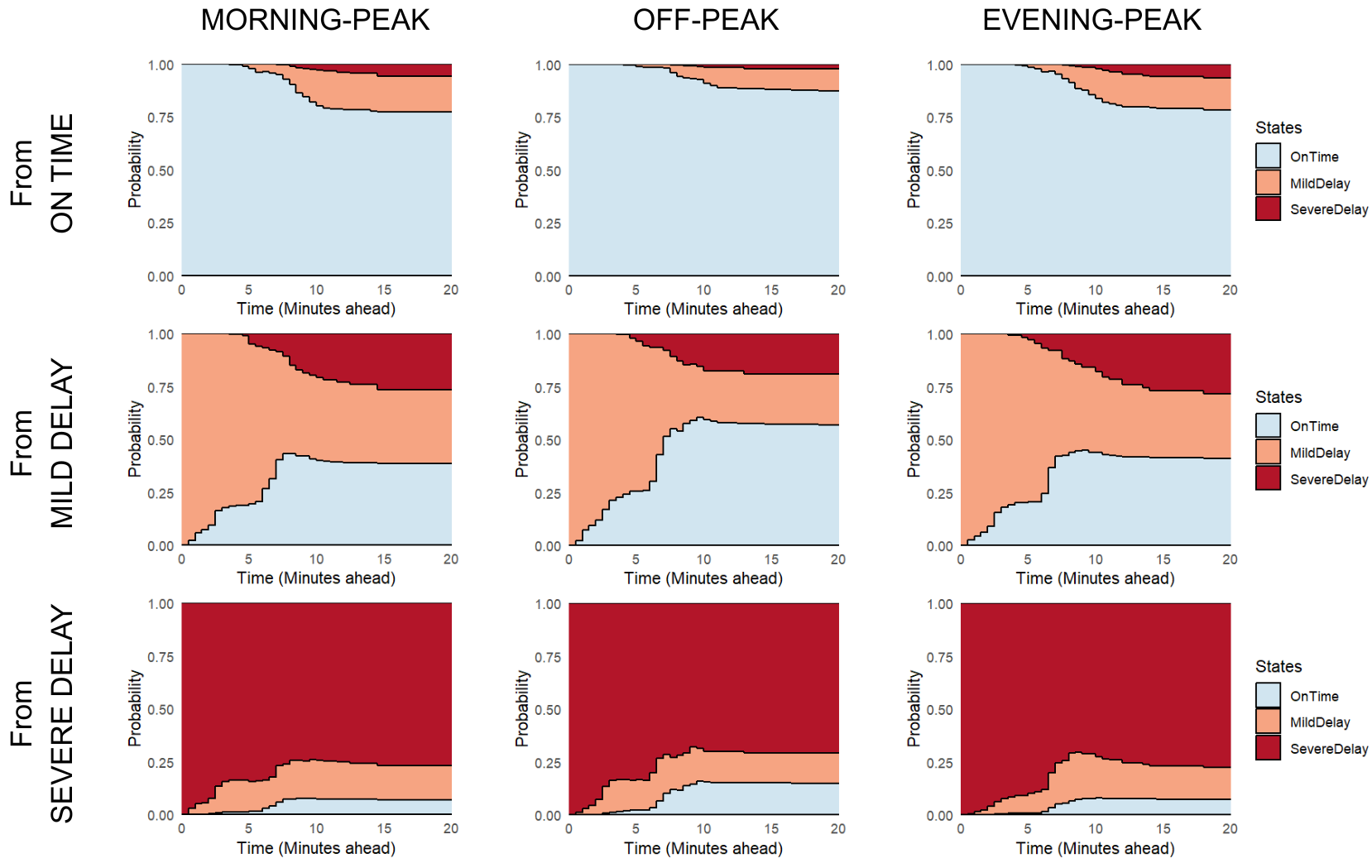}%
    }
    \caption*{(d) Zone 4 (Segrate-Treviglio).}
    \caption{Nonparametric estimates of stacked transition probabilities by time since entry into the current state, for direction Varese \(\to\) Treviglio. For each zone segmentation ((a), (b), (c), (d)), time slots are shown vertically, starting states horizontally. Color legend: light blue = On Time, orange = Mild Delay, red = Severe Delay.}
    \label{figB1}
\end{figure}

\subsection{Effect of Time Slot Covariates on Delay Transitions}
\label{app2: Timeslot}
\begin{figure}[H]
    \centering
    \hspace{-1.6cm}
    \makebox[\textwidth]{%
        \begin{subfigure}{0.48\textwidth}
            \centering
            \includegraphics[scale=0.25]{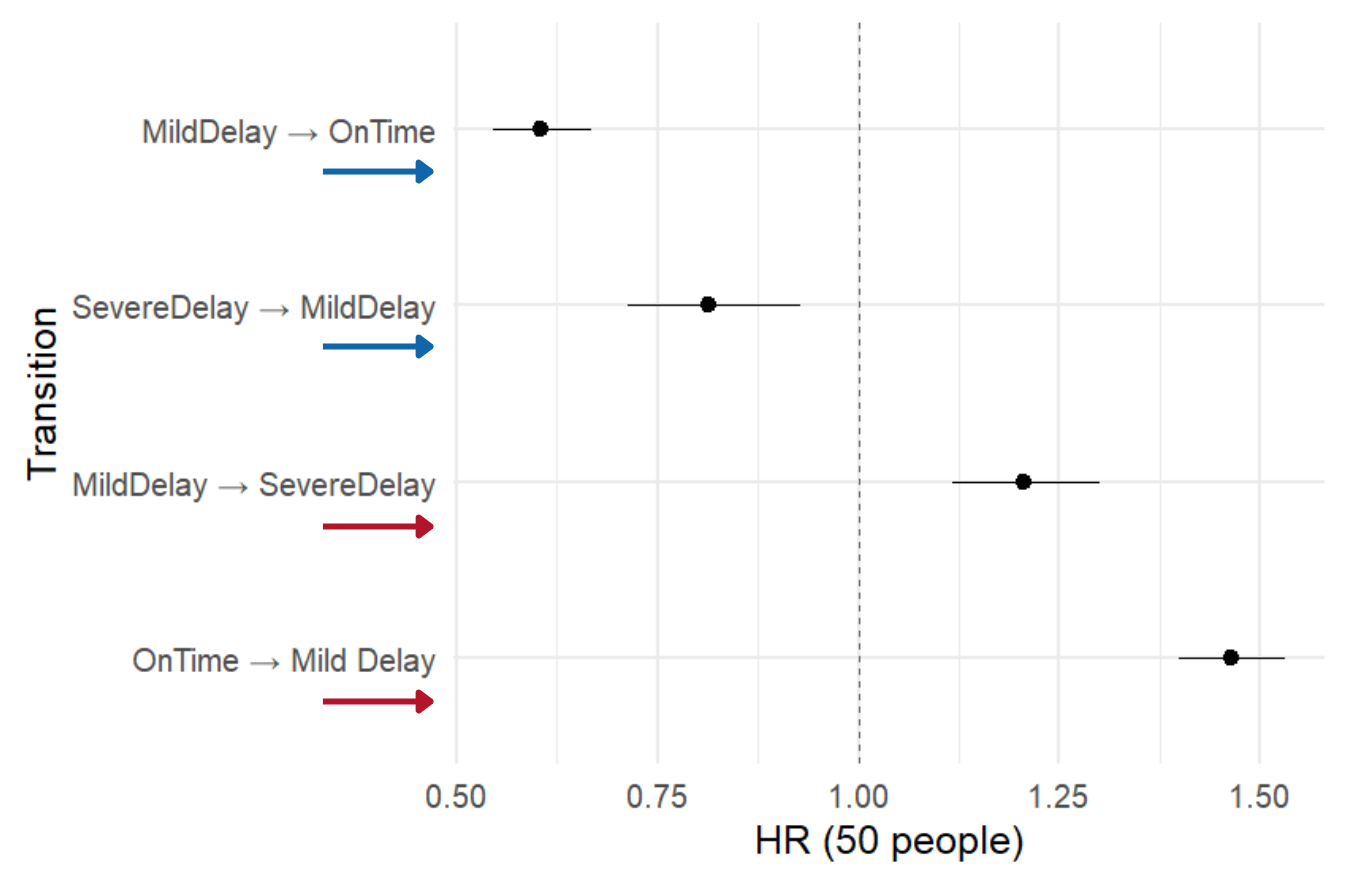}
            \caption{Boarded passengers}
        \end{subfigure}
        \hspace{0.04\textwidth} 
        \begin{subfigure}{0.48\textwidth}
            \centering
            \includegraphics[scale=0.26]{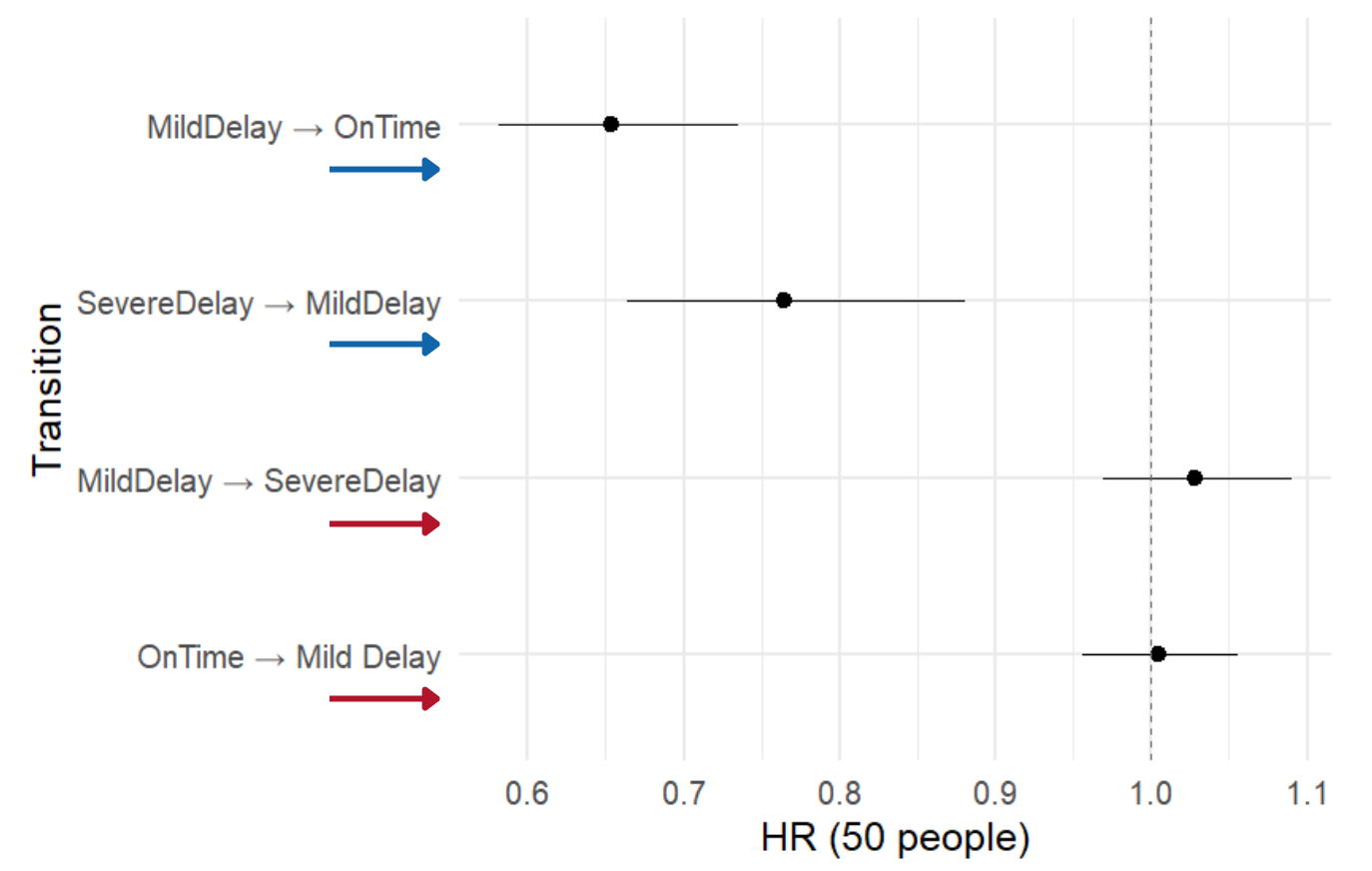}
            \caption{Alighted passengers}
        \end{subfigure}
    }

    \vspace{1em} 
    \hspace{-1.5cm}
    \makebox[\textwidth]{%
        \begin{subfigure}{0.48\textwidth}
            \centering
            \includegraphics[scale=0.26]{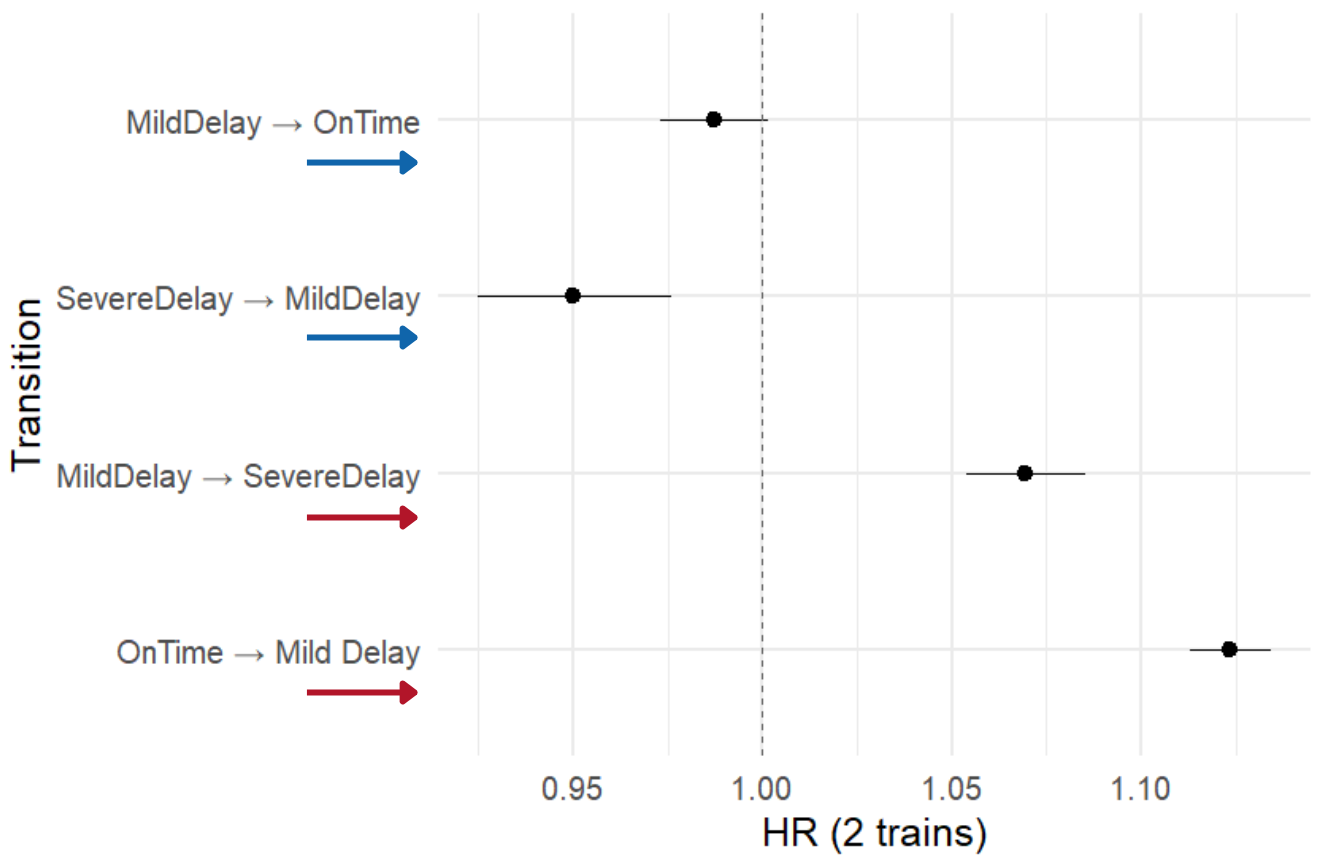}
            \caption{Train saturation}
        \end{subfigure}
        \hspace{0.04\textwidth}
        \begin{subfigure}{0.48\textwidth}
            \centering
            \includegraphics[scale=0.26]{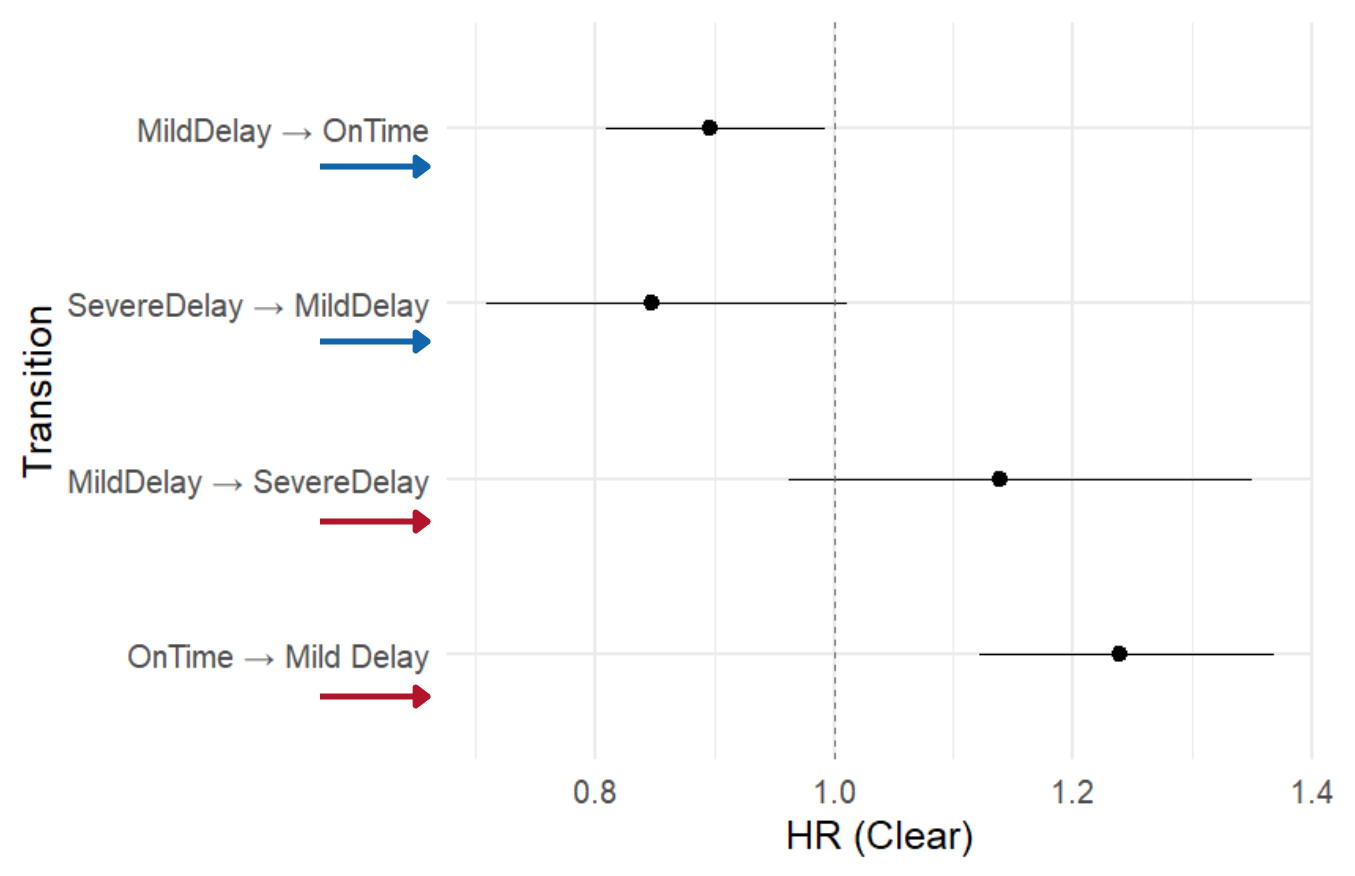}
            \caption{Adverse weather}
        \end{subfigure}
    }

    \vspace{1em}
    \hspace{-1.5cm}
    \makebox[\textwidth]{%
        \begin{subfigure}{0.48\textwidth}
            \centering
            \includegraphics[scale=0.26]{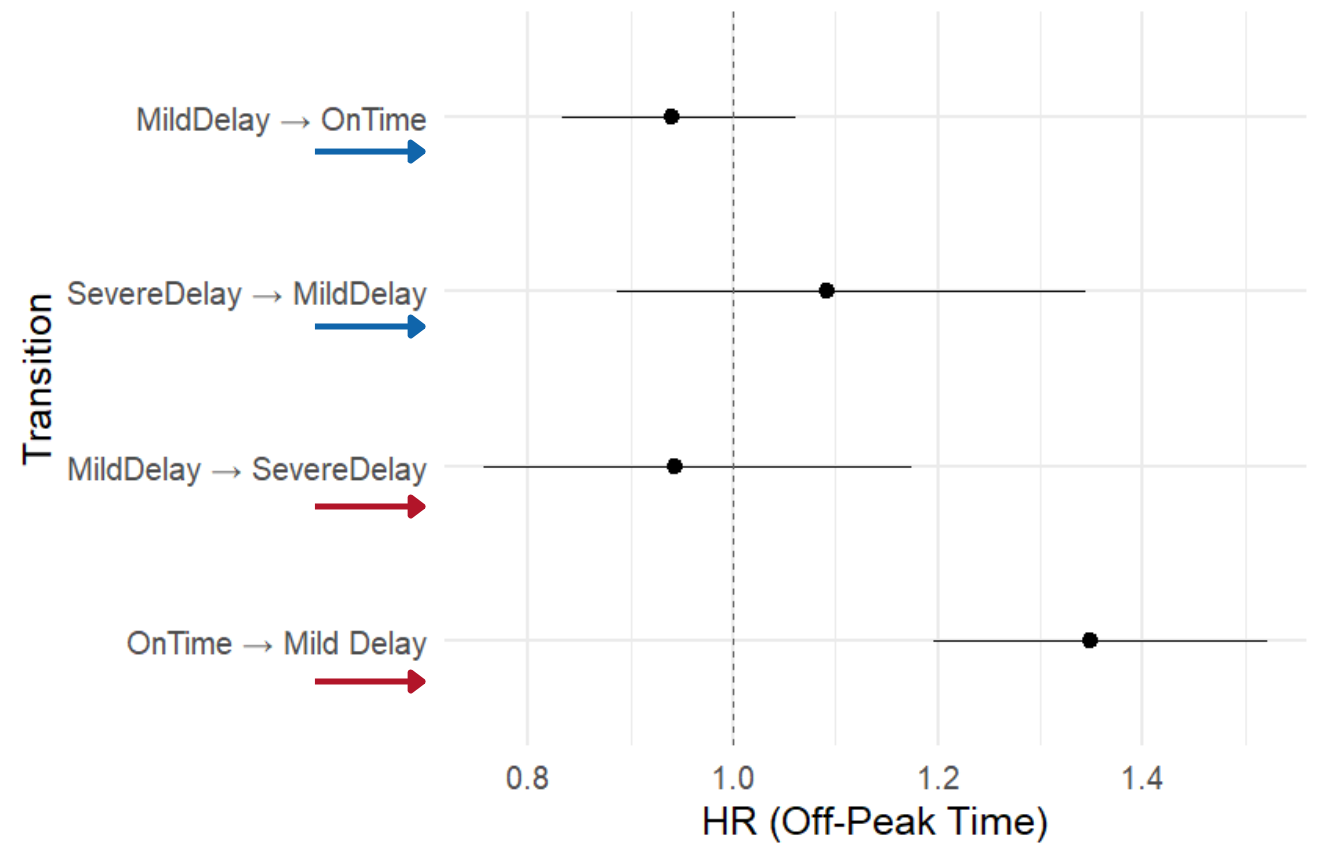}
            \caption{Morning-Peak Time}
        \end{subfigure}
        \hspace{0.04\textwidth}
        \begin{subfigure}{0.48\textwidth}
            \centering
            \includegraphics[scale=0.26]{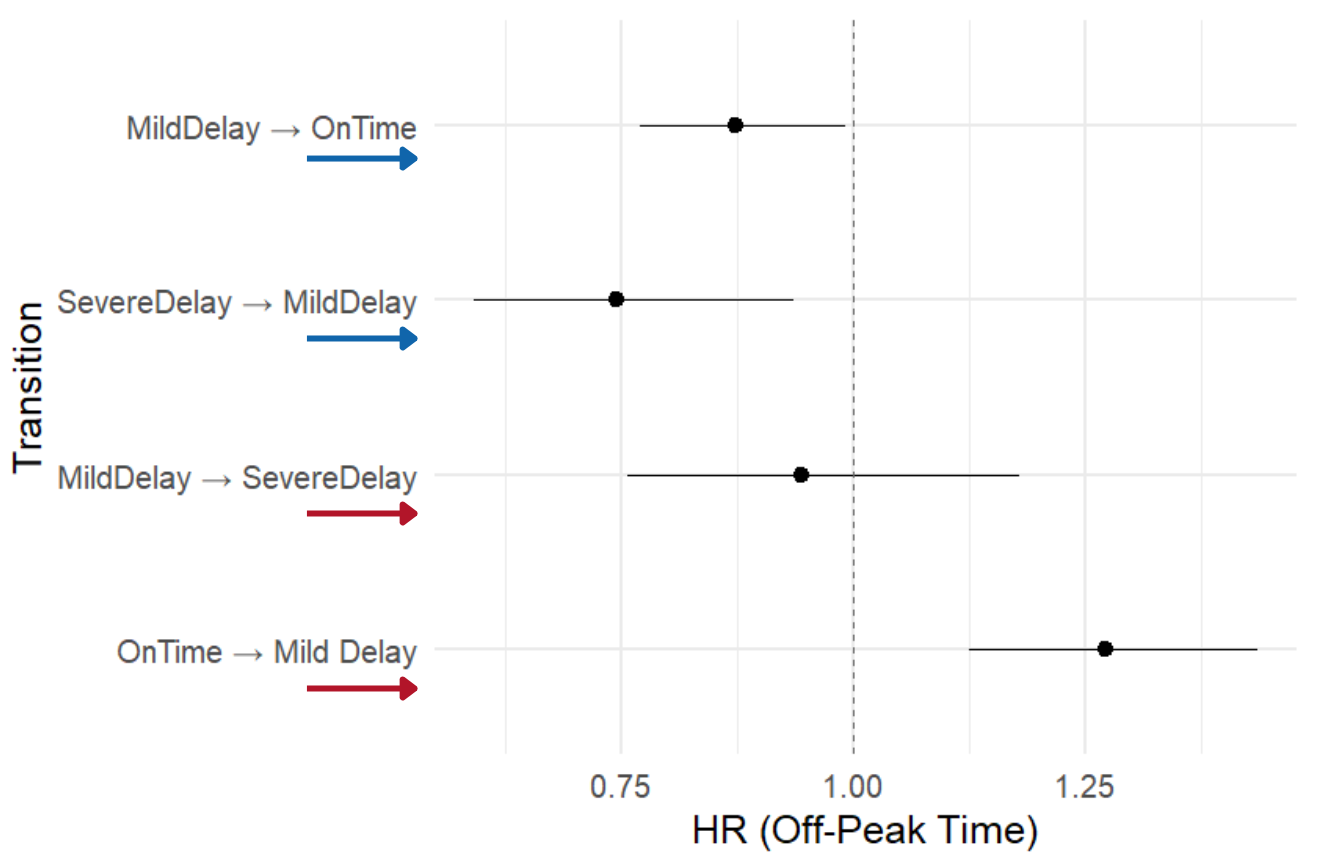}
            \caption{Evening-Peak Time}
        \end{subfigure}
    }
    \caption{Hazard ratios and 95\% confidence intervals for temporal covariates. Color legend: red = Deteriorating transitions (On Time \(\to\) Mild Delay, Mild Delay \(\to\) Severe Delay); blue = Recovering transitions (Severe Delay \(\to\) Mild Delay, Mild Delay \(\to\) On Time).}
    \label{figB5}
\end{figure}

Figure~\ref{figB5}, provide an overview of the estimated hazard ratios (HRs) and corresponding 95\% confidence intervals, illustrating how each factor influences the transition intensities between delay states. Overall, the results reveal coherent and interpretable patterns across transitions. Higher boarding volumes significantly increase the hazard of delay deterioration — especially from On Time to Mild Delay — while reducing recovery probabilities. Alighting passengers show weaker effects, mainly hindering recovery transitions. Increased train frequency similar amplifies deterioration risks and limits recovery, reflecting network congestion under high traffic density. Adverse weather exerts a smaller but still significant influence on the onset of mild delays. Temporal regularities are evident across time slots: with respect to the Off-Peak, the Morning-Peak period exhibits the strongest deterioration dynamics, while the Evening-Peak shows milder effects and slightly higher recovery potential.

Table~\ref{tabB1}, provides the full Cox model output for the set of temporal covariates.
\begin{table}[H]
\centering
\begin{tabular}{|l|c|c|c|c|c|c|}
\hline
Trans & Cov & Coef & HR & 95\% CI & Std Err & p-value \\
\hline \hline
0 → 1 & Boarded & 0.3811 & 1.4640 & [1.3998, 1.5311] & 0.0229 & < 2\(\cdot10^{-16}\) \\
 & Alighted & 0.0046 & 1.0046 & [0.9563, 1.0554] & 0.0252 & 0.8544\\
 & Train Freq & 0.1165 & 1.1235 & [1.1129, 1.3684] & 0.0252 & 0.0049\\
 & Weather & 0.2140 & 1.2387 & [1.1212, 1.1343] & 0.0049 &  < 2\(\cdot10^{-16}\)\\
 & Morning & 0.2994 & 1.3491 & [1.1960, 1.5217] & 0.0252 & 1.1\(\cdot10^{-6}\)\\
 & Evening & 0.2398 & 1.2710 & [1.1254, 1.4354] & 0.0621 & 0.0001\\
 \hline
 1 → 2 & Boarded & 0.1865 & 1.2051 & [1.1164, 1.3007] & 0.0390 & 1.7\(\cdot10^{-6}\) \\
 & Alighted & 0.0278 & 1.0282 & [0.9696, 1.0903] & 0.0300 & 0.3534\\
 & Train Freq & 0.0670 & 1.0693 & [1.0537, 1.0851] & 0.0075 &  < 2\(\cdot10^{-16}\)\\
 & Weather & 0.1303 & 1.1392 & [0.9617, 1.3494] & 0.0864 & 0.1315\\
 & Morning & -0.0586 & 0.9431 & [0.7577, 1.1737] & 0.1116 & 0.5995\\
 & Evening & -0.0568 & 0.9448 & [0.7573, 1.1786] & 0.1129 & 0.6146\\
 \hline
 2 → 1 & Boarded & -0.2066 & 0.8133 & [0.7127, 0.9282] & 0.0674 & 0.0022 \\
 & Alighted & -0.2683 & 0.7647 & [0.6639, 0.8808] & 0.0721 & 0.0002\\
 & Train Freq & -0.0511 & 0.9502 & [0.9252, 0.9759] & 0.0136 & 0.0002\\
 & Weather & -0.1660 & 0.8470 & [0.7098, 1.0107] & 0.090 & 0.0656\\
 & Morning & 0.0879 & 1.0919 & [0.8870, 1.3440] & 0.1060 & 0.4071\\
 & Evening & -0.2953 & 0.7443 & [0.5919, 0.9360] & 0.1169 & 0.0116\\
 \hline
 1 → 0 & Boarded & -0.5043 & 0.6039 & [0.5461, 0.6678] & 0.0513 &  < 2\(\cdot10^{-16}\) \\
 & Alighted & -0.4245 & 0.6541 & [0.5819, 0.7352] & 0.0597 & 1.12\(\cdot10^{-12}\)\\
 & Train Freq & -0.0128 & 0.9872 & [0.9733, 1.0014] & 0.0073 & 0.0768\\
 & Weather & -0.1092 & 0.8965 & [0.8093, 0.9932] & 0.0523 & 0.0366\\
 & Morning & -0.0610 & 0.9408 & [0.8338, 1.0614] & 0.0616 & 0.3214\\
 & Evening & -0.1350 & 0.8737 & [0.7702, 0.9911] & 0.0643 & 0.0358\\
\hline
\end{tabular}
\caption{Complete Cox model results for temporal covariates (Varese–Treviglio). In Trans column states are defined as: 0 = On Time, 1 = Mild Delay, 2 = Severe Delay.}
\label{tabB1}
\end{table}

Figures~\ref{figB11},~\ref{figB13},~\ref{figB15} display the probabilities over a 30-minute horizon, considering the one-at-time variation of passenger load, train frequency and weather condition, respectively. Each plot is obtained by analyzing the particular covariate across its observed range, while keeping all the others constant at their mean levels. These results confirm previous insights from the HRs.
\begin{figure}[H]
    \centering
     \makebox[\textwidth][c]{%
        \includegraphics[width=0.7\textwidth]{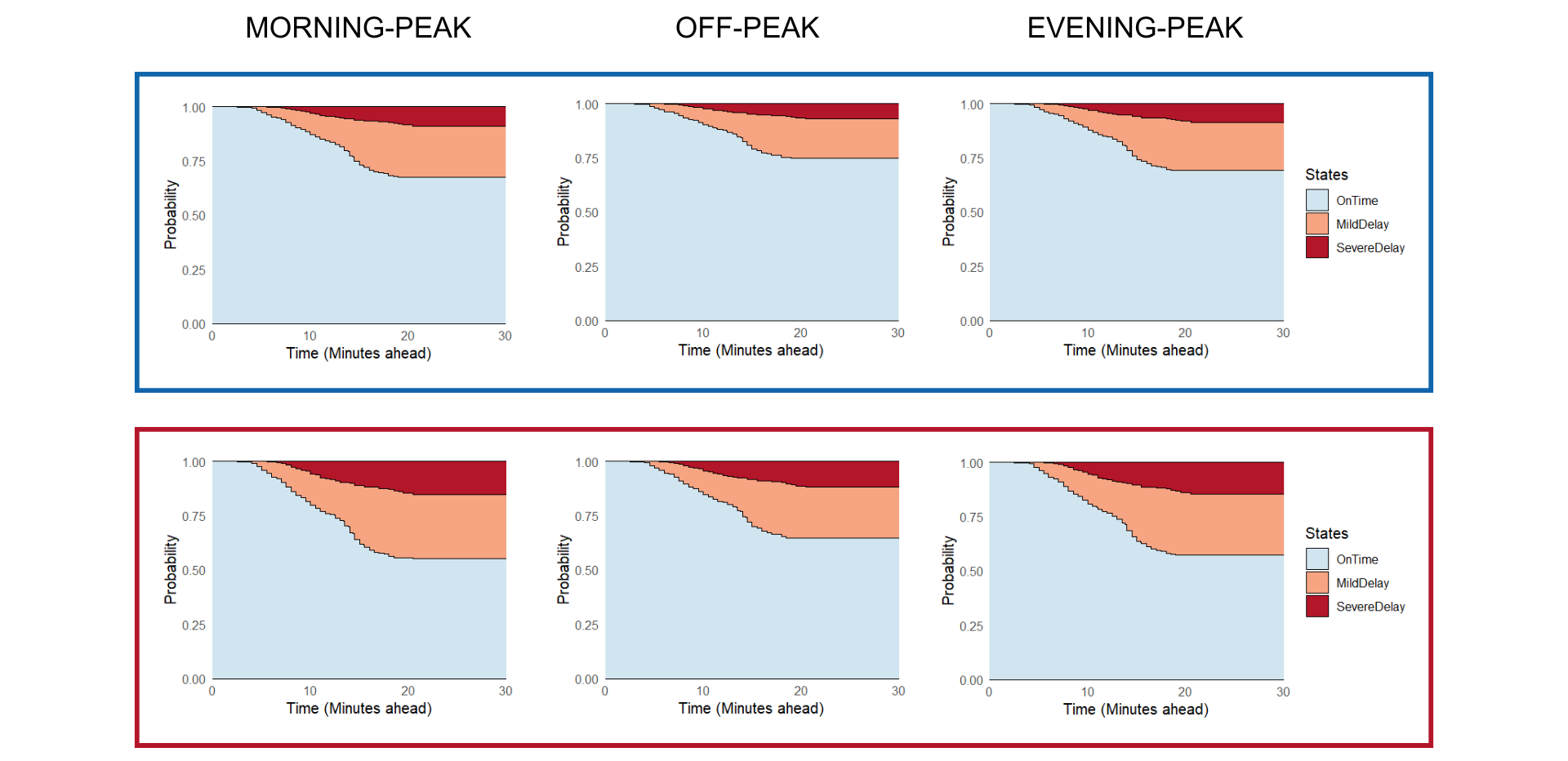}%
    }
    \caption*{(a) From On Time.}
\end{figure}

\begin{figure}[H]
    \centering
    \makebox[\textwidth][c]{%
        \includegraphics[width=0.7\textwidth]{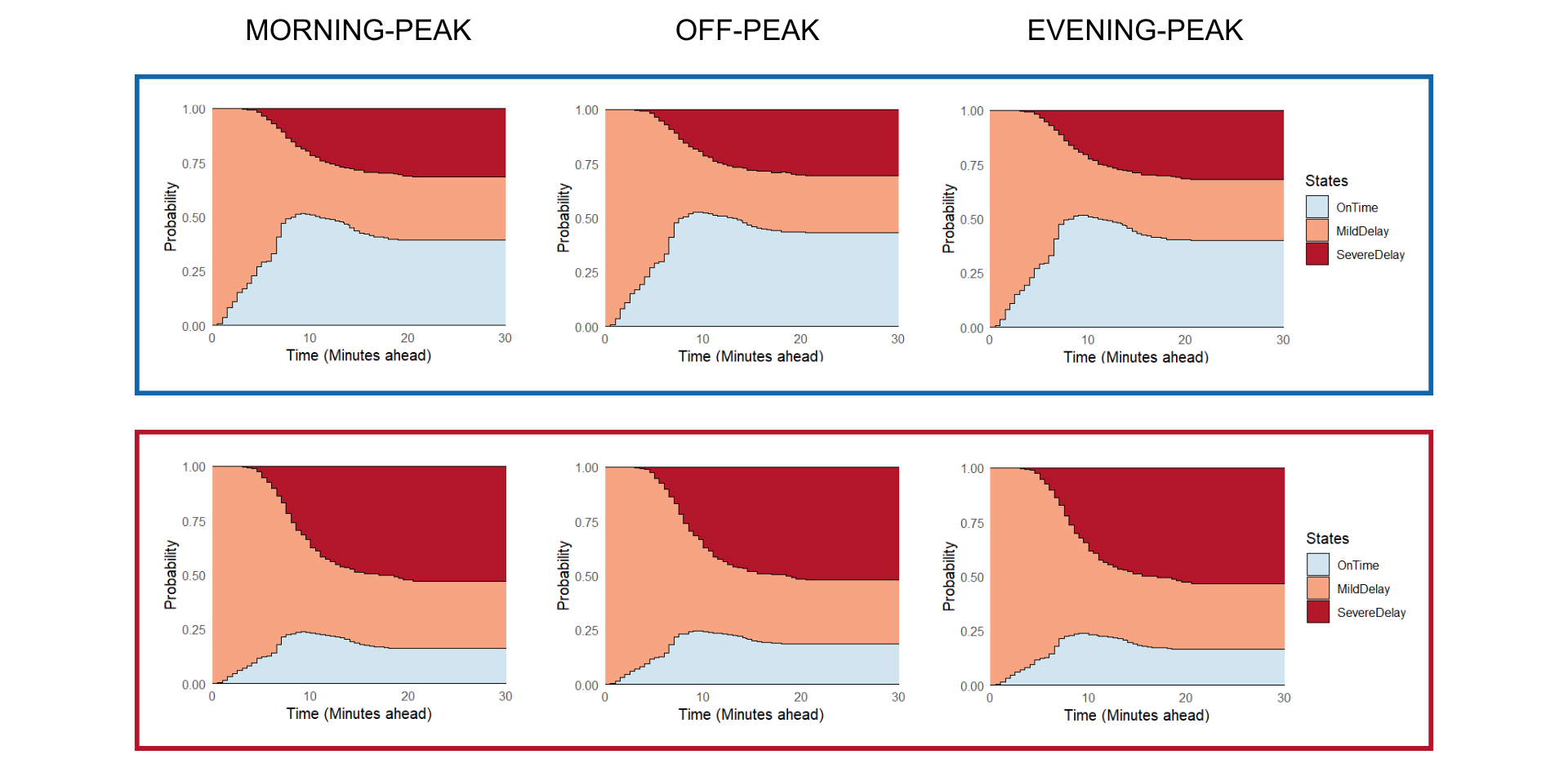}%
    }
    \caption*{(b) From Mild Delay.}
    \caption{Predicted Future State Probabilities for Boarding and Alighting Passengers per Time Slot. Blue boxes indicate low passenger volumes (15th percentile), while red boxes correspond to high volumes (85th percentile).}
    \label{figB11}
\end{figure}

\begin{figure}[H]
    \centering
     \makebox[\textwidth][c]{%
        \includegraphics[width=0.7\textwidth]{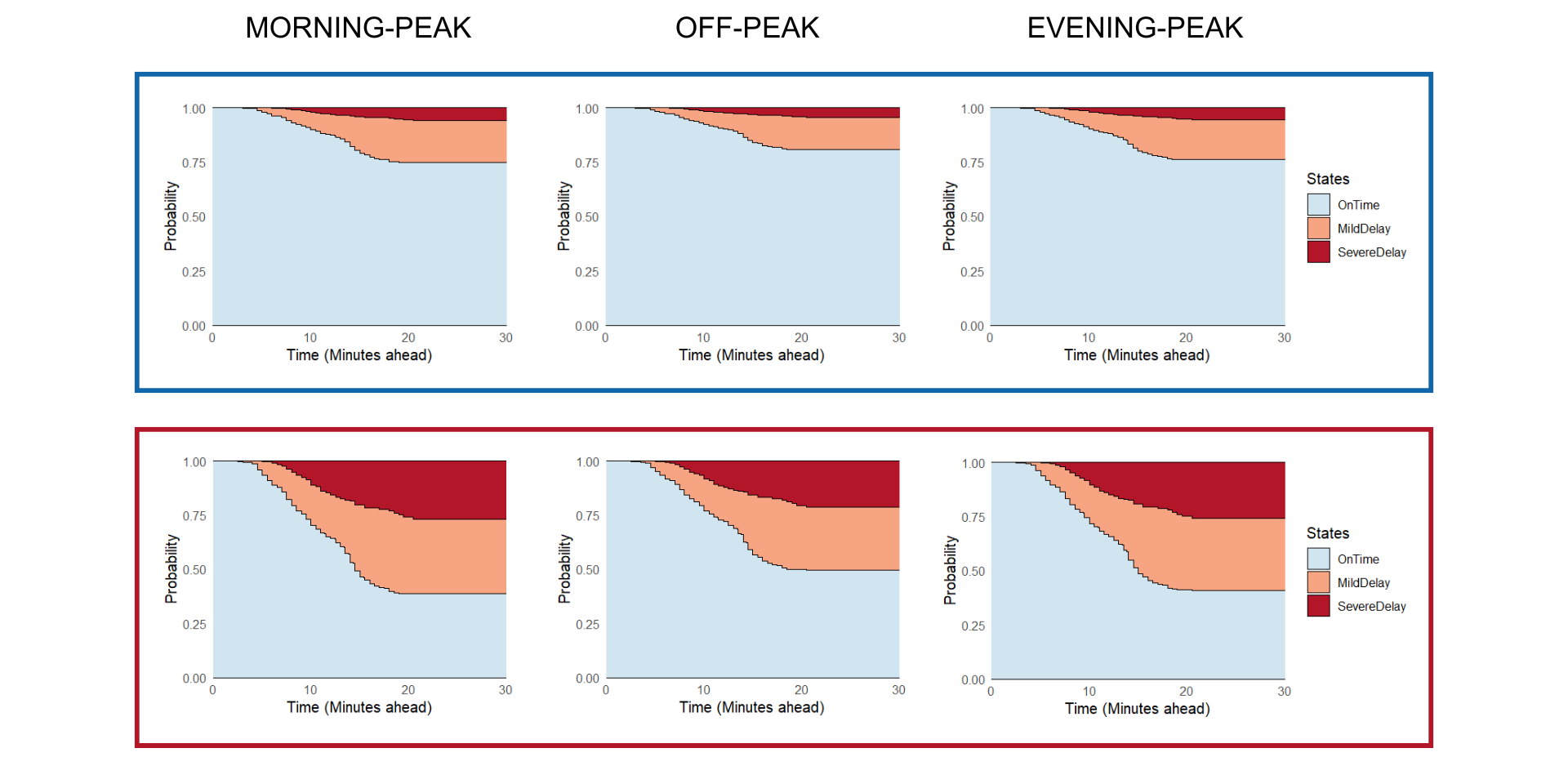}%
    }
    \caption*{(a) From On Time.}
\end{figure}

\begin{figure}[H]
    \centering
    \makebox[\textwidth][c]{%
        \includegraphics[width=0.7\textwidth]{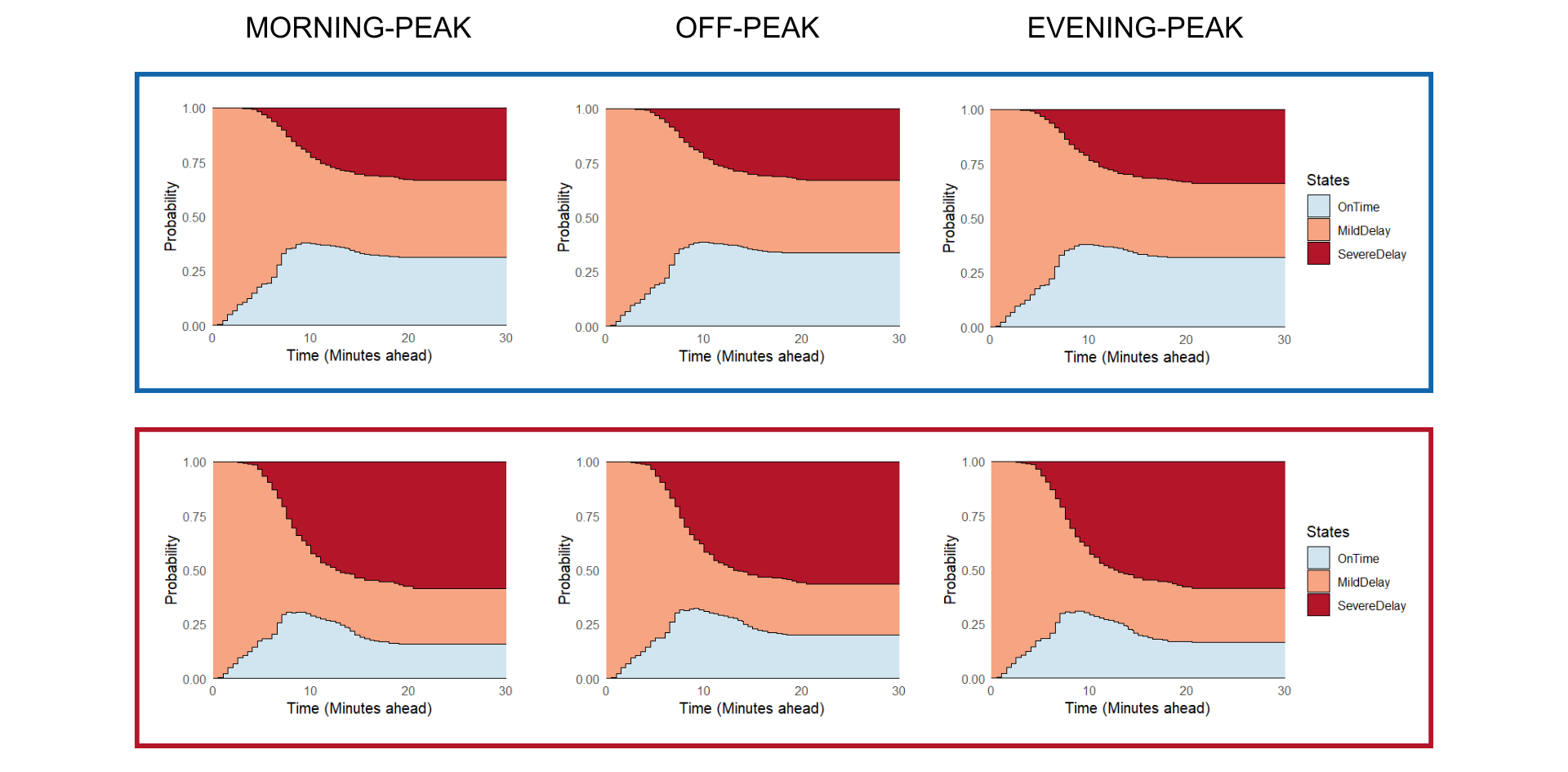}%
    }
    \caption*{(b) From Mild Delay.}
    \caption{Predicted Future State Probabilities for Train Frequency per Time Slot. Blue boxes represent low-frequency stations (4 trains/hour), while red boxes correspond to high-frequency contexts (24 trains/hour).}
    \label{figB13}
\end{figure}

\begin{figure}[H]
    \centering
     \makebox[\textwidth][c]{%
        \includegraphics[width=0.7\textwidth]{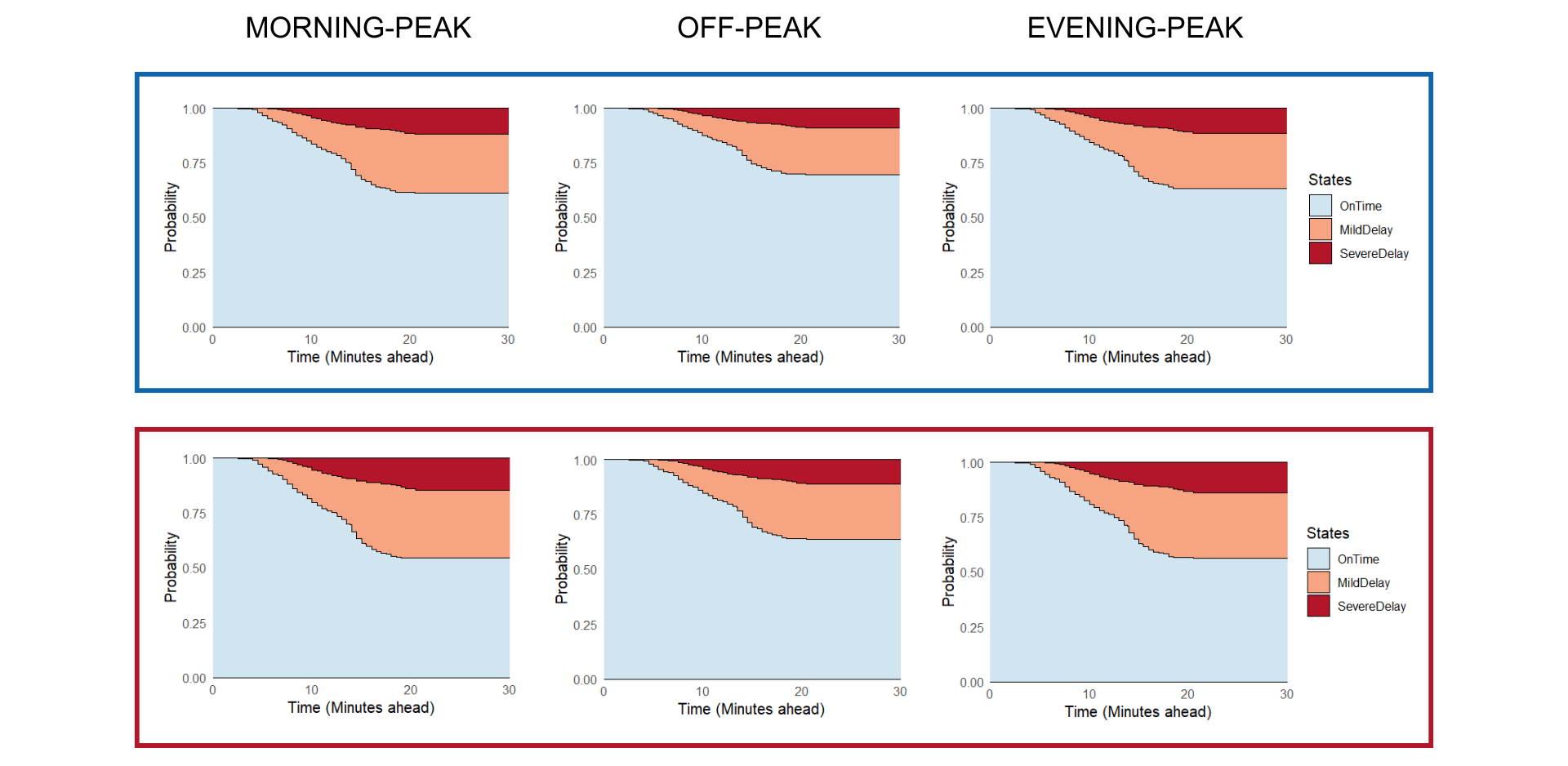}%
    }
    \caption*{(a) From On Time.}
\end{figure}

\begin{figure}[H]
    \centering
    \makebox[\textwidth][c]{%
        \includegraphics[width=0.7\textwidth]{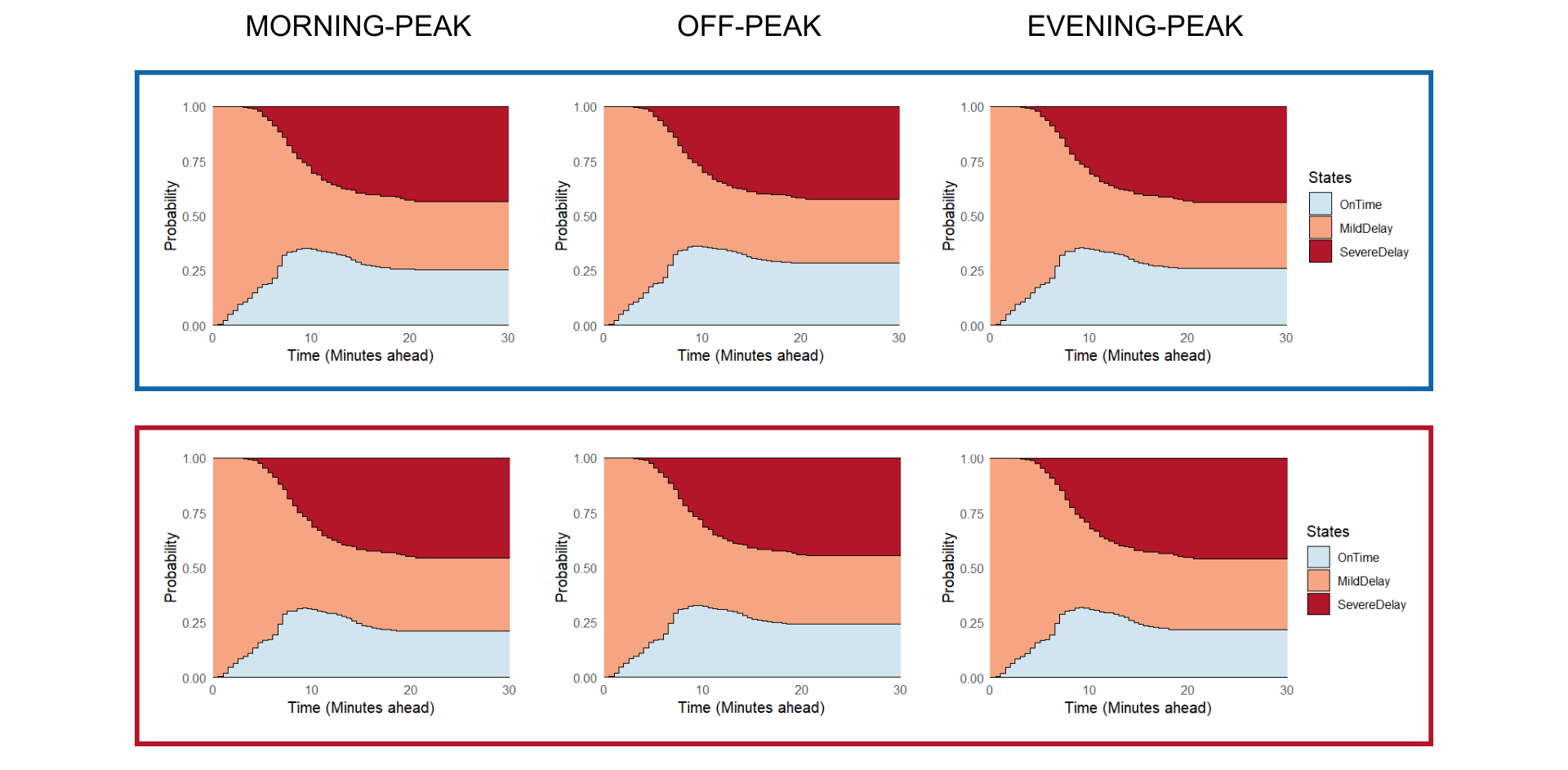}%
    }
    \caption*{(b) From Mild Delay.}
    \caption{Predicted Future State Probabilities for Weather Condition per Time Slot. Blue boxes indicate good weather (no rain/fog/thunderstorms), while red boxes represent adverse conditions.}
    \label{figB15}
\end{figure}

\subsection{Effect of Route Section Covariates on Delay Transitions}
\label{app2: Routesection}
The analysis of the estimated HRs and corresponding 95\% confidence intervals (Figure~\ref{figB17}), shows that, while boarding passengers have limited effect on delay onset, they hinder recovery: larger boarding volumes reduce the probability of returning from Mild Delay to On Time. In contrast, alighting passengers consistently increase the risk of delay escalation and impede recovery, suggesting that high-flow stations act as operational bottlenecks. Increased train frequency also amplifies deterioration risks and limits recovery, though to a lesser extent. Adverse weather exerts an overall non-significant influence, except on the onset of mild delays. This outcome is consistent with the fact that both meteorological and train frequency variables exhibit limited spatial granularity and variability across route segments, preventing meaningful differentiation. Spatial factors, on the other side, further amplify delay effects. Compared to the reference area (Zone 4: Segrate-Treviglio), central segments, particularly Zone 3 (Milano) and Zone 2 (Busto Arsizio-Rho Fiera), exhibit substantially higher risks of delay onset and progression, and lower likelihood of recovery, whereas peripheral segments (Zone 1: Varese-Gallarate) are more resilient. Although hazard ratios in urban zones may appear unusually high, they are coherent with the known operational characteristics of these sections of the line, including dense passenger flows, complex station layouts, and interactions with other rail services.

\begin{figure}[H]
    \centering
    \hspace{-1.6cm}
    \makebox[\textwidth]{%
        \begin{subfigure}{0.48\textwidth}
            \centering
            \includegraphics[scale=0.25]{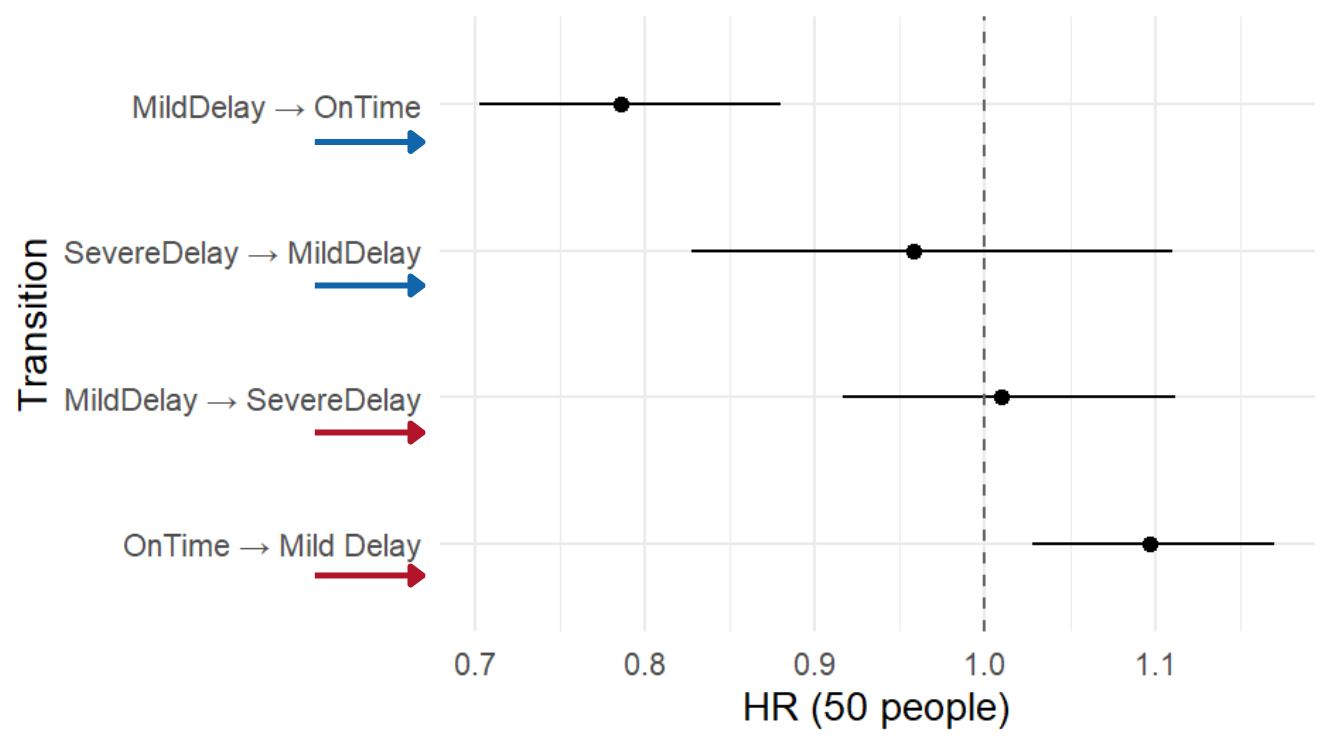}
            \caption{Boarded passengers}
        \end{subfigure}
        \hspace{0.04\textwidth}
        \begin{subfigure}{0.48\textwidth}
            \centering
            \includegraphics[scale=0.25]{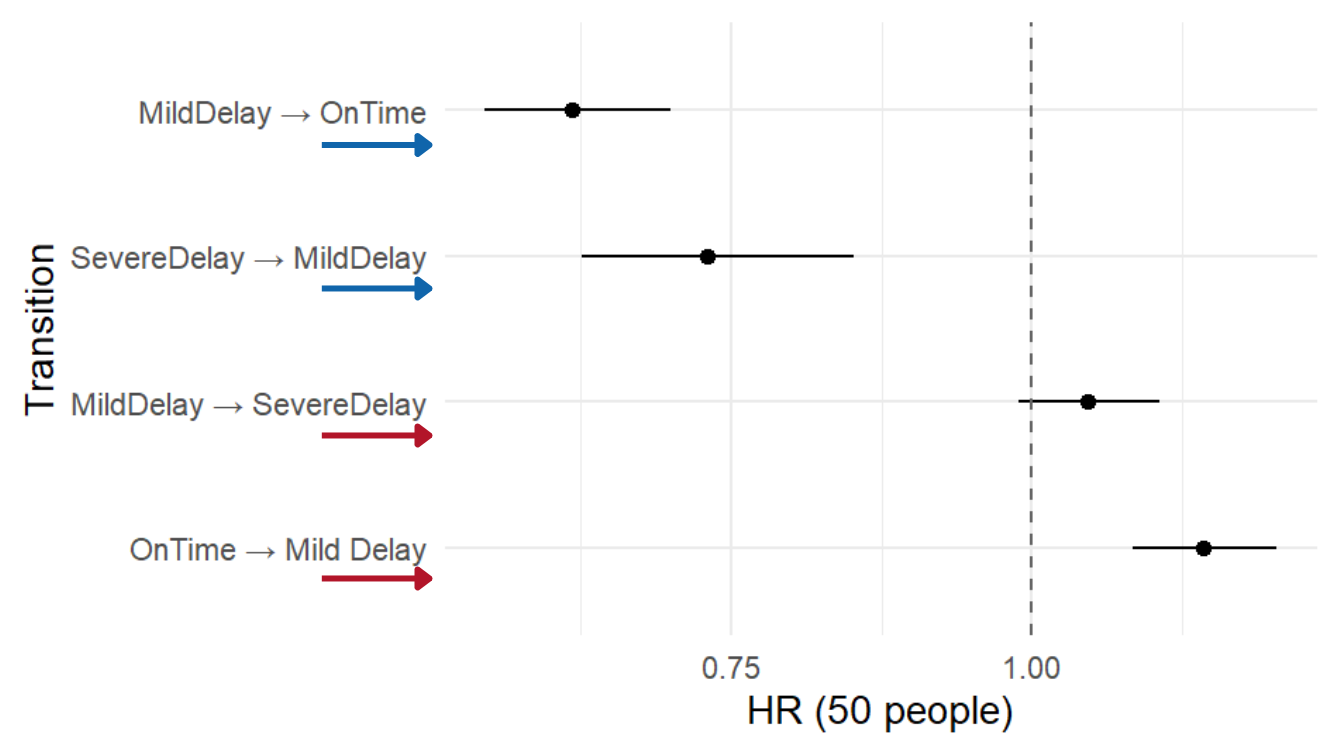}
            \caption{Alighted passengers}
        \end{subfigure}
    }

    \vspace{1em} 
    \hspace{-1.5cm}
    \makebox[\textwidth]{%
        \begin{subfigure}{0.48\textwidth}
            \centering
            \includegraphics[scale=0.25]{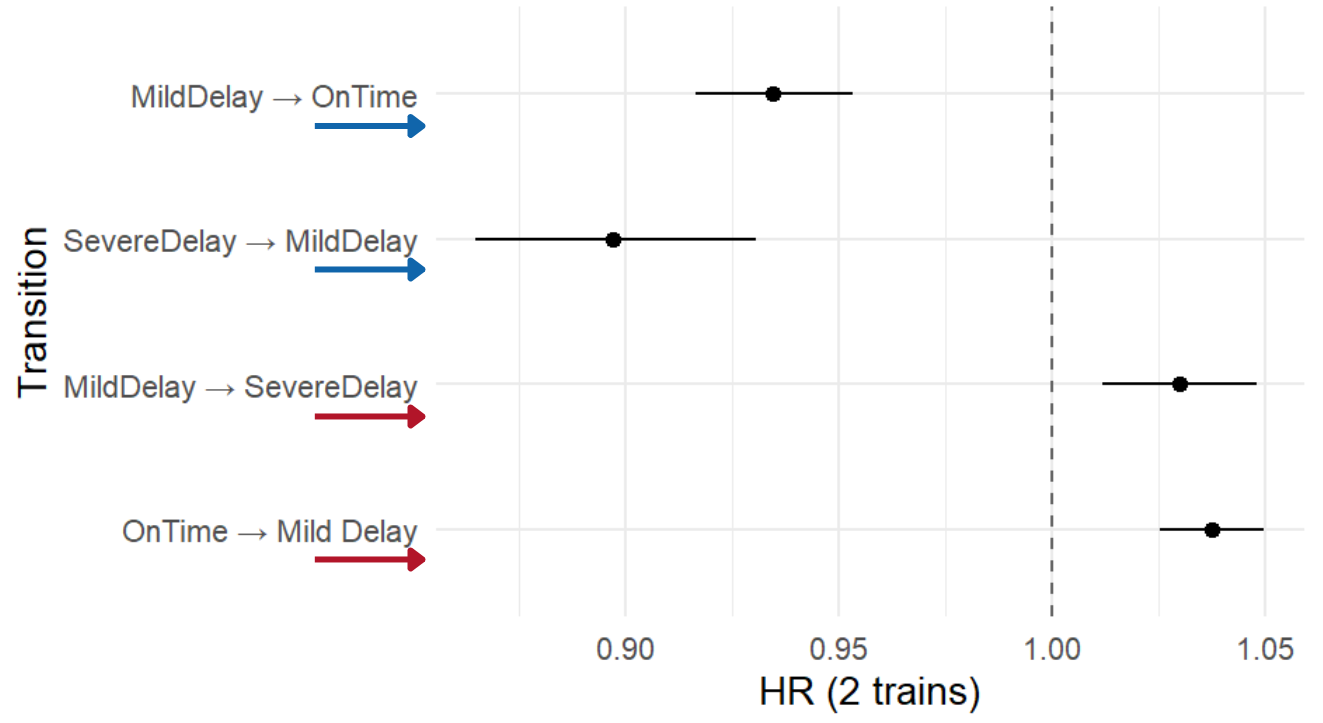}
            \caption{Train saturation}
        \end{subfigure}
        \hspace{0.04\textwidth}
        \begin{subfigure}{0.48\textwidth}
            \centering
            \includegraphics[scale=0.25]{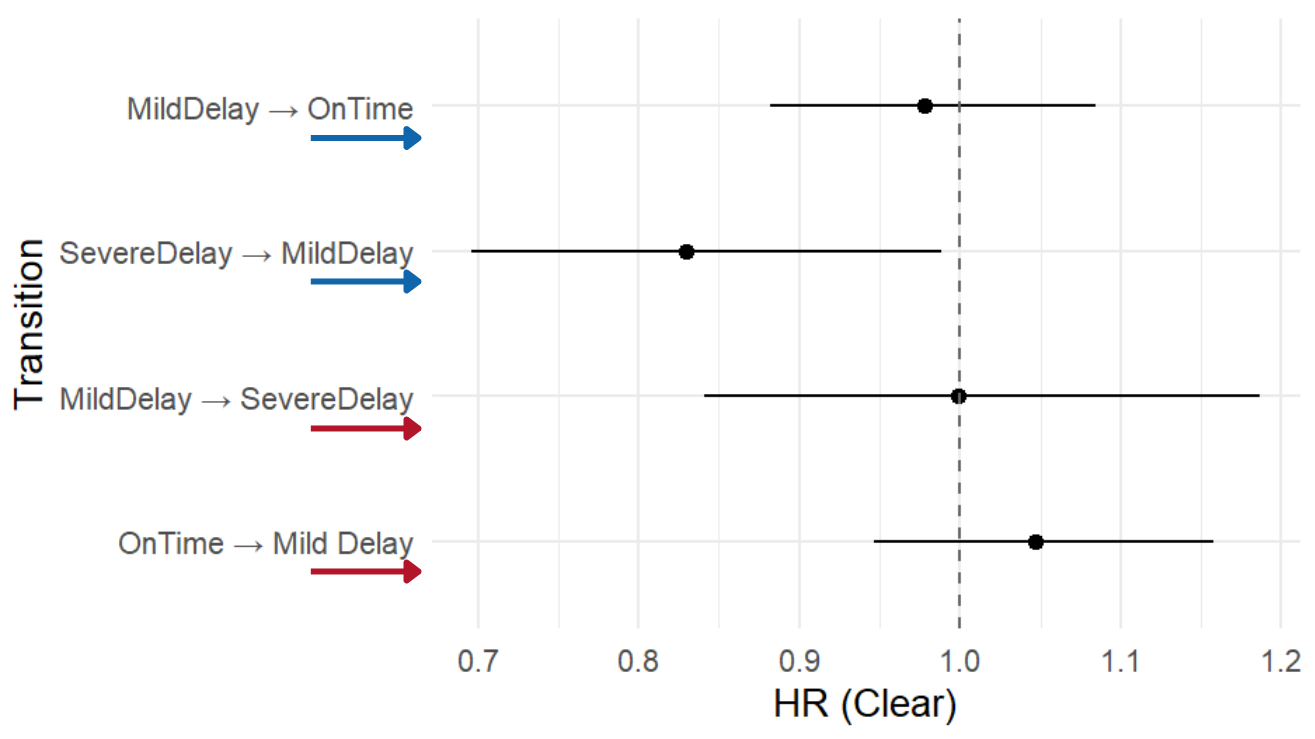}
            \caption{Adverse weather}
        \end{subfigure}
    }
    \vspace{1em} 
    \hspace{-1.5cm}
    \makebox[\textwidth]{%
        \begin{subfigure}{0.48\textwidth}
            \centering
            \includegraphics[scale=0.25]{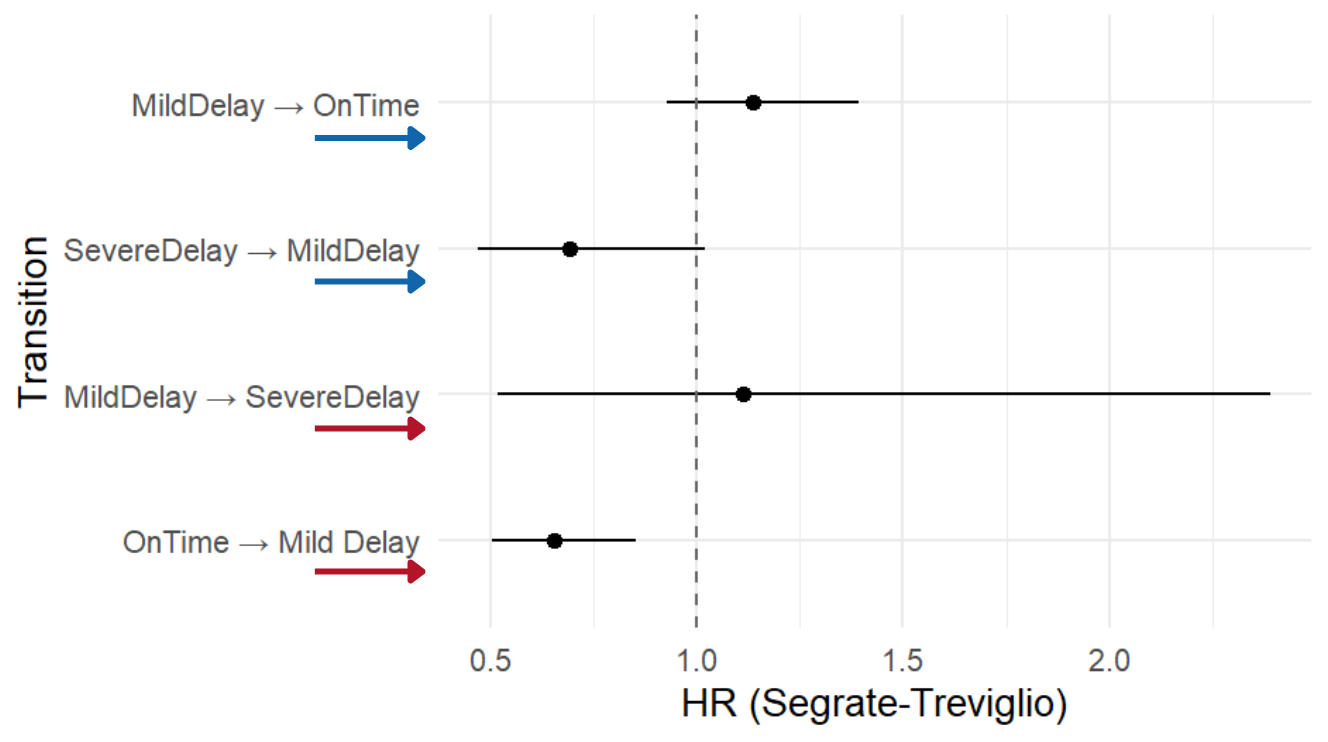}
            \caption{Zone 1: Varese-Gallarate}
        \end{subfigure}
        \hspace{0.04\textwidth} 
        \begin{subfigure}{0.48\textwidth}
            \centering
            \includegraphics[scale=0.25]{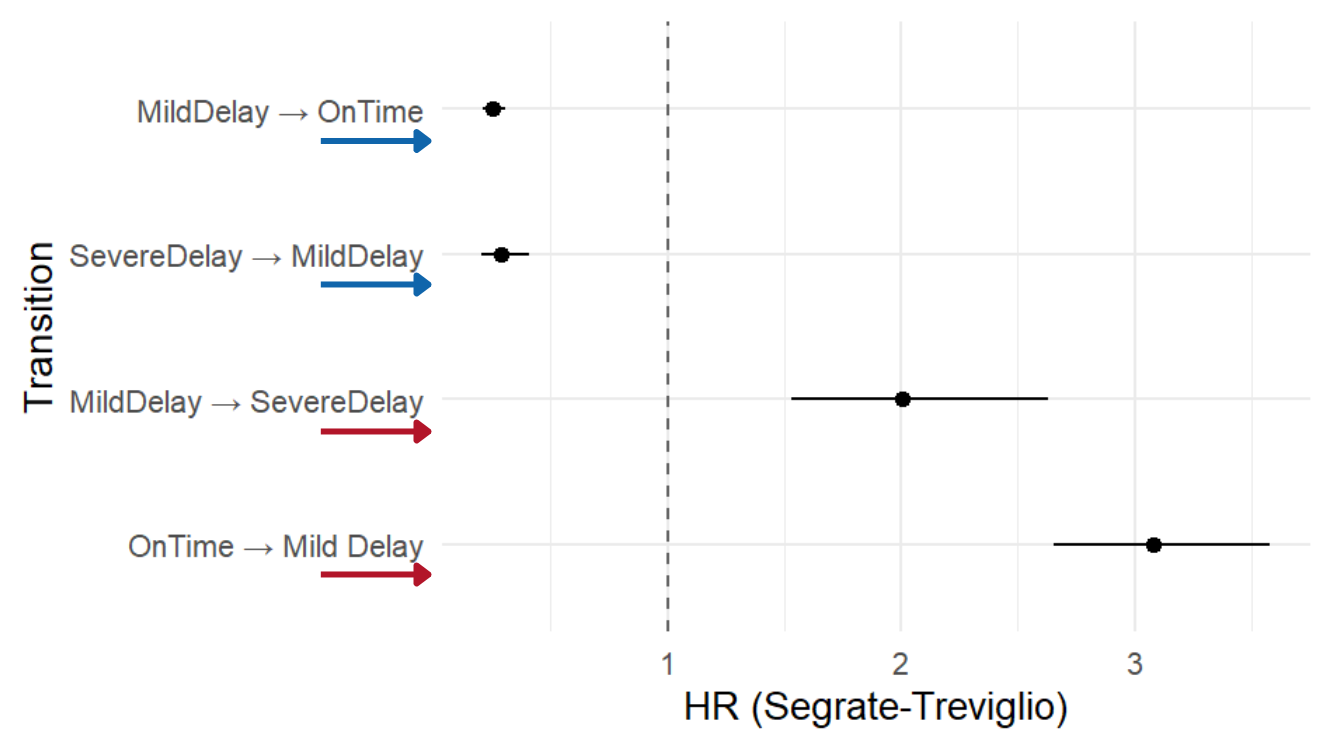}
            \caption{Zone 2: Busto Arsizio-Rho Fiera}
        \end{subfigure}
    }
\end{figure}

\begin{figure}[H]
    \centering
    \hspace{-1.6cm}
    \makebox[\textwidth]{%
        \begin{subfigure}{0.48\textwidth}
            \includegraphics[scale=0.25]{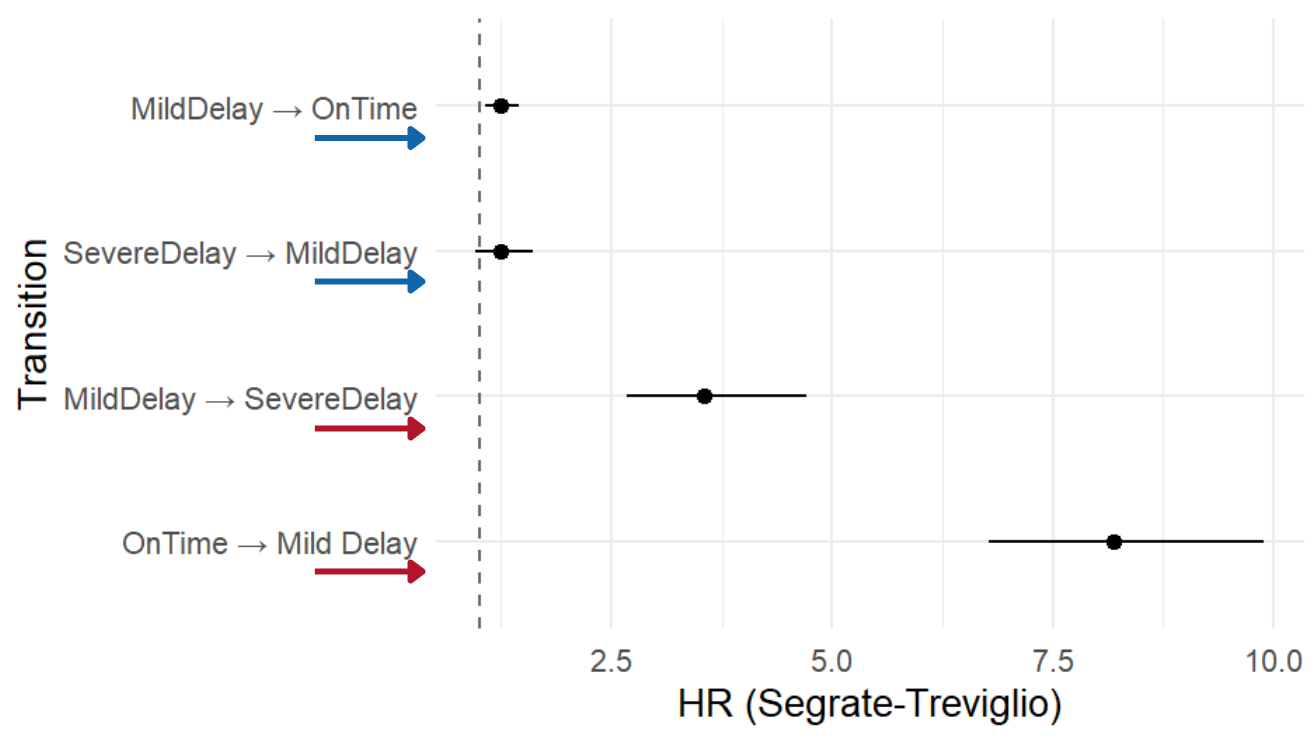}
            \caption*{(g) Zone 3: Milano Certosa-Milano Forlanini}
        \end{subfigure}
        \hspace{0.04\textwidth}
        \begin{subfigure}{0.48\textwidth}
        \end{subfigure}
    }
    \caption{Hazard ratios and 95\% confidence intervals for spatial covariates. Color legend: red = Deteriorating transitions (On Time \(\to\) Mild Delay, Mild Delay \(\to\) Severe Delay); blue = Recovering transitions (Severe Delay \(\to\) Mild Delay, Mild Delay \(\to\) On Time).}
    \label{figB17}
\end{figure}

Table~\ref{tabB2} reports the full Cox model output, for the set of spatial covariates.
\begin{table}[H]
\centering
\begin{tabular}{|l|c|c|c|c|c|c|}
\hline
Trans & Cov & Coef & HR & 95\% CI & Std Err & p-value \\
\hline \hline
0 → 1 & Boarded & 0.0925 & 1.0969 & [1.0280, 1.1705] & 0.0331 & 0.0052 \\
 & Alighted & 0.1326 & 1.1418 & [1.0831, 1.2036] & 0.0269 & 8.28\(\cdot10^{-7}\)\\
 & Train Freq & 0.0369 & 1.0375 & [1.0254, 1.0498] & 0.0060 & 8.21\(\cdot10^{-10}\)\\
 & Weather & 0.0459 & 1.0470 & [0.9465, 1.1582] & 0.0515 & 0.3726\\
 & Zone 1 & -0.4228 & 0.6552 & [0.5040, 0.8518] & 0.1339 & 0.0016\\
 & Zone 2 & 1.1260 & 3.0833 & [2.6545, 3.5813] & 0.0764 &  < 2\(\cdot10^{-16}\)\\
 & Zone 3 & 2.1026 & 8.1877 & [6.7749, 9.8951] & 0.0966 & < 2\(\cdot10^{-16}\)\\
 \hline
 1 → 2 & Boarded & 0.0096 & 1.0097 & [0.9168, 1.1119] & 0.0492 & 0.8452 \\
 & Alighted & 0.0449 & 1.0459 & [0.9890, 1.1061] & 0.0285 & 0.1156\\
 & Train Freq & 0.0294 & 1.0298 & [1.0118, 1.0482] & 0.0090 & 0.0011\\
 & Weather & -0.0008 & 0.9992 & [0.8411, 1.1870] & 0.0879 & 0.9929\\
 & Zone 1 & 0.1061 & 1.1119 & [0.5170, 2.3916] & 0.3908 & 0.7860\\
 & Zone 2 & 0.6964 & 2.0065 & [1.5321, 2.6279] & 0.1376 & 4.21\(\cdot10^{-7}\)\\
 & Zone 3 & 1.2663 & 3.5477 & [2.6692, 4.7154] & 0.1452 & < 2\(\cdot10^{-16}\)\\
 \hline
 2 → 1 & Boarded & -0.0425 & 0.9584 & [0.8274, 1.1100] & 0.0749 & 0.5703 \\
 & Alighted & -0.3141 & 0.7304 & [0.6261, 0.8522] & 0.0787 & 6.55\(\cdot10^{-5}\)\\
 & Train Freq & -0.1083 & 0.8973 & [0.8650, 0.9309] & 0.0187 & 7.14\(\cdot10^{-9}\)\\
 & Weather & -0.1868 & 0.8296 & [0.6961, 0.9888] & 0.0896 & 0.0372\\
 & Zone 1 & -0.3669 & 0.6929 & [0.4705, 1.0203] & 0.1975 & 0.0632\\
 & Zone 2 & -1.2422 & 0.2887 & [0.2049, 0.4070] & 0.1751 & 1.30\(\cdot10^{-12}\)\\
 & Zone 3 & 0.2183 & 1.2440 & [0.6909, 1.6105] & 0.1317 & 0.0975\\
 \hline
 1 → 0 & Boarded & -0.2404 & 0.7863 & [0.7028, 0.8797] & 0.0573 & 2.68\(\cdot10^{-5}\) \\
 & Alighted & -0.4813 & 0.6180 & [0.5455, 0.7000] & 0.0636 & 3.79\(\cdot10^{-14}\)\\
 & Train Freq & -0.0675 & 0.9347 & [0.9165, 0.9533] & 0.0100 & 1.72\(\cdot10^{-11}\)\\
 & Weather & -0.0223 & 0.9779 & [0.8817, 1.0846] & 0.0528 & 0.6725\\
 & Zone 1 & 0.1272 & 1.1357 & [0.9268, 1.3916] & 0.1037 & 0.2199\\
 & Zone 2 & -1.3739 & 0.2531 & [0.2087, 0.3070] & 0.0984 & < 2\(\cdot10^{-16}\)\\
 & Zone 3 & 0.2195 & 1.2455 & [1.0638, 1.4583] & 0.0805 & 0.0064\\
\hline
\end{tabular}
\caption{Complete Cox model results for spatial covariates (Varese–Treviglio). In Trans column states are defined as: 0 = On Time, 1 = Mild Delay, 2 = Severe Delay.}
\label{tabB2}
\end{table}

Figures~\ref{figB24},~\ref{figB26},~\ref{figB28} display the probabilities over a 30-minute horizon, considering the one-at-time variation of passenger load, train frequency and weather condition, respectively. Each plot is obtained by analyzing the particular covariate across its observed range, while keeping all the others constant at their mean levels. These results confirm previous insights from the HRs.
\begin{figure}[H]
    \centering
     \makebox[\textwidth][c]{%
        \includegraphics[width=0.7\textwidth]{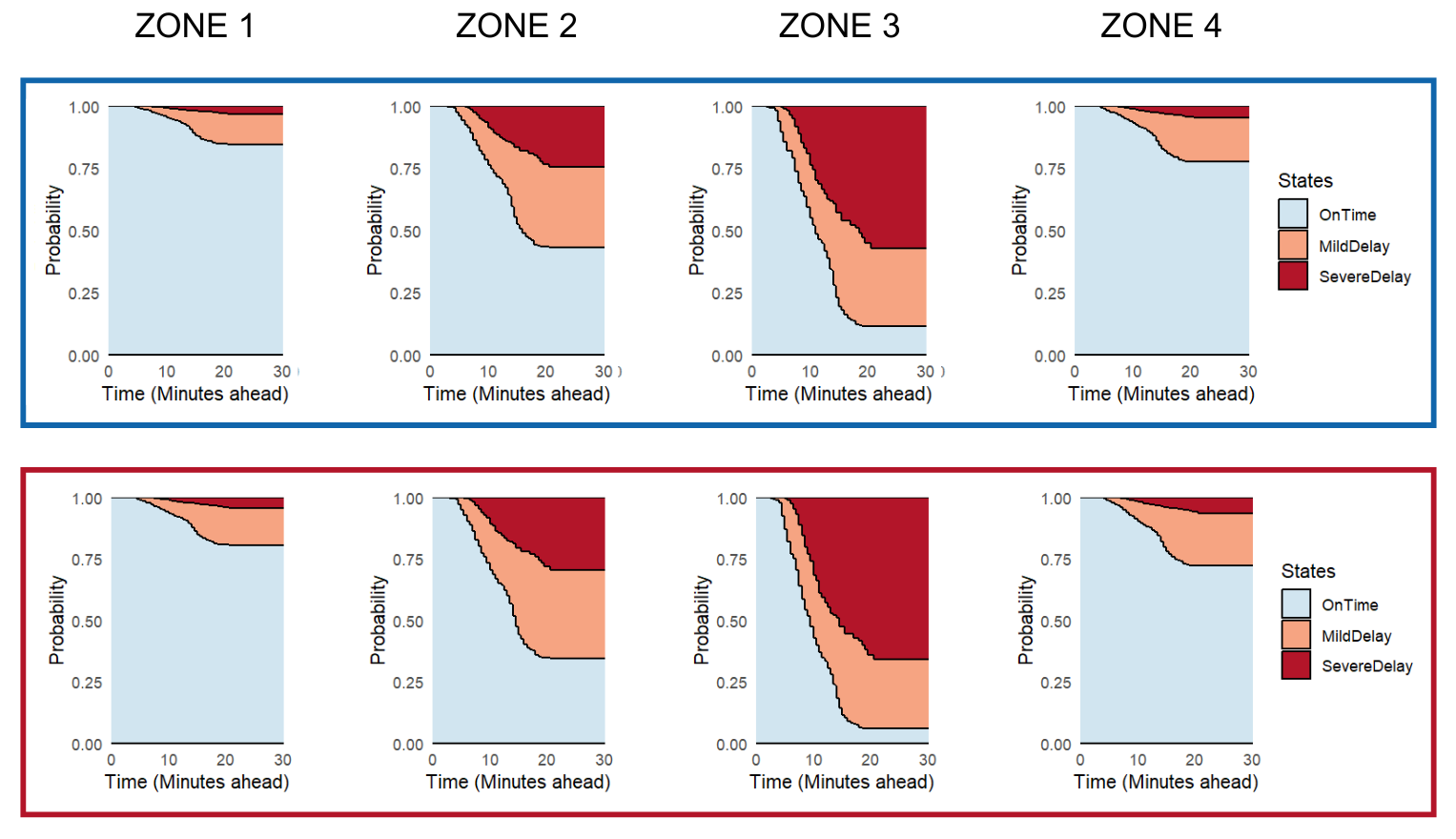}%
    }
    \caption*{(a) From On Time.}
\end{figure}

\begin{figure}[H]
    \centering
    \makebox[\textwidth][c]{%
        \includegraphics[width=0.7\textwidth]{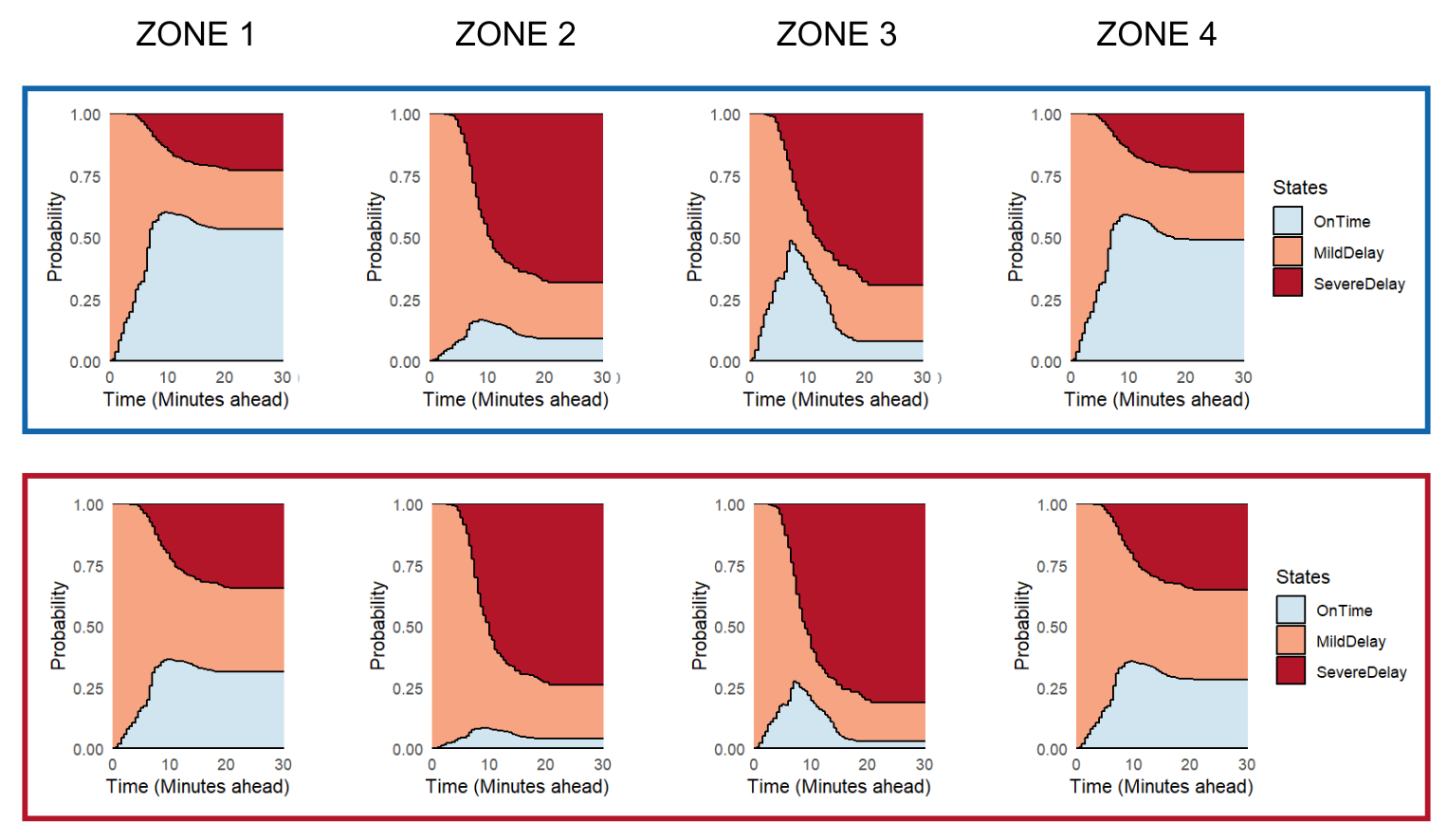}%
    }
    \caption*{(b) From Mild Delay.}
    \caption{Predicted Future State Probabilities for Boarding and Alighting Passengers per Route Section. Blue boxes indicate low passenger volumes (15th percentile), while red boxes correspond to high volumes (85th percentile). Segmentation legend: Zone 1 = Varese-Gallarate, Zone 2 = Busto Arsizio-Rho Fiera, Zone 3 = Milano Certosa-Milano Forlanini, Zone 4 = Segrate-Treviglio.}
    \label{figB24}
\end{figure}

\begin{figure}[H]
    \centering
     \makebox[\textwidth][c]{%
        \includegraphics[width=0.7\textwidth]{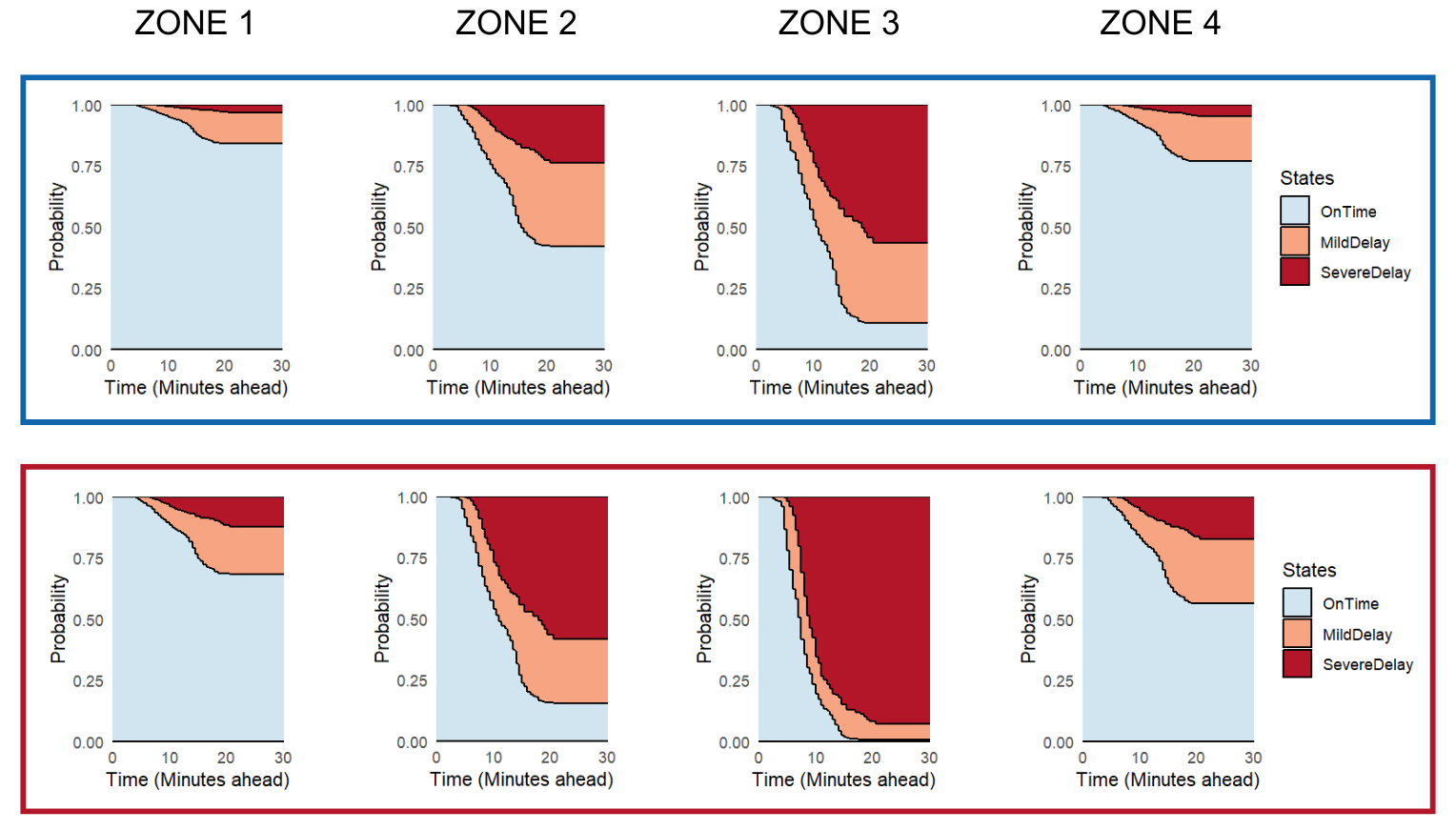}%
    }
    \caption*{(a) From On Time.}
\end{figure}

\begin{figure}[H]
    \centering
    \makebox[\textwidth][c]{%
        \includegraphics[width=0.7\textwidth]{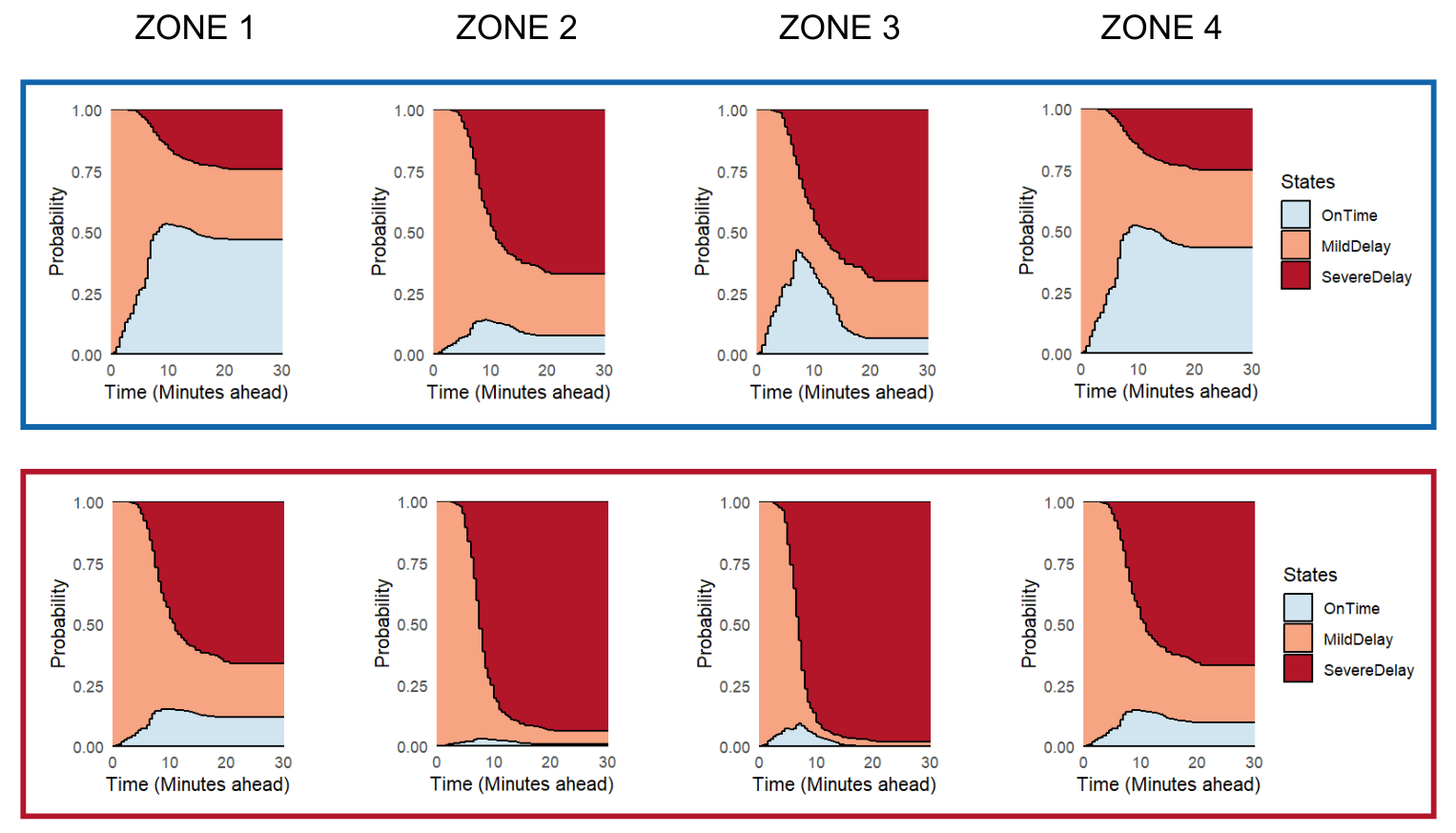}%
    }
    \caption*{(b) From Mild Delay.}
    \caption{Predicted Future State Probabilities for Train Frequency per Route Section. Blue boxes represent low-frequency stations (4 trains/hour), while red boxes correspond to high-frequency contexts (24 trains/hour). Segmentation legend: Zone 1 = Varese-Gallarate, Zone 2 = Busto Arsizio-Rho Fiera, Zone 3 = Milano Certosa-Milano Forlanini, Zone 4 = Segrate-Treviglio.}
    \label{figB26}
\end{figure}

\begin{figure}[H]
    \centering
     \makebox[\textwidth][c]{%
        \includegraphics[width=0.7\textwidth]{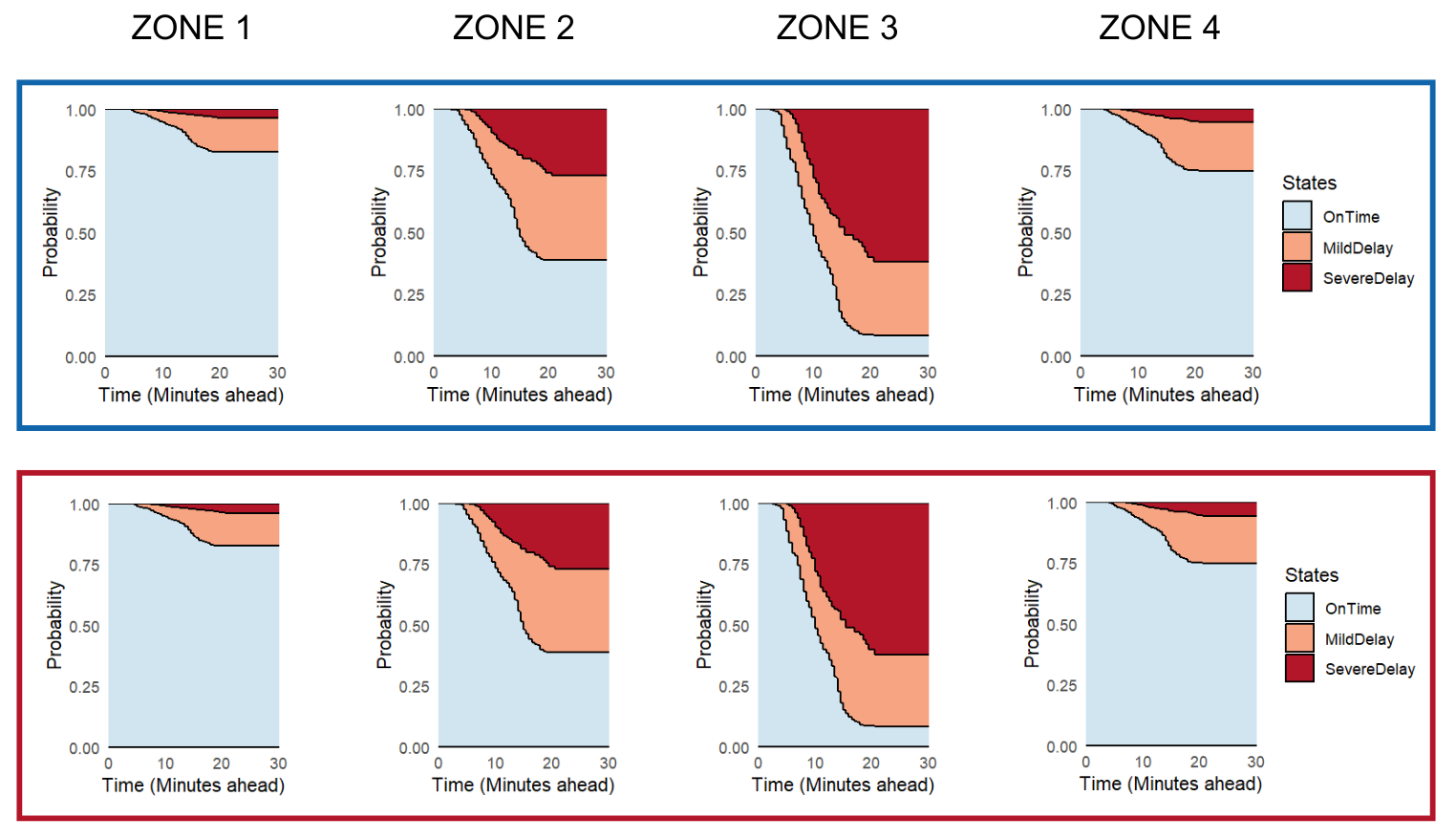}%
    }
    \caption*{(a) From On Time.}
\end{figure}

\begin{figure}[H]
    \centering
    \makebox[\textwidth][c]{%
        \includegraphics[width=0.7\textwidth]{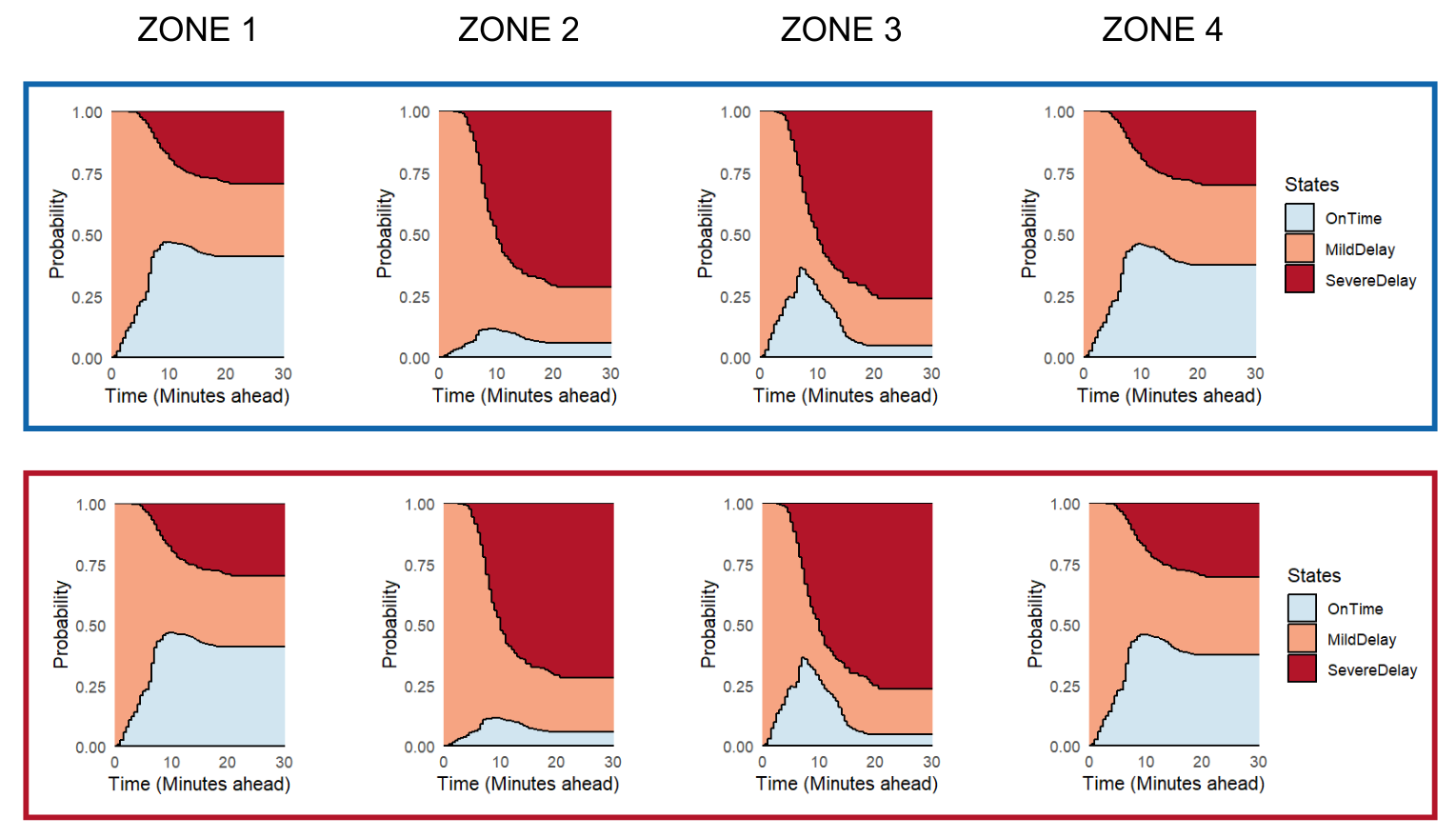}%
    }
    \caption*{(b) From Mild Delay.}
    \caption{Predicted Future State Probabilities for Weather Condition per Route Section. Blue boxes indicate good weather (no rain/fog/thunderstorms), while red boxes represent adverse conditions. Segmentation legend: Zone 1 = Varese-Gallarate, Zone 2 = Busto Arsizio-Rho Fiera, Zone 3 = Milano Certosa-Milano Forlanini, Zone 4 = Segrate-Treviglio.}
    \label{figB28}
\end{figure}

\end{document}